\documentclass[format=acmsmall, screen=true, review=false]{acmart}

\usepackage[utf8]{inputenc}
\usepackage{CONCUR2021}
\usepackage{paralist}

\title{Unifying Model Execution and Deductive Verification with Interaction Trees in Isabelle/HOL}

\acmJournal{TOSEM}

\author{Simon Foster}%
\orcid{0000-0002-9889-9514}%
\affiliation{%
  \institution{University of York}%
  \city{York}%
  \postcode{YO10 5GH}%
  \country{UK}%
}%
\email{simon.foster@york.ac.uk}%

\author{Chung-Kil Hur}%
\orcid{0000-0002-1656-0913}%
\affiliation{%
  \institution{Seoul National University}%
  \city{Seoul}%
  \postcode{08826}%
  \country{South Korea}%
}%
\email{gil.hur@sf.snu.ac.kr}%

\author{Jim Woodcock}%
\orcid{000-0001-7955-2702}%
\affiliation{%
  \institution{Southwest University}%
  \city{Chongqing}%
  \postcode{400751}%
  \country{China}%
}%
\affiliation{%
  \institution{Aarhus University}%
  \city{Aarhua}%
  \postcode{8200 Aarhus N}%
  \country{Denmark}%
}%
\affiliation{%
  \institution{University of York}%
  \city{York}%
  \postcode{YO10 5GH}%
  \country{UK}%
}%
\email{jim.woodcock@york.ac.uk}%

\usepackage{crayola}%

\date{January 2024}

\allowdisplaybreaks

\begin{document}

\begin{abstract}
  
  Model execution allows us to prototype and analyse software engineering models by stepping through their possible behaviours, using techniques like animation and simulation. On the other hand, deductive verification allows us to construct formal proofs demonstrating satisfaction of certain critical properties in support of high-assurance software engineering. To ensure coherent results between execution and proof, we need unifying semantics and automation. In this paper, we mechanise Interaction Trees (ITrees) in Isabelle/HOL to produce an execution and verification framework. ITrees are coinductive structures that allow us to encode infinite labelled transition systems, yet they are inherently executable. We use ITrees to create verification tools for stateful imperative programs, concurrent programs with message passing in the form of the CSP and \Circus languages, and abstract system models in the style of the Z and B methods. We demonstrate how ITrees can account for diverse semantic presentations, such as structural operational semantics, a relational program model, and CSP's failures-divergences trace model. Finally, we demonstrate how ITrees can be executed using the Isabelle code generator to support the animation of models.
\end{abstract}



\maketitle

\section{Introduction}
\label{sec:intro}
Model-based software engineering uses models to produce software with a high level of assurance~\cite{Kolovos2008Epsilon,Feiler2012MBE,Wei2024ACCESS}. Typically, engineers create behavioural models, such as state machines and activity diagrams, which abstractly specify a system's behaviour and can be subjected to prototyping using animation, simulation, and testing. These techniques require that models are executable, so we can step through their behaviour~\cite{Bousse2016Execution,Ciccozzi2019Execution}. When models are accompanied by suitable formal semantics, they can be further subjected to formal verification to ensure they satisfy the requirements in every possible state~\cite{Miyazawa2019-RoboChart}. The models can be refined further to produce verified code and related artefacts, creating high-integrity software with traceable links to the original requirements.

To ensure that these heterogeneous artefacts and analysis results can be applied coherently, there is a need to tie them together using unifying formal semantics to avoid semantic gaps that can introduce weaknesses~\cite{Paige1997FM-IntegratedFormalMethods,Gleirscher2018-NewOpportunitiesIntegrated}. This semantics should allow us to give a formal mathematical meaning to each model used in the development hierarchy and account for the relations between them. It should also support execution to support early-stage prototyping~\cite{Bousse2016Execution}. Moreover, the semantics must have tool support with a high level of automation to minimise the expertise engineers require. Whilst semantic frameworks exist that support such a unification, such as Hoare and He's \textit{Unifying Theories of Programming}~\cite{Hoare&98} (UTP), the models are expressive but not usually executable. For formal methods to be accessible, we, therefore, need to support models that are inherently executable and verifiable.

Theorem proving is a powerful technique for analysing models and code by automating deductive proof steps. Proof assistants like Coq~\cite{Bertot2004Coq}, Isabelle~\cite{Nipkow2002Isabelle}, and Lean~\cite{Demoura2015Lean} provide a flexible foundation for mathematical reasoning. They can be applied to many engineering paradigms, from high-level design models~\cite{foster2020formal,Foster2021-IsaSACM}, potentially including interactions with the physical environment~\cite{FosterMGS21}, to low-level code~\cite{Klein2009}. 
Moreover, theorem provers can support the verification of systems with a very large, or even infinite, state space through symbolic logic techniques and compositional reasoning. However, unlike simulation and model-checking techniques, formal proof typically has a high entry bar and requires significant investment before meaningful results can be obtained. Consequently, to harness the benefits of theorem proving in software engineering, we need to improve access with early-stage prototyping techniques, such as animation\footnote{In this context, animation refers to the interactive probing of a model's behaviour. Simulation is similar but is typically less interactive and, on the whole, more sophisticated. } and simulation, and high levels of proof automation.

The contribution of this article is an Isabelle-based framework to support model-based engineering called Isabelle/ITrees, which implements a variant of the Interaction Tree (ITree) formalism of Xia et al.~\cite{ITrees2019}. ITrees are coinductive data structures that allow us to encode labelled transition systems symbolically as trees of infinite breadth and depth. They intrinsically support mutable states, events, and communication and so can model complex infinite behaviours. Crucially, and uniquely to ITrees, these formal models are both directly executable and subject to verification by proof~\cite{Xia2022ITrees}. ITrees thus provide a natural encoding of operational semantics using coinductive techniques, where we can step through a model's behaviour in terms of its internal steps and external interactions and communications. Though ITrees are intrinsically elementary structures, they have the potential to act as a unifying semantic model for a variety of software engineering artefacts.

Our mechanisation of ITrees in Isabelle/HOL generalises the original work~\cite{ITrees2019} in Coq by using partial functions to model visible events. This allows us to support both external choice and deadlock in the style of the CSP process algebra~\cite{Brookes1984, Hoare85, Roscoe2005-TPC}, along with the algebraic semantics of these operators, which broadens our implementation's application. Our general, highly extensible framework can be applied to software engineering artefacts at various abstraction levels. We use this to provide shallow embeddings of imperative programs, communicating processes, and high-level system models in the style of the Z~\cite{Spivey89} and B~\cite{Abrial96BBook} specification languages. This is supported through results from UTP~\cite{Hoare&98, Foster2020-IsabelleUTP} to unify denotational, operational and axiomatic semantics, and in particular, the UTP semantics for the \Circus process language~\cite{Woodcock2001-Circus, Foster17c, Foster2021-JLAMP}. Our tool benefits from Isabelle's proof tools, notably the \lstinline{sledgehammer} theorem prover integration~\cite{Blanchette2016Hammers}, which we use out-of-box to automate the discharge of verification conditions and other proof obligations. Moreover, we employ Isabelle's code generator~\cite{Haftman2010-CodeGen} to provide execution of programs and animation of high-level models. Our tool thus supports a tight development cycle where animation and verification activities can be intertwined.

The structure of our paper is as follows. In \S\ref{sec:itrees}, we show how ITrees are mechanised in Isabelle/HOL, including the core operators. We show how to derive structural operational semantics from ITrees, characterise weak bisimulation, which allows the abstraction of silent events, and provide theorems for reasoning about process iteration using chains. In \S\ref{sec:rel}, we show how to model and verify imperative programs using ITrees, demonstrate a link with our previous UTP-based relation semantics, and provide program verification support using Hoare logic. In \S\ref{sec:csp-circus}, we show how deterministic CSP and \Circus processes can be semantically embedded into ITrees, including operators like external choice and parallel composition. We also link ITrees with the standard failures-divergences semantic model for CSP, which justifies their integration with other CSP-based techniques. In \S\ref{sec:animation}, we show how the code generator can be used to generate animations. In \S\ref{sec:zmachines}, we apply our library to develop a simple tool for modelling systems, similar to the B-method~\cite{Abrial96BBook}. In \S\ref{sec:related}, we consider related work, and in \S\ref{sec:concl}, we conclude.

This paper extends our previous CONCUR 2021 paper~\cite{Foster2021-ITrees}. We add results in the new section on imperative programming (\S\ref{sec:rel}), including total correctness Hoare logic and UTP-style predicative semantics, a new section on modelling with Z-Machines (\S\ref{sec:zmachines}), new theorems on iteration chains (\S\ref{sec:itrees}), and additional narrative and more minor results throughout. All our results have been mechanised and can be found in the accompanying repository\footnote{\url{https://github.com/isabelle-utp/interaction-trees}}, and clickable icon links \isalogo\xspace next to each specific result.

\paragraph{Notation} Our presentation uses both textbook-style notation and Isabelle code, though we generally prefer the former. This mixture is unavoidable, as the Isabelle code, though ultimately the single source of truth, is often more pedantic than necessary for a human reader and less accessible. We largely restrict Isabelle's code to modelling and verification examples to support the use of our tools. The reader interested in how the textbook mathematics is mechanised can follow the Isabelle links (\isalogo).

\section{Background}
\label{sec:background}
This section introduces the foundational concepts used in this paper: Isabelle/HOL and the \Circus language. \Circus is used to motivate the value of ITrees in providing formal semantics for process algebraic languages. We also use the Z mathematical toolkit in our semantic definitions, which is used in \Circus.

\subsection{Isabelle/HOL} 

Isabelle/HOL~\cite{Nipkow2002Isabelle}, at its core, is a proof assistant for Higher Order Logic (HOL)\footnote{Isabelle/HOL website: \url{https://isabelle.in.tum.de/}}. It implements a Gentzen-style natural deduction system, which can be used to prove or falsify the validity of arguments formalised using predicate logic. The language of HOL is a strongly typed polymorphic $\lambda$-calculus, which can be used to formalise mathematical theories in a functional programming style. In particular, Isabelle/HOL provides a typed set theory, arithmetic theories (natural numbers, integers, real numbers, etc.), and various data structures, such as lists and records. As in functional programming languages, programs can be specified using algebraic data types and recursive functions, with termination checks provided. These features give an expressive and extensible mathematical language in which various programming and modelling notations can be described~\cite{Wenzel2007}.

The development of theories in Isabelle is centred around \emph{theory documents}, which are used for modelling and proof. Theory documents (extension \texttt{.thy}) are written using the Isar language~\cite{Wenzel2019-Isar}, and consist of a sequence of commands manipulating Isabelle's state. Such commands may be used to define a function or start a proof, for example. 
The document model has two levels of syntax: (1) outer-syntax, which gives the syntax to individual commands, and (2) inner-syntax, which gives syntax to terms of the logic in typed $\lambda$-calculus. An example definition command is given below:

\begin{lstlisting}
definition square :: "nat ~⇒~ nat" where
"square x = x * x"
\end{lstlisting}
\noindent
The command begins with a major keyword (\lstinline{definition}), which is highlighted, followed by a type declaration for a new constant, \lstinline{square}, which is a total function from natural numbers to natural numbers (\texttt{nat} $\Rightarrow$ \texttt{nat}). Following the type declaration, there is a minor keyword (\lstinline{where}) and then the definitional equation for the function. Speech marks delimit this definitional equation since it is inner-syntax formed using the term language. The document model of Isabelle is extensible so that new commands can be implemented for bespoke modelling tasks using the meta-language Isabelle/ML~\cite{Wenzel2007}.

Several facilities complement Isabelle's modelling features for automating proof. Isabelle provides a simplifier (\lstinline{simp}) for equational rewriting of terms and a classical reasoner (\lstinline{blast}), which implements the tableaux method for automating natural deduction. Additionally, there is a resolution prover (\lstinline{metis}) for first-order predicate calculus and access to several SMT solvers, such as CVC4 and Z3, in the \lstinline{smt} method. These proof methods can be coordinated using the \lstinline{sledgehammer} tool~\cite{Blanchette2016Hammers}, which constructs proofs using external tools.

For example, with the definition above, we can pose a binomial expansion theorem:

\begin{lstlisting}
theorem square_plus: "square(x+y) = square x + square y + 2*x*y"
\end{lstlisting}
The \lstinline{theorem} command asserts that the given property can be derived by proof. In this case, we give the posed theorem a name (\texttt{square\_plus}) and the proposition that we wish to prove: \texttt{square(x+y) = square x + square y + 2*x*y}. Implicitly, this means that this equational property is valid for any natural numbers \texttt{x} and \texttt{y}. In order for Isabelle to accept this as a theorem, we need to provide a proof. In this case, the proof can be provided automatically with \lstinline{sledgehammer}. Running this calls several external automated theorem provers, and CVC4 comes back with a proof. We can then complete the proof as follows:

\begin{lstlisting}
theorem square_plus: "square(x+y) = square x + square y + 2*x*y"
  by (metis power2_eq_square power2_sum square_def)
\end{lstlisting}
The command \lstinline{by} states that the property can be proved by the following proof method. The \lstinline{sledgehammer} supplied proof uses the \lstinline{metis} resolution prover, with two theorems from the HOL theorem library, and the definitional equation for \texttt{square}. Isabelle accepts the proof, and so the property is certified as a theorem for use in further proof.

The combination of an expressive and rigorous mathematical language with extensible proof automation make Isabelle ideally suited to formal verification. This requires that the semantics of the target language first be formalised as an Isabelle theory package and a suitable proof calculus (such as Hoare logic~\cite{Nipkow2002HoareLogic}) be provided to form specifications and verify programs. Isabelle also provides a code generator~\cite{Haftman2010-CodeGen,Haftmann2013-DataRefinement} for mathematical programs, which can be used to automate code production from verified artefacts.

\subsection{Circus and Z}
\label{sec:circus}

\textsf{\textsl{Circus}}~\cite{Woodcock2001-Circus,Oliveira&09} is a formal language for modelling imperative and concurrent systems. It combines the communication primitives from the CSP process algebra with imperative programming primitives from Dijkstra's guarded command language (GCL) and rich state modelling facilities as provided by the Z notation~\cite{Spivey89}.

Z is a formal language for specifying software using set theory and relational calculus~\cite{Spivey89,Woodcock96-UsingZ}. CSP is a language for modelling concurrent systems, such as protocols. It provides several modelling primitives, including.

\begin{itemize}
\item event prefix ($a \then P$): perform event $a$ and then enable $P$;
\item guard ($B \guard P$): enable $P$ only when condition $B$ is true;
\item external choice ($P \extchoice Q$): allow the environment to choose $P$ or $Q$;
\item parallel composition ($P \parallel[A] Q$): run $P$ and $Q$ in parallel synchronising on events in $A$;
\item hiding ($P \hide A$): internalise events in set $A$.
\end{itemize}
Events are typically input events $a?x \then P(x)$ or output events $b!v \then Q$, where $a$ and $b$ are channels carrying typed data. The standard semantics for CSP is called the failures-divergences model~\cite{Roscoe2005-TPC}, a denotational semantics based on traces, which we cover in \S\ref{sec:sem-links}.

In addition to these CSP operators, \Circus also contains typical imperative programming operators from GCL like assignment ($x := e$), sequential composition ($P \relsemi Q$), and iteration ($\skey{while}~B~\skey{do}~C~\skey{od}$). From Z, it gains a mathematical toolkit, including data structures like sets, partial functions ($A \pfun B$), finite functions ($A \ffun B$), and sequences (lists), and also the ability to form abstract data types using Z schemas.

We also use the Z mathematical toolkit for partial functions in our semantic definitions. We can specify partial functions using $\lambda x \in A @ f(x)$, which restricts a function $f$ to the domain $A$\footnote{This is distinguished from the total function notation $\lambda x.\, g(x)$ by a different separator ($\bullet$)}. We can calculate the domain of a partial function $f$ with $\dom(f)$. An empty partial function $\{\mapsto\}$ has an empty domain. We also use the domain restriction ($\lhd$) and override operators ($\oplus$) from the Z mathematical toolkit, which have the following definitions:
\begin{align*}
    A \lhd f &\defs (\lambda x\in A \cap \dom(f) @ f(x)) \\
    f \oplus g &\defs (\lambda x\in \dom(f) \cup \dom(g) @ if ~ x \in \dom(g) ~then~ g(x) ~else~ f(x))
\end{align*}
\noindent
We have implemented the Z mathematical toolkit in an Isabelle library as a hierarchy of types\footnote{Z-Toolkit library: \url{https://github.com/isabelle-utp/Z_Toolkit}.}. With the associated theorems, we can use Isabelle's simplifier to automate the calculation of a partial function's domain and other properties.

As an example \Circus process, we formalise a simple reactive buffer. We introduce three channels: $Input$ to accept a new input, $Output$ to offer an output, and $State$ to show the current state of the buffer. We consider a buffer containing a sequence of natural numbers $\nat$ for simplicity. We also introduce a single variable, $buf$, which stores the values present in the buffer. The reactive behaviour of the buffer is then specified below:

\begin{example}[Unbounded Buffer in \Circus] \label{ex:buffer}
\begin{align*}
buf := [] &\relsemi \skey{while}~true~\skey{do} \\
&     \qquad \quad Input?(i) \then buf := buf \append [i] \\
&     \qquad \extchoice (length(buf) > 0) \guard Output!(hd~buf) \then buf := tl~buf \\
&     \qquad \extchoice State!(buf) \then \skey{Skip} \\
& ~~\skey{od}
\end{align*}
\end{example}
\noindent The buffer enters a reactive infinite loop after initially setting the buffer to be an empty list ($[]$). The buffer provides three options using the external choice operator ($\extchoice$). It can accept an input over $Input$, extending the buffer using the list append operator ($xs \append ys$). If the buffer is not empty (its length is non-zero), it can offer the head (i.e. first element) of the buffer over $Output$ and contract the buffer with the \textit{tl} function, which removes the head. Finally, it can display the current state of the buffer over the $State$ channel.

\section{Interaction Trees and Operational Semantics}
\label{sec:itrees}
This section introduces Interaction Trees (ITrees), develops the main theory in Isabelle/HOL, derives operational semantics, and provides several novel results. ITrees were originally mechanised in Coq by Xia et al.~\cite{ITrees2019}. Our mechanisation in Isabelle/HOL brings unique advantages, including a flexible front-end syntax, automated proof tools, and code generation to several languages.

\subsection{Interaction Trees as Codatatypes}

\begin{figure}
\xymatrix@C=10ex@R=4ex{
  & & \text{(4)}\ar[r]^{\tau} & \text{(5)} \ar[r]^{Input.y} \ar[rd]^{Output.x} \ar[rdd]_{State.[x]} & \text{(6)} \ar[r]^{\tau} & \text{(7)} \ar[r]^{Input.z} \ar[rd]^{Output.x} \ar[rdd]_{State.[x,y]} & \\
  & & & & \text{(8)} \ar[rd]^{\tau} & & \\
  \text (1) \ar[r]^{\tau} & \text{(2)} \ar[ruu]^{Input.x} \ar[r]_{State.[]} & \text{(3)} \ar[r]^{\tau} & \text{(2)} & & \text{(2)} &  \\
}

\caption{An ITree fragment for the buffer example (approximate)}
\label{fig:exitree}
\Description[%
The image shows an ITree fragment diagram for the buffer example.%
]{%
  The image shows an ITree fragment diagram for the buffer example. The nodes, represented by numbers in circles (1 to 8), correspond to states or stages in a process, while arrows between the nodes represent transitions occurring after a silent event labeled with the Greek letter $\tau$ (tau). Additionally, certain nodes contain specific labels, including ``Input.x'', ``Input.y'', ``Input.z'', ``Output.x'', and various states such as ``State.[x]'' and ``State.[x,y]''.
  \begin{itemize}
  \item Node (1) transitions to Node (2) after a $\tau$.
  \item Node (2) transitions to Node (3) with another $\tau$ and also to Node (4) based on the `Input.x`.
  \item Node (4) transitions to Node (5) after a $\tau$ and responds to `Input.x`.
  \item Node (5) and Node (6) involve states and input variables like `State.[x]`, `Input.y`, and `Output.x`.
  \item Node (6) transitions to Node (7) based on `Input.z` and `Output.x`, while Node (7) connects to the final Node (8).
  \item Node (8) also involves outputs and state variables like `State.[x,y]` and transitions back to Node (2) after a $\tau$.
  \end{itemize}
  The diagram demonstrates interactions between system inputs, states, and outputs in a buffered process, visualizing the flow of data and state transitions over time.
}%

\end{figure}
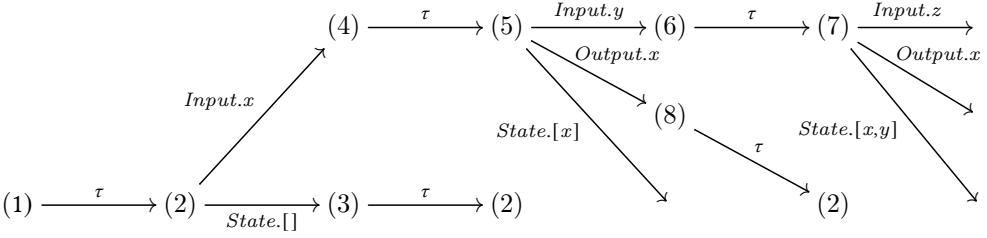

ITrees are potentially infinite trees whose edges are decorated with events, representing the interactions between a process and its environment. For intuition, an example ITree is shown in Figure~\ref{fig:exitree} for the buffer in Example~\ref{ex:buffer}. The nodes are labelled with numbers for reference, and the edges with events, including visible events, such as $Input.x$, and invisible events ($\tau$). 

From the initial node (1), a single $\tau$ event is possible, which corresponds to assigning $[]$ to the $buf$ state variable ($buf := []$). Two visible events are presented to the environment, $Input.x$ and $State.[]$. The latter event, $State.[]$ indicates that the buffer is empty, so an $Output$ event is unavailable. The former event, $Input.x$, corresponds to an infinite family of events for each possible value the channel can carry, such as $x = 0$, $x = 1$, $x = 3$ and so on. We can describe such infinite families in Isabelle/HOL symbolically as a term containing a free variable ($x$), such that ITrees can have infinite breadth.

If an input is received, we transition to node (4), from which a single $\tau$ event occurs, which corresponds to the assignment appending $x$ to the buffer ($buf := buf \append [x]$). From node (5), three events are possible: $Input.y$, $Output.x$, and $State.[x]$. At this point, the buffer contains a single value $x$, which we can output, or input another value $y$. Again, $x$ and $y$ are families of possible values carried by channels $Input$ and $Output$. If another value $y$ is input, the buffer is updated accordingly, leading to node (7). The tree continues in this manner and thus has an infinite depth. It can alternatively be considered as an unfolding of a labelled transition system.

We now describe the type in Isabelle that allows us to denote ITrees formally. ITrees are parametrised over two sorts (types): $E$ of events and $R$ of return values (or states). There are three possible interactions: (1) termination, returning a value in $R$; (2) an internal event ($\tau$) followed by a successor ITree; or (3) a choice between several visible events. In Isabelle/HOL, we encode ITrees using a codatatype~\cite{Blanchette2014BNF, Blanchette2017Coinductive}. A codatatype is similar to an algebraic datatype, having several disjoint constructors. However, the crucial difference is that whereas elements of a datatype are finite, elements of a codatatype may be infinite.

\begin{definition}[Interaction Tree Codatatype] $ $ \isalink{https://github.com/isabelle-utp/interaction-trees/blob/ff9f73f98c653b265bd9da55689715cf973499c1/Interaction_Trees.thy\#L21}

\begin{alltt}
  \isakwmaj{codatatype} ('e, 'r) itree = 
    Ret 'r | Sil "('e, 'r) itree" |  Vis "'e \(\pfun\) ('e, 'r) itree"
\end{alltt}
\vspace{-1ex}
\end{definition}
\noindent The \lstinline{codatatype} command creates a type called \texttt{itree}, with two type parameters \texttt{'e} and \texttt{'r}, and three constructors. Type parameters \texttt{'e} and \texttt{'r} encode the sorts $E$ and $R$. Constructor $\skey{Ret}$ represents a return value, and \skey{Sil} is an internal event that evolves to a further ITree. A visible event choice ($\skey{Vis}$) is represented by a partial function ($A \pfun B$) from events to ITrees, with a potentially infinite domain. For example, in Figure~\ref{fig:exitree} at node (2), a visible event choice is presented whose domain is $\{Input.x ~|~ x \in \nat\} \cup \{State.[]\}$.

The representation of visible events is the main deviation from ITrees in Coq~\cite{ITrees2019}, which has visible events composed of an output to the environment, followed by the answer. The benefit of using a partial function is to allow a straightforward encoding of deadlock and external choice, where the ITree offers several events to the environment (for a more detailed comparison, see \S\ref{sec:related}). Moreover, a side effect of this design decision is that we only need rank-1 polymorphism for the encoding, which makes the development in Isabelle possible.

We use the notation $\lambda c.x ~|~ B(x) ~\bullet~ P(x)$ for a partial function which pattern matches on events over channel $c$, and whose parameters $x$ also satisfy predicate $B$. With this notation, we can describe the main choice block of the buffer example:

\begin{example}[Buffer: Single Step ITree] $ $ \label{ex:bufbody}

$$BufBody(buf) \defs
\Vis~ \left(\begin{array}{l}
 (\lambda Input.x \bullet \Ret (buf \append [x])) \\
\mathop{\oplus}~ (\lambda Output.v ~|~ \# buf > 0 \land v = hd(buf) \bullet \Ret (tl(buf))) \\
\mathop{\oplus}~ (\lambda State.s ~|~ s = buf \bullet \Ret(buf))
\end{array}\right)
$$
\end{example}
\noindent This constructs a visible event choice over a partial function composed of three parts using the override operator ($\oplus$). Here, parameter $buf$ is a list of natural numbers, which is the current contents of the buffer. The first function accepts a value $x$ over channel $Input$ and returns the buffer with $x$ appended. The second function allows us to output a value $v$ over $Output$, but only when the buffer is non-empty ($\# buf > 0$) and $v$ is the head of the buffer. This being the case, the function returns the contracted buffer ($tl(buf)$), which effectively updates the state. The third function allows us to advertise the current values in the buffer but leaves the buffer unchanged. Since the three functions have disjoint domains, they can be commuted over the override operator. Such an example is encoded more naturally using the \Circus operators, but we defer denoting these to \S\ref{sec:csp-circus}.

We sometimes use $\ret{v}$ to denote $\Ret~v$, $\tau P$ to denote $\Sil~P$, and $\vbar\, e\!\in\!E \then P(e)$ to denote $\Vis~(\lambda e \in E @ P(e))$, which are more concise and suggestive of their process algebra equivalents. We write $e_1 \then P_1 \mathop{\vbar} \cdots \mathop{\vbar} e_n \then P_n$ for an enumerated choice with $E = \{e_1, \cdots, e_n\}$. We use $\tau^n P$ for an ITree prefixed by $n \in \nat$ internal events. We define $\skey{stop} \defs \Vis~\{\mapsto\}$, a deadlock situation where no event is possible. An example is $a \then \tau(\ret{x}) \mathop{\vbar} b \then \skey{stop}$, which can either perform the event $a$ followed by a $\tau$, and then terminate returning $x$, or perform the event $b$ and then deadlock.

We call an ITree \skey{unstable} if it has the form $\tau P$, and \skey{stable} otherwise. The ITree in Figure~\ref{fig:exitree} is stable in nodes (2), (5), and (7) and unstable in all other numbered nodes. An ITree stabilises, written $\stabilises{P}$, if it becomes stable after a finite sequence of $\tau$ events, that is $\exists n~P' @ P = \tau^n P' \land \skey{stable}(P')$. An ITree that does not stabilise is divergent, written $\divergent{P} \defs \neg (P \Downarrow)$. A divergent ITree can only perform an infinite sequence of internal $\tau$ events, and so corresponds to an infinite loop.

\subsection{ITree Combinators}

Using the constructors mentioned so far, we can only specify ITrees of finite depth. Infinite ITrees can be specified using primitive corecursion~\cite{Blanchette2014BNF}, which is the dual of recursion but allows non-terminating productive definitions. We define such an ITree below:

\begin{definition}[Divergent ITree] $ $ \label{def:div-run} \isalink{https://github.com/isabelle-utp/interaction-trees/blob/ff9f73f98c653b265bd9da55689715cf973499c1/ITree_Divergence.thy\#L77}
\begin{alltt}
  \isakwmaj{primcorec} div :: "('e, 's) itree" \isakwmin{where} "div = \(\tau\) div"
\end{alltt}
\end{definition}
\noindent The \isakwmaj{primcorec} command creates a typed constant which obeys several corecursive equations (following the \isakwmin{where}). Each definition requires that a constructor guards every corecursive call on the right-hand side of an equation, ensuring it is productive. This means that, though the definition does not terminate, it is always possible to strip off the next constructor.

ITree \skey{div} represents the divergent ITree that does not terminate and only performs internal activity. Though self-referential and non-terminating, its definition is productive since we can always remove the next $\tau$. Since \skey{div} never stabilises, it is divergent, $\divergent{\skey{div}}$. Moreover, we can show that $\skey{div}$ is the unique fixed-point of $\tau^{n+1}$ for any $n \in \nat$, $\tau^{n+1} P = P \iff P = \skey{div}$, and consequently $\skey{div}$ is the only divergent ITree: $\divergent{P} \implies P = \skey{div}$.

We call an ITree \skey{pure} if it has the form of $\tau^n P$, where $P$ has the form of either $\ret{x}$, $\skey{stop}$, or $\skey{div}$. The external environment cannot influence a pure ITree, which must either terminate, deadlock (abort) or diverge. Intuitively, a pure ITree can perform all the computation it wants without needing to communicate. We give another infinite ITree below:

\begin{definition}[Run ITree] $ $

\begin{alltt}
  \isakwmaj{primcorec} run :: "'e set \(\Rightarrow\) ('e, 's) itree" \isakwmin{where}
  "run E = Vis (map_pfun (\(\lambda\) x. run E) (pId_on E))"
\end{alltt}
\end{definition}

\noindent Here, the type \lstinline{'e set} denotes the set of all subsets of type \lstinline{'e}. ITree $\skey{run}~E$ can repeatedly perform any $e \in E$ without ceasing. It has the equivalent definition of $\skey{run}~E \defs \vbar e \in E \then \skey{run}~E$, an ITree that can repeatedly choose any event in $E$. It also has the case $\skey{run}~\emptyset = \skey{stop}$. The formulation above uses the function $\texttt{map\_pfun} :: (\texttt{'b}\!\Rightarrow\!\texttt{'c}) \Rightarrow (\texttt{'a}\!\pfun\!\texttt{'b}) \Rightarrow (\texttt{'a}\!\pfun\!\texttt{'c})$ which maps a total function over every output of a partial function. The function \texttt{pId\_on E} is the identity partial function with domain \texttt{E}. This formulation is required to satisfy the syntactic guardedness requirements. For the sake of readability, we omit these details in the following definitions.

Corecursive definitions can have several equations ordered by priority, like a recursive function. Using such a set of equations, we specify a monadic bind operator for ITrees~\cite{ITrees2019}.

\begin{definition}[Interaction Tree Bind] \label{def:mbind} We fix $P, P' : (E, R)\skey{itree}$,
  $K : R \to (E, S)\skey{itree}$, $r : R$, and $F : E \pfun (E, S)\skey{itree}$. Then, $P \mbind K$
  is defined corecursively by the equations \isalink{https://github.com/isabelle-utp/interaction-trees/blob/ff9f73f98c653b265bd9da55689715cf973499c1/Interaction_Trees.thy\#L96}
  \begin{align*}
   \ret{r} \mbind K &= K~r \\ 
   \tau P' \mbind K &= \tau(P' \mbind K) \\ 
   \Vis~F \mbind K &= \Vis~(\lambda e \in \dom(F) @ F(e) \mbind K)
  \end{align*}
\end{definition}
\noindent The intuition of $P \mbind K$ is to execute $P$, and whenever it terminates ($\tick_r$), pass the given value $r$ on to the continuation $K$, yielding $K~r$. If the first ITree can perform a $\tau$ event, this is performed first, and the remaining ITree is bound to $K$. If the first ITree can perform a visible event $e \in \dom(F)$, then we perform $e$, pass this on to $F$, and bind the result to $K$.

We term $K$ a Kleisli tree~\cite{ITrees2019}, or KTree since it is a Kleisli lifting of an ITree. KTrees are important for defining processes that depend on a previous state. For this, we define the type synonym $(E, S)\skey{htree} \defs (S \Rightarrow (E, S)\skey{itree})$ for a homogeneous KTree. For example, $BufBody$ is homogeneous Kleisli tree of type $\skey{int}~\skey{list} \Rightarrow (E, \skey{int}~\skey{list})\skey{itree}$. Intuitively, the construction $K(s)$ can be read as ``the Klesli tree $K$ started in the initial state $s$''. We define the Kleisli composition operator $P \fatsemi Q \defs (\lambda x.\, P~x \mbind Q)$, symbolised because it is used as a sequential composition. Bind satisfies several algebraic laws:

\begin{theorem}[Interaction Tree Bind Laws] \label{thm:bind-laws} $ $ \isalink{https://github.com/isabelle-utp/interaction-trees/blob/ff9f73f98c653b265bd9da55689715cf973499c1/Interaction_Trees.thy\#L178} 
  
    \begin{minipage}{.5\linewidth}
    \begin{align*}
        \Ret~x \mbind K &= K~x \\
        P \mbind \Ret &= P \\
        P \mbind (Q \fatsemi R) &= (P \mbind Q) \mbind R \\
        \skey{div} \mbind K &= \skey{div}
    \end{align*}
    \end{minipage}
    \begin{minipage}{.5\linewidth}
    \begin{align*}
        \Ret \fatsemi K &= K \\
        K \fatsemi \Ret &= K \\
        K_1 \fatsemi (K_2 \fatsemi K_3) &= (K_1 \fatsemi K_2) \fatsemi K_3 \\
        \skey{run}~E \mbind K &= \skey{run}~E
    \end{align*}
    \end{minipage}
\end{theorem}
\noindent Bind satisfies the three monad laws: it has $\Ret$ as left and right units and is essentially associative. Moreover, both $\skey{div}$ and $\skey{run}$ are left annihilators for bind since they do not terminate. The monad laws show that $(\fatsemi, \Ret)$ also forms a monoid; $\fatsemi$ is commutative and has $\Ret$ as its left and right units. 

The laws of \cref{thm:bind-laws} are proved by coinduction, using the following derivation rule.

\begin{theorem}[ITree Coinduction] \label{thm:coind} We fix a relation $\mathcal{R} : (E, R)\skey{itree} \rel (E, R)\skey{itree}$. Then, given $(P, Q) \in \mathcal{R}$ we can deduce $P = Q$ provided that the following conditions hold: \isalink{https://github.com/isabelle-utp/interaction-trees/blob/ff9f73f98c653b265bd9da55689715cf973499c1/Interaction_Trees.thy\#L63}
  \begin{enumerate}
    \item $\forall (P', Q')\!\in\!\mathcal{R}.\, \skey{is\_Ret}(P') = \skey{is\_Ret}(Q') \land \skey{is\_Sil}(P') = \skey{is\_Sil}(Q') \land \skey{is\_Vis}(P') = \skey{is\_Vis}(Q')$;
    \item $\forall (x, y).\, (\Ret~x, \Ret~y) \in \mathcal{R} \implies x = y$;
    \item $\forall (P', Q')\, (\Sil~P', \Sil~Q') \in \mathcal{R} \implies (P', Q') \in \mathcal{R}$;
    \item $\forall (F, G)\, (\Vis\,F, \Vis\,G)\!\in\! \mathcal{R} \implies (\dom(F) = \dom(G) \land (\forall e\!\in\!\dom(F) \bullet (F(e), G(e)) \in \mathcal{R}))$.
    \end{enumerate}
  \end{theorem}
\noindent To show that $P = Q$, we need to construct a (strong) bisimulation relation $\mathcal{R}$, which intuitively relates two ITrees, and show that $(P, Q) \in \mathcal{R}$. There are four provisos to show that $\mathcal{R}$ is a bisimulation. The first requires that only ITrees of the same kind are related; that is, a $\Ret$ is only related to a $\Ret$, a $\Sil$ with a $\Sil$, and a $\Vis$ with a $\Vis$. Here, $\skey{is\_Ret}$, $\skey{is\_Sil}$, and $\skey{is\_Vis}$ distinguish the three cases. The second proviso states that if $(\ret{x}, \ret{y}) \in \mathcal{R}$ then $x = y$, two related ITrees must return equal values. The third proviso states that internal events must yield bisimilar continuations: $(\tau P, \tau Q) \in \mathcal{R} \implies (P, Q) \in \mathcal{R}$. The final proviso states that for two visible interactions, the two functions must have the same domain ($\dom(F) = \dom(G)$), and every event $e \in \dom(F)$ must lead to bisimilar continuations. Most of our ITree proofs in Isabelle apply this law and then use a mixture of equational simplification and automated reasoning with \skey{sledgehammer} to generate proofs that discharge the resulting provisos.

Next, we define an operator for iterating ITrees in the style of a while-loop: \isalink{https://github.com/isabelle-utp/interaction-trees/blob/ff9f73f98c653b265bd9da55689715cf973499c1/ITree_Divergence.thy\#L352}
\begin{definition}[Iteration] $ $%
\label{def:iteration}
  
\begin{alltt}
  \isakwmaj{corec} while :: "('s \(\Rightarrow\) bool) \(\Rightarrow\) ('e, 's) htree \(\Rightarrow\) ('e, 's) htree" \isakwmin{where}
  "while b P s = (if (b s) then Sil (P s \(\mbind\) while b P) else Ret s)."
\end{alltt}
\end{definition}
\noindent This is not primitively corecursive since the corecursive call uses $\mbind$, and so we define it using the \isakwmaj{corec} command~\cite{Blanchette2015ExtCorec, Blanchette2017Corec} instead of \isakwmaj{primcorec}. This requires us to show that $\mbind$ is a ``friendly'' corecursive function~\cite{Blanchette2017Corec}: it consumes at most one input constructor to produce one output constructor. A while loop iterates whilst the condition $b$ is satisfied by state $s$. In this case, a $\tau$ event is followed by the loop body and the corecursive call. If the condition is false, the current state is returned. We introduce the exceptional cases $\skey{loop}~F \defs \skey{while}~(\lambda s.\, \skey{True})~F$ and $\skey{iter}~P \defs \skey{loop}~(\lambda s.\, P)~()$, which represent infinite loops with and without state, respectively. We can show that $\skey{iter}~(\ret{()}) = \skey{div}$ since it never terminates and has no visible behaviour.

With \skey{while}, we can easily complete the definition of the buffer: $Buffer \defs \skey{while}~BufBody~[]$. This iterates the buffer body in Example~\ref{ex:bufbody} over and over to provide the complete ITree shown in Figure~\ref{fig:exitree}. The initial empty state of the buffer is provided with the parameter $[]$.

\subsection{Structural Operational Semantics and Weak Bisimulation}

The ITree model allows us to describe structural operational semantics for our abstract language naturally. We give big-step operational semantics to ITrees using an inductive predicate.

\begin{definition}[Big-Step Operational Semantics] \label{def:opsem} $ $ \isalink{https://github.com/isabelle-utp/interaction-trees/blob/4bdea2d0a52341e7a19abc3950a3bcdd4b65e7fd/Interaction_Trees.thy\#L310}
\vspace{-2ex}

$$\begin{array}{ccc}
  \begin{array}{c}
     - \\[.5ex] \hline
     P \xrightarrow{[]} P
  \end{array}
&~~
  \begin{array}{c}
  P \xrightarrow{tr} P' \\ \hline
  \tau P \xrightarrow{tr} P'
  \end{array}
&~~
  \begin{array}{c}
  e \in E \quad F(e) \xrightarrow{tr} P' \\ \hline
  \left(\vbar\, x \in E @ F(x)\right) \xrightarrow{e \# tr} P'  
  \end{array}
\end{array}$$

\vspace{-2ex}

\end{definition}
\noindent The relation $P \xrightarrow{tr} Q$ means that $P$ can perform the trace of visible events contained in the list $tr : E~\skey{list}$ and evolve to the ITree $Q$. This relation skips over $\tau$ events. The first rule states that any ITree may perform an empty trace ($[]$) and remain in the same state. We sometimes omit the trace and write $P \xrightarrow{} P'$. The second rule states that if $P$ can evolve to $P'$ by performing $tr$, then so can $\tau P$. The final rule states that if $e$ is an enabled visible event, and $P(e)$ can evolve to $P'$ by doing $tr$, then the event choice can evolve to $P'$ via $e \# tr$, which is $tr$ with $e$ inserted at the head. This inductive predicate is different from the trace predicate (\texttt{is\_trace\_of}) in \cite{ITrees2019}, since $P \xrightarrow{tr} P'$ records both the trace and the continuation ITree. It is, therefore, more general and provides the foundation for characterising structural operational and denotational semantics.

We next prove some important theorems of the transition relation.

\begin{theorem}[Transition Relation Properties]
  \begin{align*}
    (P \xrightarrow{tr_1} Q \land Q \xrightarrow{tr_2} R) &\implies (P \xrightarrow{tr_1 \append tr_2} R) & \textnormal{(sequential transitions)} \\
    (P \xrightarrow{tr} \Vis~F \land P \xrightarrow{tr \append [e]} P') &\implies e \in \dom(F) & \textnormal{(events resolve choices)} \\
    (P \xrightarrow{tr} \Ret~x \land P \xrightarrow{tr} \Ret~y) &\implies x = y & \textnormal{(termination is deterministic)} \\
    (P \xrightarrow{tr_1} Q \land tr_2 \le tr_1) &\implies (\exists R.\, P \xrightarrow{tr_2} R) & \textnormal{(prefix closure)}
  \end{align*}
\end{theorem}

\noindent A pair of sequential transitions can be combined by appending the two traces, $tr_1$ and $tr_2$. Whenever an event $e$ follows a visible choice over $F$, that event must have been enabled by $F$. If we can reach two return ITrees by the same trace, then the two values returned must be equal -- termination is deterministic. Finally, whenever $P$ can reach $Q$ by performing $tr_1$, every prefix of $tr_2$ must also have an intermediate successor ITree $R$.

With these laws, we can prove the usual operational laws for sequential composition as theorems:

\begin{theorem}[Sequential Operational Semantics] $ $ \isalink{https://github.com/isabelle-utp/interaction-trees/blob/4bdea2d0a52341e7a19abc3950a3bcdd4b65e7fd/Interaction_Trees.thy\#L420}
    \vspace{-2ex}

  $$
  \begin{array}{ccc}
  \begin{array}{c}
     - \\[.5ex] \hline
     \skey{skip} \rightarrow \ret{()}
   \end{array}
  &
  \begin{array}{c}
     P \xrightarrow{tr} P' \\ \hline
     (P \mbind Q) \xrightarrow{tr} (P' \mbind Q)
   \end{array}
  &
  \begin{array}{c}
     P \xrightarrow{tr_1} \ret{x} \quad Q(x) \xrightarrow{tr_2} Q' \\ \hline
     (P \mbind Q) \xrightarrow{tr_1 \mathop{\text{@}} tr_2} Q'
   \end{array}
  \end{array}$$
\end{theorem}
\noindent The $\skey{skip}$ process immediately terminates, returning $()$. If the left-hand side $P$ of $\mbind$ can evolve to $P'$ performing the events in $tr$, the overall bind evolves similarly. If $P$ can terminate after doing $tr_1$, returning $x$, and the continuation $Q(x)$ can evolve over $tr_2$ to $Q'$ then the overall $\mbind$ can also evolve over the concatenation of $tr_1$ and $tr_2$, $tr_1 \append tr_2$, to $Q'$.

Strong bisimulation is a useful equivalence, but we often wish to abstract over $\tau$s. We, therefore, also introduce weak bisimulation, $P \approx Q$, as a coinductive-inductive predicate. Given a relation $\mathcal{R}$, we define $\approx_\mathcal{R}$ inductively:

\begin{definition}[Weak Bisimulation] $ $ \isalink{https://github.com/isabelle-utp/interaction-trees/blob/77053abf1830ddc5202a5c737607088529c4b929/ITree_Weak_Bisim.thy}

 $$\begin{array}{cccc}
 \begin{array}{c}
 - \\ \hline \ret{x} \approx_\mathcal{R} \ret{x}
 \end{array} &
 \begin{array}{c}
 P \approx_\mathcal{R} Q \\ \hline \tau P \approx_\mathcal{R} Q
 \end{array} &
 \begin{array}{c}
 P \approx_\mathcal{R} Q \\ \hline P \approx_\mathcal{R} \tau Q
 \end{array} &
 \begin{array}{c}
\forall e \in E @ \mathcal{R}(F(e), G(e)) \\ \hline
\left(\vbar\, x \in E @ F(x)\right) \approx_\mathcal{R} \left(\vbar\, x \in E @ G(x)\right)
 \end{array}
 \end{array}
 $$

\centering
  
${(\approx) \defs \bigcup \{\mathcal{R} | \mathcal{R} \subseteq \{(\skey{div}, \skey{div})\} \cup (\approx_{\mathcal{R}})\}}$
\end{definition}

It requires us to construct a relation $\mathcal{R}$ such that whenever $(P, Q)$ in $\mathcal{R}$ both stabilise, all their visible event continuations are also related by $\mathcal{R}$. For example, $\tau^m~P \approx \tau^n~Q$ whenever $P \approx Q$.  We have proved that $\approx$ is an equivalence relation, and $P \approx \skey{div} \implies P = \skey{div}$. \isalink{https://github.com/isabelle-utp/interaction-trees/blob/4bdea2d0a52341e7a19abc3950a3bcdd4b65e7fd/ITree_Weak_Bisim.thy}

\subsection{Iteration Chains}
\label{subsec:iterchains}

To reason about iteration ($\skey{while}~b~\skey{do}~P~\skey{od}$), as usual, we need to characterise iteration chains. This, for example, is necessary for us to verify the properties of the buffer example. A chain is typically a sequence of states reached during an iteration's successive stages. For ITrees, we also need to consider the events that occur during iteration.

We adopt the notation $s \vdash P \chainto{chn} s'$ to mean that state $s' :: S$ can be reached when the loop body $P :: (E, S)\skey{htree}$ is started in state $s :: S$, by following the chain $chn$. Here, $chn :: (E~\skey{list} \times S)~\skey{list}$ is a list of trace and state pairs, each element of which denotes a single terminating execution of $P$. The formal definition of an iteration chain is given using the inductive predicate below.

\begin{definition}[Iteration Chains] $ $ \isalink{https://github.com/isabelle-utp/interaction-trees/blob/77053abf1830ddc5202a5c737607088529c4b929/ITree_Iteration.thy\#L117}%
  \vspace{2ex}

  \centering
  \begin{tabular}{cc}
    \AxiomC{\vphantom{$\xrightarrow{tr}$}---}
    \UnaryInfC{$s \vdash P \chainto{[]} s$}
    \DisplayProof
    &
    \AxiomC{$P(s) \xrightarrow{tr} \ret{s_0}$}
    \AxiomC{$s_0 \vdash P \chainto{chn} s_1$}
    \BinaryInfC{$s \vdash P \chainto{(tr, s_0) \# chn} s_1$}  
    \DisplayProof
  \end{tabular}
  
\end{definition}

\noindent This does not yet consider the loop condition, which will be added subsequently. The first rule states that $P$ can complete execution at state $s$ by performing zero iterations starting in $s$. This occurs when the condition of a loop is false initially. The second rule allows a chain extension by a single execution of $P$. If $P$, when started in state $s$, terminates in the intermediate state $s_0$ having performed the trace $tr$, and $P$ can further transition to $s_1$, when started from $s_0$ via chain $chn$, then we can prefix $chn$ with the element $(tr, s_0)$. For example, $s_0$ may result from the loop body's first iteration with the trace $tr$, and then $chn$ characterises all subsequent iterations.

Next, we use chains to define a partial iteration $(b, s) \vdash P \tchainto{tr} s'$, which intuitively means that $P$ is executed several times, starting in $s$ and reaching $s'$, whilst yielding trace $tr$. Moreover, in each intermediate state, the condition $b$ remains satisfied. We define this operator directly using iteration chains:

\begin{definition}[Partial Iteration] $ $ \isalink{https://github.com/isabelle-utp/interaction-trees/blob/77053abf1830ddc5202a5c737607088529c4b929/ITree_Iteration.thy\#L181}%
  
  $(b, s) \vdash P \tchainto{tr} s' \iff \exists (chn, s_0, tr_0) @ \left(\begin{array}{l} b(s) \land s \vdash P \chainto{chn} s_0 \land (\forall s\in states(chn) @ b(s)) \\ \land P(s_0) \traceto{tr_0} \ret{s'} \land tr = trace(chn) \append tr_0 \end{array}\right)$
\end{definition}

\noindent Here, the function $\textit{states}$ extracts the set of all states a chain encounters, and $\textit{trace}$ is the concatenated trace described by the whole chain. Using this definition, we can now state the main theorem for reasoning about terminating loops.

\begin{theorem}[Terminating Loops] \label{thm:terminating-loops} $ $ \isalink{https://github.com/isabelle-utp/interaction-trees/blob/77053abf1830ddc5202a5c737607088529c4b929/ITree_Iteration.thy\#L509}%

\vspace{1ex}
  
  $(\skey{while}~b~\skey{do}~P~\skey{od})(s) \traceto{tr} \ret{s'} ~~\iff~~ (\neg b(s) \land s = s' \land tr = []) \lor \left(b(s) \land (b, s) \tchainto{tr} s' \land \neg b(s')\right)$
\end{theorem}

\noindent If a loop terminates in state $s'$, when started in initial state $s$, then there are two possibilities. Firstly, $s$ does not satisfy the condition, so $s'$ is the same as $s$, and an empty trace is emitted. Secondly, $s$ does satisfy the condition; there is a partial iteration from $P$ to $s'$ emitting $tr$, and $s'$ does not satisfy the condition. In other words, the loop is executed several times, with each intermediate satisfying $b$, and ends in a state that exits the loop. A consequence of this theorem is that a chain leads to the existing terminating state whenever a loop terminates. This theorem equips us to reason about the partial and total correctness of programs in \S\ref{sec:rel}.

The proof of this theorem is complex and requires induction on the structure of the transition relation in \cref{def:opsem}. Our approach is to show that every transition of an iteration leads to an ITree of the form $Q \mbind \skey{while}~b~\skey{do}~P~\skey{od}$, that is a prefixed iteration, where the prefix $Q$ is a partial execution of the loop body. The interested reader is directed to our proofs in Isabelle/HOL, which total about 300 lines of Isar.

We have now completed the foundational mechanisation of ITrees. In the next section, we will apply our theory to the modelling and verification of imperative programs before further considering reactive and concurrent programs in \S\ref{sec:csp-circus}.

\section{Imperative Programs and Axiomatic Semantics}
\label{sec:rel}
This section builds on ITrees to develop a theory of Dijkstra-style imperative programs and an associated Hoare logic for partial and total correctness, which can be used to verify programs. The language is implemented as a ``shallow embedding''~\cite{Boulton1993,Gibbons2014} in Isabelle/ITrees. This essentially means that we reuse Isabelle's modelling and proof facilities in the language semantics, which maximises the scope for proof automation. We also develop the weakest precondition calculus and a link with a UTP-style predicative semantics~\cite{Hoare&98}, which provides the basis for a refinement calculus.

\subsection{Modelling Imperative Programs}
\label{sec:model-imper}

Imperative programs can be modelled as homogeneous Kleisli trees, $\src \to (E, \src)\skey{itree}$, where $\src$ is the program's store type. Programs are typically pure for every initial state, meaning they depend only on their internal store for computation. An exception is nondeterministic programs, which we model using a special event to resolve any internal choices (see \S\ref{sec:nondet}).

The store of an imperative program consists of a finite set of mutable state variables. In our work~\cite{Foster17c,Foster2020-IsabelleUTP,Foster2021-JLAMP}, each state variable is modelled as a lens~\cite{Foster09}, $x :: \view \lto \src$, where $\view$ is the variable's type, and $\src$ is the store type. A lens is a pair of functions $\lget :: \view \to \src$ and $\lput :: \src \to \view \to \src$, which query and update the variables present in state $\src$, and satisfy intuitive algebraic laws~\cite{Foster2020-IsabelleUTP}. They allow an abstract representation of stores, where no explicit model is required to support the laws of programming~\cite{Hoare87}. Lenses can be designated as independent, $x \lindep y$, meaning they refer to different regions of $\src$.

An expression or assertion over the state variables is a function $e :: \src \to \view$, where $\view$ is the return type. For example, if $x$ and $y$ are state variables, then the expression $x + y$ is denoted by $\lambda s.\, \lget_x~s + \lget_y~s$. This function retrieves the values of $x$ and $y$ from the state $s$ and adds them together. We can check whether an expression $e$ uses a lens $x$ using unrestriction, written $x \unrest e$. If $x \unrest e$, then $e$ does not use $x$ in its valuation, for example $x \unrest\,(y + 1)$, when $x \lindep y$. Updates to variables can be expressed as a sequence of maplets using the notation $[x_1 \leadsto e_1, x_2 \leadsto e_2, \cdots]$, with $x_i :: \view_i \lto \src$ and $e_i :: \src \to \view_i$, which represents a function $\src \to \src$. 

Creation of program store types is facilitated by a new command called \lstinline{zstore}, which we have implemented on top of our lens library~\cite{Foster2020-IsabelleUTP}: \isalink{https://github.com/isabelle-utp/Shallow-Expressions/blob/main/Z/Z_Store.thy}
\begin{lstlisting}
  zstore S = x~${}_1$~::T1~${}_1$~  x~${}_2$~::T~${}_2$~ ... x~${}_n$~::T~${}_n$~ where "P(x~${}_1$~, x~${}_2$~, ..., x~${}_n$~)"
\end{lstlisting}

\noindent This generates a set of lenses $x_1 \cdots x_n$, which have type $T_i \lto S$, for $i \in \{1..n\}$. An independence property is also generated for each pair of lenses: $x_i \lindep x_j$ where $i \neq j$. Our expression parser automates the lifting of terms containing such lenses so that expressions like $x + y$ are semantically interpreted as $\lambda s.\, \lget_x~s + \lget_y~s$. Optional invariant predicates following the \lstinline{where} clause can also accompany store types, which is useful to facilitate invariant-based reasoning (see \S\ref{sec:zmachines}). Internally, a store is compiled into a record type $S$ with a collection of lenses and an invariant assertion $S\_inv : S \to \bool$.

We can now denote the operators of an idealised imperative programming language. Sequential composition is modelled by Kleisli composition ($P \fatsemi Q$). The remaining operators are given below:

\begin{definition}[Imperative Program Operators] \isalink{https://github.com/isabelle-utp/interaction-trees/blob/master/UTP/ITree_Circus.thy}
\begin{align*}
\conditional{C_1}{P}{C_2} &~\defs~ (\lambda s.\, if~P(s)~then~C_1(s)~else~C_2(s)) \\
\assigns{\sigma} &~\defs~ (\lambda s.\, \Ret(\sigma(s))) \\
x := e &~\defs~ \assigns{[x \leadsto e]} \\
\skey{Skip} &~\defs~ \assigns{[\leadsto]} \\
\skey{Stop} &~\defs~ (\lambda s.\, \skey{stop}) \\
\skey{Div} &~\defs~ (\lambda s.\, \skey{div}) \\
\utest{P} &~\defs~ \conditional{\skey{Skip}}{P}{\skey{Stop}}
\end{align*}
\end{definition}

\noindent $\conditional{C_1}{P}{C_2}$ is our algebraic notation for a conditional statement (if-then-else), where $P$ is the condition. Operator $\assigns{\sigma}$ lifts a function $\sigma : \src \to \src$ to a KTree. It is principally used to represent assignments, which can be constructed using our maplet notation, such that a single assignment $x := e$ is $\assigns{[x \leadsto e]}$. Since substitutions can assign multiple variables, they can also represent simultaneous assignment, $(x, y) := (e, f)$. Similarly, the vacuous $\skey{Skip}$ statement is denoted by an empty assignment. $\skey{Stop}$ is simply a Kleisli-lifted version of the ITree $\skey{stop}$, which deadlocks (or aborts) in any initial state, and $\skey{Div}$ similarly diverges in every initial state. Finally, $\utest{P}$ is a test operator, which deadlocks when $P$ is false and otherwise has no effect. These operators satisfy all the usual laws of programming~\cite{Hoare87}, a small selection of which is shown below. These laws give equational algebraic semantics for imperative programs.

\begin{theorem}[Laws of programming] \label{thm:laws-of-programming} \isalink{https://github.com/isabelle-utp/interaction-trees/blob/master/UTP/ITree_Circus.thy\#L103}
\begin{align*}
  \skey{Skip} \fatsemi C = C \fatsemi \skey{Skip} &~=~ C \\
  x := e \fatsemi y := f &~=~ y := f \fatsemi x := e & \text{if } x \lindep y, x \unrest f, y \unrest e \\
  \assigns{\sigma} \fatsemi \assigns{\rho} &~=~ \assigns{\rho \circ \sigma} \\
  x := e \fatsemi (\conditional{C_1}{P}{C_2}) & ~=~ \conditional{(x := e \fatsemi C_1)}{P[e/x]}{(x := e \fatsemi C_2)} \\
  \conditional{C_1}{P}{(\conditional{C_2}{P}{C_3})} & ~=~ \conditional{C_1}{P}{C_3}
\end{align*}
\end{theorem}

\noindent \skey{Skip} is the unit of sequential composition. Two variable assignments commute provided their variables are independent ($x \lindep y$), and their respective expressions do not depend on the adjacent variable. More generally, the sequential composition of two state updates $\sigma$ and $\rho$ entails their functional composition. Assignment can be pushed into a conditional by first substituting the assignment into the condition $P$. Finally, an outer conditional masks an inner one, meaning that $C_2$ is an unreachable branch. Such laws can be used for symbolic execution and optimisation of imperative programs.

\subsection{Concrete and Symbolic Execution}

A particular benefit of our ITree-based semantics is that imperative programs can be directly executed. A non-divergent and non-aborting pure ITree reduces to the form of $\tau^n\left(\ret{s'}\right)$, for $n \in \nat$, where $s'$ is the final state of the program. This is a particular case of a stable ITree. Consequently, an imperative program can be executed by supplying an initial state $s$ and stripping off all the $\tau$s (internal steps) until $s'$ is reached. If the program is divergent (i.e., non-terminating) it will never get a final state, so execution will hang.

To aid the modelling of programs in our tool, we provide the following command:

\begin{definition}[Program command] \isalink{https://github.com/isabelle-utp/interaction-trees/blob/aec70ffed4631666f42ed604d35360da00728bab/UTP/VCG/ITree_VCG.thy\#L48}$ $%
\vspace{1ex}
  
$\left\llbracket
  \begin{array}{l}
    \textbf{program}~Pr (x_1 :: T_1, \cdots, x_n :: T_n) \\
    ~~ \textbf{over}~\mathcal{S} = Body(x_1, \cdots, x_n)
  \end{array}\right\rrbracket = \left(\begin{array}{l} Pr :: T_1 \times \cdots \times T_n \Rightarrow \src \to (E, \src)\skey{htree}) \\ Pr \defs (\lambda (x_1, \cdots, x_n).\, Body(x_1, \cdots, x_n)) \end{array}\right)$
\end{definition}

\noindent A program takes a tuple of parameters $(x_1, \cdots, x_n)$ and operates over a store $\src$. The program's body is a parametric ITree in $x_1 \cdots x_n$. These parameters are not program variables (lenses) but logical variables. Intuitively, they are constants that cannot be written to.

As an example, below is the definition of a simple imperative program for reversing a list:

\noindent\begin{example}[List Reversal Program] $ $ \isalink{https://github.com/isabelle-utp/interaction-trees/blob/6be22531983c2020595fba6b67fb1910399d601f/UTP/VCG/examples/List_Reversal.thy} \label{ex:listev}%

\begin{lstlisting}
  program reverse (xs :: int list) over state =
  "ys := []; i := 0; 
   while i < length xs
   do 
      ys := xs~!~i # ys; 
      i := i + 1 
   od"
  \end{lstlisting}
\end{example}
\noindent We define the program \texttt{reverse}, with input parameter \texttt{xs :: int list}. It operates over the store type \texttt{state}, containing the variables \texttt{i :: nat} and \texttt{ys :: int list}. The program iterates through the input \texttt{xs}, pushing each element on \texttt{ys}, with the result that \texttt{xs} is reversed.

Next, we define a command \lstinline{execute} that executes an ITree-based program with given argument. For example, we can execute \texttt{reverse} with the following command:

\begin{lstlisting}
  execute "reverse [1,2,3,4,5]"},
\end{lstlisting}
This executes \texttt{reverse} with the input list \texttt{[1,2,3,4,5]}. The command generates code using the aforementioned ITree semantics, executes it, and then reports termination with the final state $[y \leadsto [5,4,3,2,1], i \leadsto 5]$.

A terminating program with no nondeterminism is denoted by a pure ITree of the form $\tau^n\,\ret{v}$, where $v$ is the final state or return value. Since we do not generally know how many $\tau$ events a program may perform, execution involves stripping off all the $\tau$ events until a return value is encountered. Since programs in general can fail to terminate, it is necessary to place an upper bound on the number of $\tau$ events that can be skipped. We therefore provide a global constant called $\textit{MAX\_SIL\_STEPS} :: \skey{nat}$, which acts as a timeout for execution. Then, we use the function $\textit{un\_Sils\_n} :: \skey{nat} \Rightarrow (E, \src)\skey{itree} \Rightarrow (E, \src)\skey{itree}$ to strips a number $n$ of $\tau$ events of an ITree, with $n \le \textit{MAX\_SIL\_STEPS}$.

We can finally realise program execution using Isabelle's evaluation mechanism, as present in the \lstinline{value} command, which evaluates an executable term~\cite{Haftman2010-CodeGen,Haftmann2012NBE}. The evaluator can perform concrete execution using the SML code generator, when the program contains no free variables (as in the above example), and symbolic execution using normalisation by evaluation or the simplifier. The former is most efficient, and so is the default behaviour for \lstinline{execute}.

An execution can produce one of four possible results: (1) termination with a final state, (2) an abort, (3) a visible event, and (4) a timeout. A timeout occurs if \textit{MAX\_SIL\_STEPS} $\tau$ events have occurred without producing a return or visible event. For example, \lstinline{execute "Div"} results in a timeout. A termination results if execution encounters a $\Ret$ before reaching the maximum number of $\tau$s. This being the case, the interface displays the final state of each variable. 

Abortion occurs when an empty visible event is encountered (i.e. $\skey{stop})$, following a finite number of $\tau$ events. Thus, if the execute command encounters a $\Vis$ constructor, it checks whether the choice function is empty. If it is empty, then the execution has aborted. Otherwise, it indicates that an event choice was encountered and goes no further. For ITrees that use visible events, we typically cannot use such a non-interactive execution. We must instead rely on animation (\S\ref{sec:animation}), which allows further user input when a visible event is presented.

\subsection{Nondeterminism}
\label{sec:nondet}

The operators given so far allow us to model only deterministic programs, which typically reduce to pure functions on the state. However, nondeterminism is useful both as a specification device and where design choices are deferred. Nondeterministic decisions can be encoded by introducing a special channel $nd$, which the environment can conceptually use to resolve, acting as an oracle. Here, $I$ is an index type, which denotes the maximum cardinality of any choices. Whilst, in theory, $I$ can be any type, we can typically only animate countable choices, and therefore, for now, we set $I \subseteq \nat$. We can now use this to define the internal choice operator.

\begin{definition}[Countable nondeterminism] $ $ \isalink{https://github.com/isabelle-utp/interaction-trees/blob/77053abf1830ddc5202a5c737607088529c4b929/UTP/ITree_Countable_Nondeterminism.thy\#L1}%

  Assume a channel $nd$ carrying a value of type $\nat$ and a set $I \subset \nat$ exists. Then, we encode nondeterministic choice as $$\bigsqcap_{i \in I} @ C(i) ~\defs~ 
  \Vis~(\lambda nd.i ~|~ i \in I @ P(i)).$$
\end{definition}

\noindent This constructs a visible event choice over $nd$ events parameterised by the elements of $I$. The particular index chosen is passed to $P$ as a parameter. We can then define a binary nondeterministic choice as $C_1 \sqcap C_2 \defs \left(\bigsqcap_{i\in\{0,1\}} @ \conditional{C_1}{i = 0}{C_2}\right)$. Programs containing nondeterministic choices cannot be directly executed using the \lstinline{execute} command, as the events must be resolved using animation (see \S\ref{sec:animation}).

\subsection{Predicative Semantics and Refinement}
\label{sec:pred-sem-ref}

We now focus on a predicative semantic interpretation for ITrees, which allows us to link with the established UTP predicative semantics for imperative programs~\cite{Hoare&98, Cavalcanti04}. This semantics has many uses, but one particular use is to provide a notion of refinement for nondeterministic imperative programs.

UTP uses predicate calculus as a unifying language for programs and specifications. Dijkstra-style programs can be denoted as alphabetised relations, predicates that relate the initial values of variables to their final values. For example, assuming a store with three integer variables $x$, $y$, and $z$, an assignment $x := x + 1$ can be denoted by the predicate $x' = x + 1 \land y' = y \land z' = z$, where $x$ is the initial value of $x$ and $x'$ is its final value.

Central to UTP is a notion of refinement $P \refinedby Q$ for alphabetised relations $P$ and $Q$, which means that $Q$ is more deterministic or concrete than $Q$. For example, we can write the specification $x' > x$, which means that in the final state, $x$ should be strictly greater than its initial value. Then, using refinement, we can demonstrate that $x' > x \refinedby x := x + 1$, meaning that the program on the right implements the specification on the left. The refinement order induces a complete lattice of alphabetised relations. It gives rise to fixed-point operators $\mu F$ (least fixed point) and $\nu F$ (greatest fixed point), which can specify iterative and recursive behaviour. UTP provides the predicative semantics for the \Circus language~\cite{Oliveira&09}.

To relate our ITree-based semantics with such predicative semantics, we must distinguish a program's terminating states from divergence. We can reason about termination and divergence with our transition relation, $P \xrightarrow{tr} Q$. Terminating imperative programs are characterised by pure ITrees that eventually reach a $\Ret$. We define the set of return values of an ITree using the following function:

\begin{definition}[Return Values] $\retvals{P} = \{x ~|~ \exists tr.\, P \xrightarrow{tr} \ret{x}\}$. \isalink{https://github.com/isabelle-utp/interaction-trees/blob/77053abf1830ddc5202a5c737607088529c4b929/Interaction_Trees.thy\#L664}
\end{definition}
  
\noindent $\retvals{P}$ induces the set of possible values a process $P$ may return, whenever $P$ terminates. In other words, 
 $\retvals{P}$ is the set of reachable final states. If $\retvals{P} = \emptyset$, then $P$ can never terminate. $\retvals{P}$ abstracts over all possible traces through existential quantification, and therefore, it does not distinguish return values that arise from different event interactions. All events are, therefore, effectively treated as nondeterminism in this semantic interpretation. Below, we give the valuations of $\retvals{P}$ for the main ITree constructors.

\begin{theorem}[Return Values for ITree Constructors] \isalink{https://github.com/isabelle-utp/interaction-trees/blob/77053abf1830ddc5202a5c737607088529c4b929/Interaction_Trees.thy\#L675}
\begin{align*}
  \retvals{\ret{x}} &= \{x\} \\
  \retvals{\tau P} &= \retvals{P} \\
  \retvals{\Vis(F)} &= \bigcup \{\retvals{P} ~|~ P \in \ran(F)\} \\
  \retvals{P \mbind Q} &= \bigcup \{\retvals{Q(x)} ~|~ x \in \retvals{P}\} \\
  \retvals{\skey{stop}} &= \retvals{\skey{div}} = \emptyset 
\end{align*}
\end{theorem}

\noindent A $\Ret$ returns a single value. A $\Sil$ returns the values following the $\tau$ event. A visible event ($\Vis$) returns all possible values returned by the continuation ITrees, $P \in \ran(F)$. If we view the ITree as a transition graph, we take the values returned on all paths. A bind $P \mbind Q$ first calculates the return values of $P$, then uses these as the possible inputs for $Q$, and calculates all the resulting return values. Neither $\skey{stop}$ or $\skey{div}$ have any return values because they do not successfully terminate. We can now use this function to provide a predicative semantic interpretation for ITrees.

\begin{definition}[Predicative semantics] $\psem{P} = (\lambda (s, s').\, s' \in \retvals{P(s)})$ \isalink{https://github.com/isabelle-utp/interaction-trees/blob/77053abf1830ddc5202a5c737607088529c4b929/UTP/ITree_Relation.thy\#L11}
\end{definition}

\noindent The function $\psem{P}$ induces a predicate of type $\src \times \src \Rightarrow \bool$ for the homogeneous ITree $P$, which corresponds to a binary relation. Thus, $\psem{P} (s, s')$ holds whenever $s'$ is reachable from the start state $s$. With this function, we can show that our imperative programs respect a UTP-style predicative semantics~\cite{Hoare&98}.

\begin{theorem}[Predicative semantics of loop-free imperative programs] \label{thm:predsem} \isalink{https://github.com/isabelle-utp/interaction-trees/blob/6be22531983c2020595fba6b67fb1910399d601f/UTP/ITree_Relation.thy}

  \begin{align*}
    \psem{\assigns{\sigma}}(s, s') &~=~ (s' = \sigma(s)) \\
    \psem{P \fatsemi Q}(s, s') &~=~ \left(\exists s_0 @ \psem{P}(s, s_0) \land \psem{Q}(s_0, s')\right) \\
    \psem{\conditional{P}{B}{Q}} &~=~ \left(\left(B(s) \land \psem{P}(s, s')\right) \lor \left(\neg B(s) \land \psem{Q}(s, s')\right)\right) \\
    \psem{\skey{Stop}}(s, s') ~=~ \psem{\skey{Div}}(s, s') &~=~ False \\
    \psem{P \sqcap Q}(s, s') &~=~ \left(\psem{P}(s, s') \lor \psem{Q}(s, s')\right)
  \end{align*}
  
\end{theorem}

\noindent A state update applies the update function to the initial state $s$ to obtain the final state $s'$. The semantics of an assignment $x := e$ is a special case, conceptually $x' = e(s) \land y' = y$, for all other variables $y$ in $\src$. The predicative semantics for $P \fatsemi Q$ yields the usual definition of relational composition: an intermediate state $s_0$, a final state for $P$ and an initial state for $Q$. Conditional behaves as $P$ when $B$ is true in the initial state and $Q$ otherwise. Both $\skey{Stop}$ and $\skey{Div}$ have the same interpretation $False$, as this semantics cannot distinguish between deadlock and divergence. Finally, a state pair is satisfied by a nondeterministic choice $P \sqcap Q$ if it is satisfied by either $P$ or $Q$, which is the usual UTP interpretation of nondeterministic choice as disjunction~\cite{Hoare&98}.

Next, we consider the predicative semantics of iteration. First of all, we note the following corollary of Theorem~\ref{thm:terminating-loops}:

\begin{corollary}[Iteration return values] \isalink{https://github.com/isabelle-utp/interaction-trees/blob/6be22531983c2020595fba6b67fb1910399d601f/ITree_Iteration.thy\#L524}
$$\retvals{(\skey{while}~b~\skey{do}~P~\skey{od})(s)} = \{s' ~|~ (\neg P(s) \land s = s') \lor (\exists tr.\, B(s) \land (B, s) \tchainto{tr} s' \land \neg B(s'))\}$$    
\end{corollary}
\noindent
The return values for a loop started in state $s$ is precisely the set of states for which there is a number of iterations of $P$ yielding some trace $tr$. Whilst we could now express the predicative semantics in these terms, it is more convenient and concise to do this in terms of the reflexive transitive closure operation $R^{*}$. We first reiterate a result of the Isabelle/HOL standard library:

\begin{theorem}[Reflexive Transitive Closure paths]
$$R^{*} (s, s') \iff s = s' \lor (\exists xs. \, \forall i<length(xs).\, R((s \# xs) ! i, xs ! i) \land x' = last(xs))$$
\end{theorem}
\noindent A pair of states $(s, s')$ are related by $R^{*}$ either when $s = s'$, or there is a path $xs$ leading from $s$ to $s'$ through several iterations of $R$. Here, the path is a list of states, where each consecutive pair of states, starting from $s$ and ending with $s'$, are related by $R$. With this result, we can now express the predicative semantics of iteration:

\begin{theorem}[Predicative semantics of iteration] \isalink{https://github.com/isabelle-utp/interaction-trees/blob/6be22531983c2020595fba6b67fb1910399d601f/UTP/ITree_Relation.thy\#L76}
    $$\psem{\skey{while}~B~\skey{do}~C~\skey{od}} ~=~ (\utest{B} \mathop{;} \psem{C})^{*} \mathop{;} \!\utest{\neg B}$$
\end{theorem}

\noindent For conciseness, the predicate semantics for \skey{while} is expressed point-free. The notation $\utest{P}$ denotes a test, i.e. $\lambda (s, s'). \, P(s) \land s' = s$, which skips states that satisfy $P$. The semicolon operator ($P \mathop{;} Q$) denotes relational composition, such that $\psem{P \fatsemi Q} = (\psem{P} \mathop{;} \psem{Q})$. In this relational context, a \skey{while} loop iterates $C$ when $B$ is true and ends when $B$ is false. This corresponds to the usual Kleene algebra interpretation of iteration~\cite{Armstrong2015, Gomes2016}.

The predicative interpretation in Theorem~\ref{thm:predsem} induces a homomorphism between the ITree semantics and the relational semantics for each of the imperative programming operators ($:=$, $\fatsemi$, $\conditional{}{b}{}$, etc.). This homomorphism is not only of theoretical interest but also practical benefit. Using the equations as code equations for the Isabelle/HOL code generator~\cite{Haftman2010-CodeGen} allows us to employ the ITree semantics as a means to generate code for and execute relational imperative programs (see \S\ref{sec:animation}). We can also use our predicative semantics to obtain a notion of refinement for ITrees. We first recall the usual definition of refinement for relational programs in UTP:

\begin{definition}[Predicative refinement]
  $(P \refinedby Q) \defs (\forall (s, s').\, Q(s, s') \implies P(s, s'))$
\end{definition}

\noindent This is the usual UTP definition of refinement as a universally closed reverse implication. Specifically, $P$ is refined by $Q$ ($P \refinedby Q$) if $Q$ contains no more observable behaviours than $P$. Since we can interpret an ITree as a predicate, we can define $(P \refinedby_{p} Q) \defs \psem{P} \refinedby \psem{Q}$. In particular, we can now use refinement to reduce nondeterminism: $P \sqcap Q \refinedby_{p} {P}$. This refinement relation forms a preorder, but it is not antisymmetric. This is because the predicative semantics is too coarse and does not, for example, distinguish $\skey{Stop}$ and $\skey{Div}$, which are both $False$. For antisymmetry, we would need a finer predicative interpretation, such as the UTP theory of designs~\cite{Cavalcanti04} or reactive designs~\cite{Foster17c}, but this is out of the scope of this paper.

\subsection{Hoare logic and Weakest Preconditions}
\label{subsec:hoarelogic}

We can now use our predicative interpretation of ITrees to define a partial correctness Hoare logic.

\begin{definition}[Partial Correctness Hoare Logic] \label{def:hoare-logic} \isalink{https://github.com/isabelle-utp/interaction-trees/blob/6be22531983c2020595fba6b67fb1910399d601f/UTP/ITree_Hoare.thy\#L27}
  $$\lamporttriple{P}{C}{Q} \defs (\forall (s, s', tr) @ P(s) \land C(s) \traceto{tr} \ret{s'} \implies Q(s'))$$
\end{definition}

\noindent Whenever $P$ is satisfied by initial state $s$, and $C$ when started in $s$ terminates in final state $s'$, it follows that $Q$ is satisfied by $s'$. This is partial correctness because we do not commit if $C$ aborts or does not terminate. We can handle these additional aspects separately through deadlock-freedom and termination checks or by a total correctness Hoare logic. Our definition of the Hoare triple can alternatively be characterised directly using refinement in the UTP style, as the following theorem demonstrates.

\begin{theorem} $\lamporttriple{P}{C}{Q} ~\text{if and only if}~ (\upre{P} \implies \upost{Q}) \refinedby \psem{C}$ \isalink{https://github.com/isabelle-utp/interaction-trees/blob/6be22531983c2020595fba6b67fb1910399d601f/UTP/ITree_Hoare.thy\#L33}

\end{theorem}

Here, $\upre{P}$ and $\upost{Q}$ are shorthands for $\lambda (s, s').\, P(s)$ and $\lambda (s, s'). Q(s')$ lift these predicate expressions to pre- and postconditions. We construct a relational specification for the program and then use it to assert a refinement. This allows us to obtain all the laws of Hoare logic for straight-line programs (cf. \cite{Foster2020-IsabelleUTP}), for example:

\begin{theorem}[Hoare logic laws] \label{thm:hl-laws} \isalink{https://github.com/isabelle-utp/interaction-trees/blob/6be22531983c2020595fba6b67fb1910399d601f/UTP/ITree_Hoare.thy\#L113} $ $%

  \vspace{2ex}
  
  \begin{tabular}{ccc}
  \AxiomC{$P \implies Q[e/x]$}
  \UnaryInfC{$\lamporttriple{P}{x := e}{Q}$}
  \DisplayProof
  &
  \AxiomC{$\lamporttriple{P}{C_1}{Q}$}
  \AxiomC{$\lamporttriple{Q}{C_2}{R}$} 
  \BinaryInfC{$\lamporttriple{P}{C_1 \fatsemi C_2}{R}$}
  \DisplayProof 
  &
  \AxiomC{$\lamporttriple{P}{C_1}{Q}$}
  \AxiomC{$\lamporttriple{P}{C_2}{Q}$}
  \BinaryInfC{$\lamporttriple{P}{C_1 \sqcap C_2}{Q}$}
  \DisplayProof  
  \end{tabular}

\end{theorem}

For while loops, using the construct introduced in \cref{def:iteration}, there is a little more work to be done. Recall the partial correctness law for Hoare logic:

\begin{theorem}[Partial Correctness While law] \label{thm:hl-loop} $ $ \isalink{https://github.com/isabelle-utp/interaction-trees/blob/6be22531983c2020595fba6b67fb1910399d601f/UTP/ITree_Hoare.thy\#L273}%

\centering
\vspace{1ex}
\AxiomC{$\lamporttriple{P \land B}{C}{P}$}
\UnaryInfC{$\lamporttriple{P}{\skey{while}~B~\skey{do}~C~\skey{od}}{\neg B \land P}$}
\DisplayProof
\vspace{1ex}
\end{theorem}

\noindent Here, $P$ is the loop invariant, which must remain true whenever the body $C$ is executed. We outline the mechanised proof below, which uses \cref{thm:terminating-loops}.

\begin{proof}
From \cref{def:hoare-logic}, we need to show that given an initial state $s$ satisfying $P$, whenever $\skey({while}~B~\skey{do}~S~\skey{od})(s) \traceto{} \ret{s'}$, then it follows that $s'$ satisfies $\neg B$ and $P$ (partial correctness). From \cref{thm:terminating-loops}, we know that the loop terminates immediately or executes several times. Suppose it terminates immediately, then clearly $P(s)$ and $\neg B(s)$. Suppose it executes, a chain $chn$ leads to $s'$ such that $\neg B(s')$. The premises of the loop invariant law tell us that for any state $s_0$, such that $P(s_0)$ and $B(s_0)$, whenever $C(s_0) \traceto{} \ret{s_1}$ then also $P(s_1)$. As a result, we can deduce that any $s_0 \in \textit{states}(chn)$ and subsequent state $s_1$ must maintain the invariant. This being the case, it also particularly follows that $P(s')$, since $s'$ is such an $s_1$ state. This completes the proof. 
\end{proof}

In addition to Hoare logic, we can also characterise the weakest (liberal) preconditions:

\begin{definition}[Weakest Preconditions] \isalink{https://github.com/isabelle-utp/interaction-trees/blob/6be22531983c2020595fba6b67fb1910399d601f/UTP/ITree_WP.thy\#L9}
  \begin{align*}
    \skey{wp}~C~P \defs (\lambda s.\, \exists s'.\, \psem{C} (s, s') \land P(s')) \\
    \skey{wlp}~C~P \defs (\lambda s.\, \forall s'.\, \psem{C} (s, s') \implies P(s'))
  \end{align*}
\end{definition}

\noindent The weakest precondition $\skey{wp}~C~P$ obtains the weakest precondition required for $C$ to reach a state satisfying $P$. It formally requires that for any initial state $s$, there is a final state $s'$, such that $P(s')$. In particular, we can use the weakest precondition to calculate the domain or ``precondition'' of a program: $\skey{pre}(C) \defs \skey{wp}~C~\skey{true}$. For ITrees, this is the set of initial states that do not lead to deadlock or divergence. For imperative programs specifically, this can be considered the initial states for which the program terminates. The weakest liberal precondition is similar, but for any final state $s'$ of $C$ that $P(s')$ holds, it does not require such an $s'$ exists. Both of these laws satisfy the standard laws~\cite{Dijkstra75}, which we have previously presented for Isabelle/UTP~\cite{Foster2020-IsabelleUTP}.

As usual, we can use the simplifier to calculate the weakest preconditions for a program in Isabelle/HOL equationally. Moreover, we also prove the following standard theorem linking Hoare logic and $\skey{wlp}$:

\begin{theorem} $\lamporttriple{P}{C}{Q} \iff (P \implies \skey{wlp}~C~Q)$ \label{thm:hoare-wlp} \isalink{https://github.com/isabelle-utp/interaction-trees/blob/6be22531983c2020595fba6b67fb1910399d601f/UTP/ITree_Hoare.thy\#L423}
\end{theorem}

\noindent We can prove a Hoare triple by calculating the $\skey{wlp}$, and then proving the precondition $P$ satisfies the resulting predicate. Finally, we can use $\skey{wp}$ to define the total correctness Hoare triple:

\begin{definition}[Total Correctness Hoare Logic] \isalink{https://github.com/isabelle-utp/interaction-trees/blob/6be22531983c2020595fba6b67fb1910399d601f/UTP/ITree_THoare.thy\#L11} $ $%
$$\tlamporttriple{P}{C}{Q} \defs \left(\lamporttriple{P}{C}{Q} \land (P \implies \skey{pre}(C)\right)$$
\end{definition}
\noindent This follows the usual intuition of \textit{total correctness = partial correctness + termination}. Here, $P \implies \skey{pre}(C)$ means that the precondition is a sufficient condition to ensure that $C$ terminates. With this definition, we can obtain the corresponding laws to those in \ref{thm:hl-laws}, and also the total correctness law for loops, which requires a decreasing variant expression $V$:

\begin{theorem}[Total Correctness While law] \label{thm:thl-loop} $ $ \isalink{https://github.com/isabelle-utp/interaction-trees/blob/6be22531983c2020595fba6b67fb1910399d601f/UTP/ITree_THoare.thy\#L121}%

\centering
\vspace{1ex}
\AxiomC{$\tlamporttriple{P \land B \land V = z}{C}{P \land V < z}$}
\UnaryInfC{$\tlamporttriple{P}{\skey{while}~B~\skey{do}~C~\skey{od}}{\neg B \land P}$}
\DisplayProof
\vspace{1ex}
\end{theorem}

\noindent The proof of this depends on \cref{thm:terminating-loops}. %

We will make further use of the weakest preconditions when we develop our Z-Machine tool in Section~\ref{sec:zmachines}. For now, we are turning our attention to the automation of program verification.

\subsection{Verification Condition Generation}

Automation of program verification is conducted, as usual, through a verification condition generator (VCG). Our VCG method repeatedly applies Hoare logic laws to obtain a collection of verification condition predicates, following a standard approach~\cite{Armstrong2015,Gomes2016}. The resulting predicates can often be discharged by Isabelle's automated proof methods, like \skey{blast}, \skey{metis}, and \skey{smt}, with the help of \skey{sledgehammer}. If verification fails, we can find errors using counterexample finders, like \skey{nitpick} and \skey{quickcheck}.

For automated reasoning, we need laws that avoid the introduction of meta-variables, as these can introduce backtracking and hamper automation. For example, the general sequential composition law in \cref{thm:hl-laws} and iteration law in \cref{thm:hl-loop} introduce variables that only appear in the hypotheses, and a suitable witness must be supplied. Instead, we specialise the Hoare logic theorems to avoid this. In particular, we introduce the following two corollaries for assignment.

\begin{corollary}[Forward and Backward Assignment Laws] $ $ \isalink{https://github.com/isabelle-utp/interaction-trees/blob/6be22531983c2020595fba6b67fb1910399d601f/UTP/ITree_Hoare.thy\#L118}%
  \vspace{1.5ex}

  \centering
  \begin{tabular}{ccc}
  \AxiomC{$\lamporttriple{x = e[x_0/x] \land P[x_0/x]}{C}{Q}$}
  \RightLabel{\scriptsize$x_0 \notin fv(e,P)$}  
  \UnaryInfC{$\lamporttriple{P}{x := e \fatsemi C}{Q}$}
  \DisplayProof
  &\quad&
  \AxiomC{$\lamporttriple{P}{C}{Q[e/x]}$}
  \UnaryInfC{$\lamporttriple{P}{C \fatsemi x := e}{Q}$}
  \DisplayProof
  \end{tabular}
  
\end{corollary}

\noindent The forward law allows us to push the effect of the assignment into the precondition. We introduce a new fixed logical variable, $x_0$, which stands for the initial value of $x$ before the assignment occurred. We substitute $x$ for $x_0$ in the assigned expression $e$ and the precondition $P$. The backward law similarly applies the assignment to the postcondition.

VCG, as usual, depends on the annotation of loops with invariants. We adopt the approach of Armstrong et al.~\cite{Armstrong2015} and introduce the syntax $\skey{while}~B~\skey{invariant}~I~\skey{do}~C~\skey{od}$, which annotates with the invariant $I$. This annotation is semantically vacuous and exists only to help proof automation using the following derived law.

\begin{theorem}[Loop Invariant Annotation] $ $ \isalink{https://github.com/isabelle-utp/interaction-trees/blob/6be22531983c2020595fba6b67fb1910399d601f/UTP/ITree_Hoare.thy\#L389}%
  \vspace{1.5ex}

  \centering
  \AxiomC{$\lamporttriple{I \land B}{C}{I}$}
  \AxiomC{$P \implies I$}
  \AxiomC{$\neg B \land I \implies Q$}
  \TrinaryInfC{$\lamporttriple{P}{\skey{while}~B~\skey{invariant}~I~\skey{do}~C~\skey{od}}{Q}$}
  \DisplayProof
  
\end{theorem}

\noindent This requires we prove that $I$ is an invariant of the loop body, $I$ weakens precondition $P$, and $I$ strengthens postcondition $Q$ when the loop condition does not hold. The proof combines the consequence law and the partial correctness law for while loops. Similarly, we derive a corresponding total correctness law, which uses a variant annotation: $\skey{while}~B~\skey{invariant}~I~\skey{variant}~V~\skey{do}~C~\skey{od}$, where $V$ is the variant expression.

Finally, we implement the \texttt{vcg} proof method, which implements the following steps:

\begin{enumerate}
\item Atomise assignments and conditionals where possible, using the \cref{thm:laws-of-programming};
\item Repeatedly apply specialised Hoare logic laws as introduction rules to decompose goal;
\item Evaluate substitutions in resulting expressions and convert them to HOL proof obligations.
\end{enumerate}

The result is a set of VCs for which discharge can be attempted. We can now annotate our imperative list reversal program from Example~\ref{ex:listev} with an invariant and a variant to allow its verification:

\noindent\begin{example}[Annotated List Reversal Program] $ $ \isalink{https://github.com/isabelle-utp/interaction-trees/blob/6be22531983c2020595fba6b67fb1910399d601f/UTP/VCG/examples/List_Reversal.thy} \label{ex:listrev-annot}%

\begin{lstlisting}
program reverse (xs :: int list) over state =
"ys := []; i := 0; 
 while i < length xs
 invariant ys = rev (take i xs)
 variant length xs - i
 do 
    ys := xs~!~i # ys; 
    i := i + 1 
 od"
\end{lstlisting}
\end{example}
\noindent We supply the invariant $\texttt{ys = rev(take i xs)}$, since \texttt{ys} is always the first \texttt{i} elements of \texttt{xs} in reverse. The functions \texttt{take} and \texttt{rev} are defined in Isabelle/HOL. The variant \texttt{length xs - i} counts down from the length of \texttt{xs} to zero. 

With this, we want to prove the Hoare triple $\tlamporttriple{True}{reverse(xs)}{ys = rev(xs)}$: the imperative program satisfies the functional specification provided by the $rev$ function. Applying the \texttt{vcg} method to this proof goal yields a single verification condition:

\begin{center}
  \texttt{xs ! i \# rev (take i xs) = rev (take (i + 1) xs)} for \texttt{i < length(xs)}.
\end{center}

\noindent This states that taking the first \texttt{i + 1} elements of \texttt{xs} and then reversing it can be achieved by appending the \texttt{i}th element of \texttt{xs} at the beginning of the reversed \texttt{i} elements. This can be discharged by \skey{sledgehammer} using the built-in laws from Isabelle/HOL. The variant proof is straightforward and discharged simply by the simplifier.

Although this is a trivial example, we have verified more substantial benchmark examples, such as sorting algorithms, with a high level of automation provided by \skey{sledgehammer}.

\section{Reactive and Concurrent Programming}
\label{sec:csp-circus}
In this section, we move on from imperative programs and give an ITree semantics to deterministic fragments of the CSP~\cite{Brookes1984, Hoare85} and \Circus~\cite{Woodcock2001-Circus, Oliveira&09} process languages. Our deterministic CSP fragment is consistent with the one identified by Roscoe~\cite[Section~10.5]{Roscoe2010-UCS}. The standard CSP denotational semantics is provided by the failures-divergences model~\cite{Brookes1984, Roscoe2010-UCS}, and we provide preliminary results on linking to this in \S\ref{sec:sem-links}.

\subsection{CSP}

CSP processes are parametrised by an event alphabet ($\Sigma$), which specifies the possible ways a process communicates with its environment. For ITrees, $\Sigma$ is provided by the type parameter $E$. Whilst the event sort of an ITree $E$ is typically infinite, in process algebraic languages, like CSP, it is usually expressed with a finite set of channels, which can carry data of various types. Here, we characterise channels abstractly using prisms~\cite{Pickering2017-Optics}, a concept well known in the functional programming world:

\begin{definition}[Prisms] A prism is a quadruple $(\view, \Sigma, \pmatch, \pbuild)$ where $\view$
  and $\Sigma$ are non-empty sets. Functions $\pmatch : \Sigma \pfun \view$ and
  $\pbuild : \view \to \Sigma$ satisfy the following laws:
$$\pmatch(\pbuild~x) = x \qquad\quad y \in \dom(\pmatch) \implies \pbuild~(\pmatch~y) = y$$ We write $X : V \pto E$ if $X$ is a prism with $\Sigma_X = E$ and $\view_X = V$.
\end{definition}
\noindent Intuitively, a prism abstractly characterises a datatype constructor, $E$, taking a value of type $\mathcal{V}$. Then, $\pbuild$ is the constructor, and $\pmatch$ is the destructor, which is partial due to the possibility of several disjoint constructors. For CSP, each prism models a channel in $E$ carrying a value of type $\view$. We have created a command \isakwmaj{chantype}, which automates the creation of prism-based event alphabets. Technically, this is achieved by creating an algebraic data type with a constructor for each channel and a corresponding prism for each constructor.

CSP processes typically do not return data, though their components may, and so they are typically denoted as ITrees of type $(E, ())\skey{itree}$, returning the unit type $()$. An example is $\skey{skip} \defs \Ret~()$, which is a degenerate form of $\Ret$. We now define the basic CSP operators.

\begin{definition}[Basic CSP Constructs] \label{def:basic-constructs} \isalink{https://github.com/isabelle-utp/interaction-trees/blob/ff9f73f98c653b265bd9da55689715cf973499c1/ITree_CSP.thy\#L7}
  \vspace{-1ex}
  
  \begin{align*}
    \skey{inp} &:: (V \pto E) \Rightarrow V~\skey{set} \Rightarrow (E, V)\skey{itree} \\
    \skey{inp}~c~A &\defs \Vis~(\lambda e \in \dom(\pmatch_c) \cap \pbuild_c \limg A \rimg @ \Ret~(\pmatch_c~e))
  \end{align*}

  \vspace{-1ex}
  
  \begin{minipage}{.5\linewidth}
  \begin{align*}
  \skey{outp} &:: (V \pto E) \Rightarrow V \Rightarrow (E, ())\skey{itree} \\[-.5ex]
    \skey{outp}~c~v &\defs \Vis~\{\skey{build}_c~v \mapsto \Ret~()\}
  \end{align*}
  \end{minipage}
  \begin{minipage}{.5\linewidth}
  \begin{align*}
    \skey{guard}~b &:: \mathbb{B} \Rightarrow (E, ())\skey{itree} \\[-.5ex]
    \skey{guard}~b &\defs (\textit{if}~b~\textit{then}~\skey{skip}~\textit{else}~\skey{stop})
  \end{align*}
\end{minipage}

\end{definition}
\noindent An input event ($\skey{inp}~c~A$) permits any event over the channel $c$, that is $e \in \dom(\pmatch_c)$, provided that its parameter is in $A$ ($e \in \pbuild_c \limg A \rimg$). It returns the value received for use by a continuation. It corresponds to the \texttt{trigger} construct in \cite{ITrees2019}. With this and monadic bind, the usual CSP input prefix can be denoted as $$c?x \then P(x) \defs (\skey{inp}~c~\textit{UNIV} \mbind P)$$ where $UNIV$ is the set of all values of a particular type. The input prefix receives any value over $c$ and then passes it on to $P$.

An output event ($\skey{outp}~c~v$) permits a single event, $v$, on channel $c$ and returns a null value of type $()$. We can then denote the standard CSP output prefix as $$c!v \then Q \defs (\skey{outp}~c~v \mbind (\lambda x.\, Q)$$ We also define the special case $\skey{sync}~e \defs \skey{outp}~e~()$ for a basic event $e :: () \pto E$. A $\skey{guard}~b$ behaves as $\skey{skip}$ if $b = true$ and otherwise deadlocks. It corresponds to the guard in CSP, which can be defined as $b \guard P \defs (\skey{guard}~b \mbind (\lambda x.\, P))$.

Using the monadic ``do'' notation, which boils down to applications of $\mbind$, we can now write simple reactive
programs such as $\skey{do} \{ x \leftarrow \skey{inp}~c; \skey{outp}~d~(2 \cdot x); \Ret~x \}$, which inputs $x$ over
channel $c : \nat \pto E$, outputs $2 \cdot x$ over channel $d$, and finally terminates, returning $x$.

Next, we define the external choice operator, $P \extchoice Q$, where the environment resolves the choice with an initial event of $P$ or $Q$. In CSP, $\extchoice$ can also introduce nondeterminism; for example, $(a \then P) \extchoice (a \then Q)$ introduces an internal choice since the $a$ event can lead to $P$ or $Q$, and is equal to $a \then (P \intchoice Q)$. Since we explicitly wish to avoid introducing such nondeterminism, we make a design choice to exclude this possibility by construction. There are other possibilities for handling nondeterminism in ITrees, which we consider in \S\ref{sec:concl}. As for $\mbind$, we define external choice corecursively using a set of ordered equations.

\begin{definition}[External choice] \label{def:extchoice} $P \extchoice Q$, is defined by the following set of equations: \isalink{https://github.com/isabelle-utp/interaction-trees/blob/ff9f73f98c653b265bd9da55689715cf973499c1/ITree_CSP.thy\#L75}

\begin{minipage}{.35\linewidth}
\begin{align*}
    (\Vis~F) \extchoice (\Vis~G) &= \Vis~(F \odot G) \\
    (\Sil~P') \extchoice Q &= \Sil~(P' \extchoice Q) \\
    P \extchoice (\Sil~Q') &= \Sil~(P \extchoice Q')
\end{align*}
\end{minipage}
\begin{minipage}{.65\linewidth}
\begin{align*}
    (\Ret~x) \extchoice (\Vis~G) &= \Ret~x \\
    (\Vis~F) \extchoice (\Ret~y) &= \Ret~y \\
    (\Ret~x) \extchoice (\Ret~y) &= (\textit{if}~x = y~\textit{then}~(\Ret~x)~\textit{else}~\skey{stop})
\end{align*}
\end{minipage}

\vspace{1ex}

where $F \odot G \defs (\dom(G) \ndres F) \oplus (\dom(F) \ndres G)$
\end{definition}
An external choice between two functions, $F$ and $G$, essentially combines all the choices presented using $F \odot G$. The caveat is that if the domains of $F$ and $G$ overlap, then any events in common are excluded. Thus, $\odot$ restricts the domain of $F$ to maplets $e \mapsto P$ where $e \notin \dom(G)$, and vice-versa. This has the effect that $(a \then P) \extchoice (a \then Q) = \skey{stop}$, for example. In the special case that $\dom(F) \cap \dom(G) = \emptyset$, $P \odot Q = P \oplus Q$. We chose this behaviour to ensure that $\extchoice$ is commutative, though we could alternatively bias one side.

Internal steps on either side of $\extchoice$ are greedily consumed. Due to the equation order, $\tau$ events have the highest priority, following a maximal progress assumption~\cite{Hennessy1995TPL}. Return events also have priority over visible events. If two returns are present, then they must agree on the value. Otherwise, they deadlock. External choice satisfies several essential properties:
\begin{theorem}[External Choice Properties] \isalink{https://github.com/isabelle-utp/interaction-trees/blob/df092d827c91393ea5b29a0cece4567380a8c931/ITree_CSP.thy\#L231}
  $$P \extchoice Q = Q \extchoice P \quad \skey{stop} \extchoice P = P \quad \skey{div} \extchoice P = \skey{div} \quad P \extchoice (\tau^n~Q) = (\tau^n~P) \extchoice Q = \tau^n (P \extchoice Q)$$

  \vspace{-3ex}
  
  $$(\Vis~F \extchoice \Vis~G) \mbind H = (\Vis~F \mbind H) \extchoice (\Vis~G \mbind H)$$
\end{theorem}
The external choice is commutative and has \skey{stop} as a unit. It has \skey{div} as an annihilator because the $\tau$ events mean no other activity is chosen. A finite number of $\tau$ events on the left or right can be extracted to the front. Finally, bind distributes from the left across a visible event choice. We prove these properties using coinduction (\cref{thm:coind}), case analysis on stability of constituent processes, followed by several invocations of \skey{sledgehammer} to discharge the resulting provisos.

Using the operators defined so far, we can implement the buffer from Examples~\ref{ex:buffer} using a monadic syntax: \isalink{https://github.com/isabelle-utp/interaction-trees/blob/7695143479e6f604209545c500db1c6ee6d25faa/examples/ITree_CSP_Examples.thy\#L23}
  
\begin{alltt}
\isakwmaj{chantype} Chan = Input::int  Output::int  State::"int list"

\isakwmaj{definition} buffer :: "int list \(\Rightarrow\) (Chan, int list) itree" \isakwmin{where}
"buffer = loop (\(\lambda\) s. 
                 do \{ i \(\leftarrow\) inp Input \{0..\}; Ret (s @ [i]) \}
               \(\extchoice\) do \{ guard(length s > 0); outp Output (hd s); Ret (tl s) \}
               \(\extchoice\) do \{ outp State s; Ret s \})"
\end{alltt}
We first create a channel type \texttt{Chan}, which has channels (prisms) for inputs and outputs and to view the current buffer state. We define the buffer process as a simple loop with a choice of three branches inside. The variable \texttt{s::int list} denotes the state. The first branch allows a value to be received over \texttt{Input}, and then returns \texttt{s} with the new \texttt{i} value appended, and then iterates. The second branch is only active when the buffer is not empty. It outputs the head on \texttt{Output} and returns the tail. The final branch outputs the current state. In \S\ref{sec:animation}, we will see how such an example can be animated.

Next, we tackle parallel composition. The objective is to define the usual CSP operator $P \parallel[E] Q$, which requires that $P$ and $Q$ synchronise on the events in $E$ and can otherwise evolve independently. We first define an auxiliary operator for merging choice functions.
\begin{align*}
   merge_E(F, G) &= (\lambda e \in \dom(F) \setminus (\dom(G) \cup E) @ \skey{Left}(F(e))) \\
                 &\,\oplus (\lambda e \in \dom(G) \setminus (\dom(F) \cup E) @ \skey{Right}(G(e))) \\
                 &\,\oplus (\lambda e \in \dom(F) \cap \dom(G) \cap E @ \skey{Both}(F(e), G(e))
\end{align*}
Operator $merge_E(F, G)$ merges two event functions, which are being offered by two parallel composed ITrees. Each event is tagged depending on whether it occurs on the $\skey{Left}$, $\skey{Right}$, or $\skey{Both}$ sides of a parallel composition. An event in $\dom(F)$ can occur independently when not in $E$ or $\dom(G)$. The latter proviso is required, like for $\extchoice$, to prevent nondeterminism by disallowing the same event from occurring independently on both sides. An event in $\dom(G)$ can occur independently through the symmetric case for $\dom(F)$. An event can synchronise provided it is in the domain of choice functions and the set $E$. We use this operator to define the generalised parallel composition. For the sake of presentation, we present partial functions as sets.

\begin{definition} $P \parallel_E Q$ is defined corecursively by the following equations: \isalink{https://github.com/isabelle-utp/interaction-trees/blob/7695143479e6f604209545c500db1c6ee6d25faa/ITree_CSP.thy\#L321}
\begin{align*}
    (\Vis~F) \parallel_E (\Vis~G) &= 
        \Vis\left(\begin{array}{l}
            \{e \mapsto (P' \parallel_E (\Vis~G)) | (e \mapsto \skey{Left}(P')) \in merge_A(F, G)\} \\
            \oplus~ \{e \mapsto ((\Vis~F) \parallel_E Q') | (e \mapsto \skey{Right}(Q')) \in merge_E(F, G)\} \\
            \oplus~ \{e \mapsto (P' \parallel_E Q') | (e \mapsto \skey{Both}(P', Q')) \in merge_E(F, G)\}
        \end{array}\right) \\
    (\Sil~P') \parallel_E Q &= \Sil~(P' \parallel_E Q) \qquad P \parallel_E (\Sil~Q') = \Sil~(P \parallel_E Q') \\
    (\Ret~x) \parallel_E (\Ret~y) &= \Ret~(x, y) \\
    (\Ret~x) \parallel_E (\Vis~G) &= \Vis~\{e \mapsto \Ret~x \parallel_E Q' | (e \mapsto Q') \in G\} \\
    (\Vis~F) \parallel_E (\Ret~y) &= \Vis~\{e \mapsto P' \parallel_E \Ret~y | (e \mapsto P') \in F\}
\end{align*}

\end{definition}
The most complex case is for $\Vis$, which constructs a new choice function by merging $F$ and $G$. Three partial functions again represent the three cases. The first two allow the left and right to evolve independently to $P'$ and $Q'$, respectively, using one of their enabled events, leaving their opposing side, $\Vis~G$ and $\Vis~F$, respectively, unchanged. The third case allows them both to evolve simultaneously on a synchronised event.

The $\Sil$ cases allow $\tau$ events to happen independently and with priority. If both sides can return a value, $x$ and $y$, respectively, then the parallel composition returns a pair, which can later be merged if desired. The final two cases show what happens when only one side has a return value and the other has visible events. In this case, the $\Ret$ value is retained and pushed through the parallel composition until the other side also terminates.

We use $\parallel_E$ to define two special cases for CSP: $P \parallel[E] Q \defs (P \parallel_E Q) \mbind (\lambda (x, y).\, \Ret~())$ and $P \interleave Q \defs P \parallel[\emptyset] Q$.  As usual in CSP, these operators do not return any values, and so $P, Q :: (E, ())\skey{itree}$. The $P \parallel[E] Q$ operator is similar to $\parallel_E$, except if both sides terminate, any resultant values are discarded, and a null value is returned. This is achieved by binding to a simple merge function. $P$ and $Q$ do not return values, so this does not affect the behaviour, just the typing. The interleaving operator $P \interleave Q$, where there is no synchronisation, is defined as $P \parallel[\emptyset] Q$. We prove several algebraic laws:
\isalink{https://github.com/isabelle-utp/interaction-trees/blob/7695143479e6f604209545c500db1c6ee6d25faa/ITree_CSP.thy\#L476}
  $$(P \parallel_E Q) = (Q \parallel_E P) \mbind (\lambda (x, y).\, \Ret~(y, x)) \quad \skey{div} \parallel_E P = \skey{div}$$
  $$P \parallel[E] Q = Q \parallel[E] P \quad P \interleave Q = Q \interleave P \quad \skey{skip} \interleave P = P$$
Parallel composition is commutative, except that we must swap the outputs, and so $\parallel[E]$ and
$\interleave$ are commutative as well. Parallel has $\skey{div}$ as an annihilator for similar reasons to
$\extchoice$. For $\interleave$, $\skey{skip}$ is a unit since there is no possibility of communication and no values
are returned.

The final operator we consider is hiding, $P \hide A$, which turns the events in $A$ into $\tau$s:

\begin{definition}[Hiding] $P \hide A$ is defined corecursively by the following equations: \isalink{https://github.com/isabelle-utp/interaction-trees/blob/7695143479e6f604209545c500db1c6ee6d25faa/ITree_CSP.thy\#L571}
\begin{align*}
    \Vis(F) \hide A &= 
        \begin{cases}
            \Sil~ (F(e) \hide A) & \text{if } A \cap \dom(F) = \{e\} \\
            \Vis~ \{(e, P \hide A) | (e, P) \in F\} & \text{if } A \cap \dom(F) = \emptyset\\
            \skey{stop} & \text{otherwise}
        \end{cases} \\
    \Sil(P) \hide A & = \Sil(P \hide A) \qquad \Ret~x \hide A = \Ret~x
\end{align*}

\end{definition}
\noindent We consider a restricted version of hiding where only one event can be hidden at a time to avoid nondeterminism. When hiding the events of $A$ in the choice function $F$, there are three cases: (1) there is precisely one event $e \in A$ enabled, in which case it is hidden; (2) no enabled event is in $A$, in which case the event remains visible; (3) more than one $e \in A$ is enabled, and so we deadlock. We again impose maximal progress here so that an enabled event to be hidden is prioritised over other visible events: $(a \then P \mathop{\vbar} b \then Q) \hide \{a\} = \tau P$, for example. Despite the significant restrictions on hiding, it supports the typical pattern where one output event is matched with an input event. Moreover, a priority can be placed on the order in which events are hidden, rather than deadlocking, by sequentially hiding events. Hiding can introduce divergence, as the following theorem shows: $(\skey{iter}~(\skey{sync}~e)) \hide {e} = \skey{div}$.

\subsection{Circus}

While CSP processes can be parametrised to allow modelling states, there is no support for explicit state operators like assignment. The $do$ notation somewhat allows variables, but these are immutable and are not preserved across iterations. \Circus~\cite{Woodcock2001-Circus, Oliveira&09} is a CSP extension allowing state variables.

We can characterise \Circus through a Kleisli lifting of CSP processes that return values so that \Circus actions are homogeneous KTrees. Then, thanks to the compositionality of our ITree-based semantics, we can use the operators defined in \S\ref{sec:rel}, such as assignment $x := e$, to allow manipulation of the state. Then, we define the core operators for concurrency:
\begin{definition}[Circus Operators] \label{def:circus-ops} \isalink{https://github.com/isabelle-utp/interaction-trees/blob/4bdea2d0a52341e7a19abc3950a3bcdd4b65e7fd/ITree_Circus.thy\#L7}
\begin{align*}
  c?x{:}A \then F(x) &\defs (\lambda s.\, \skey{inp}~c~A \mbind (\lambda x.\, F(x)~s)) \\
  c!e \then P &\defs (\lambda s.\, \skey{outp}~c~(e~s) \mbind (\lambda x.\, P~s)) \\
  P \extchoice Q &\defs (\lambda s.\, P(s) \extchoice Q(s)) \\
  P \sfpar{ns_1}{E}{ns_2} Q &\defs \left(\lambda s.\, (P(s) \parallel_E Q(s)) \mbind (\lambda (s_1, s_2).\, \lovrd{\lovrd{s}{s_1}{ns_1}}{s_2}{ns_2})\right)
\end{align*}
\end{definition}
\noindent The operators are defined by the lifting of their CSP equivalents. An output $c!e \then P$ carries an expression $e$ rather than a value, which can depend on the state variables. The main complexity is the \Circus parallel operator, $P \sfpar{ns_1}{E}{ns_2} Q$, which allows $P$ and $Q$ to act on disjoint portions of the state, characterised by the name sets $ns_1$ and $ns_2$. We represent $ns_1$ and $ns_2$ as independent lenses, $ns_1 \lindep ns_2$, though they can be thought of as sets of variables with $ns_1 \cap ns_2 = \emptyset$. The definition of the operator first lifts $\parallel_E$ and composes this with a merge function. The merge function constructs a state consisting of the $ns_1$ region from the final state of $P$, the $ns_2$ region from $Q$, and the remainder from the initial state $s$. This is achieved using the lens override operator $\lovrd{s_1}{s_2}{X}$, which extracts the region described by $X$ from $s_2$ and overwrites the corresponding region in $s_1$, leaving the complement unchanged.

We can now model the buffer from Example~\ref{ex:buffer} with these operator definitions. Given a state variable \texttt{buf::int list}, the buffer can be expressed in Isabelle/HOL as follows:

\begin{example}{Buffer in ITree-based \Circus} \isalink{https://github.com/isabelle-utp/interaction-trees/blob/9bfc25ecdce8ccb5d400c71f89921c145bd00e63/examples/Buffer_Circus.thy}
\begin{lstlisting}
  buf := []; 
  loop ((Input?(i) ~→~ buf := buf @ [i])
        ~□~ (length(buf) > 0 & Output!(hd buf) ~→~ buf := tl buf)
        ~□~ State!(buf) ~→~ Skip)
\end{lstlisting}
\end{example}
\vspace{-3ex}

\noindent We update the state with assignments threaded through sequential composition.

Our \Circus operators satisfy several standard laws~\cite{Oliveira&09,Foster2021-JLAMP}, beyond the CSP laws, for example: \isalink{https://github.com/isabelle-utp/interaction-trees/blob/4bdea2d0a52341e7a19abc3950a3bcdd4b65e7fd/ITree_Circus.thy\#L44}
  \begin{align*}
    \assigns{\sigma} \fatsemi (P \extchoice Q) &~=~ (\assigns{\sigma} \fatsemi P) \extchoice (\assigns{\sigma} \fatsemi Q) \\
    P \sfpar{ns_1}{E}{ns_2} Q &~=~ Q \sfpar{ns_2}{E}{ns_1} P & \text{if } ns_1 \lindep ns_2
  \end{align*}
State updates are distributed through external choice from the left. \Circus parallel composition is commutative, provided that we also switch the name sets.

\subsection{Denotational Semantics}
\label{sec:sem-links}

Next, we show how ITrees are related to the standard failures-divergences semantics of CSP~\cite{Brookes1984}. The utility of this link is to both allow symbolic verification of ITrees and allow them to act as a target of step-wise refinement. In this way, we can use the existing mechanisation of the CSP set-based and relational semantics~\cite{Taha2020CSP-Isabelle, Foster2021-JLAMP} to capture and reason about nondeterministic specifications and use ITrees to provide executable implementations. %

In the failures-divergences model, a process is characterised by two sets: $F :: (E^\tick~\skey{list} \, \times \, \textit{E}~\skey{set})~\skey{set}$ and $D :: \power (E~\skey{list})$, which are, respectively, the set of failures and divergences. A failure is a trace of events plus a set of events that can be refused at the end of the interaction. A divergence is a trace of events that leads to divergent behaviour. A distinguished event $\tick \in E$ is used as the final element of a trace to indicate that this is a terminating observation.

For example, consider the process $a \then c \then \skey{skip} \extchoice b \then \skey{div}$, which initially permits an $a$ or $b$ event, and following $a$ permits a $c$ event. It exhibits the failure $([], \{c\})$ since before any events are performed, the event $c$ is being refused. A second failure is $([a], \{a, b\})$, since after performing an $a$, only $c$ is enabled, and the other events are refused. A third failure is $([a, c, \tick], \{a, b, c\})$, which represents successful termination, after which all events are refused. This process also has a divergence trace $[b]$ since the process diverges after performing event $b$. If a divergent state is unreachable, then $D$ is empty. Here, we show how to extract $F$ and $D$ from any ITree, and thus processes constructed from the operators of \S\ref{sec:csp-circus}.

In CSP, one likes to show that there are no divergent states, a property called divergence freedom. The following inductive-coinductive definition captures it:

\begin{definition}[Divergence Freedom] \isalink{https://github.com/isabelle-utp/interaction-trees/blob/4bdea2d0a52341e7a19abc3950a3bcdd4b65e7fd/ITree_Divergence.thy\#L30}
  $$\begin{array}{cccc}
  \begin{array}{c}
     - \\ \hline \ret{x} \SEarrow \mathcal{R}
  \end{array} &
  \begin{array}{c}
     P \SEarrow \mathcal{R} \\ \hline \tau P \SEarrow \mathcal{R}
  \end{array} &
  \begin{array}{c}
    \ran(F) \subseteq \mathcal{R} \\ \hline \Vis~F \SEarrow \mathcal{R}
  \end{array} &
    \skey{div-free} \defs \bigcup \, \{ \mathcal{R} | \mathcal{R} \subseteq \{ P | P \SEarrow \mathcal{R} \} \}
    \end{array}
  $$  
\end{definition}
Predicate $P \SEarrow \mathcal{R}$ is defined inductively. It requires that $P$ stabilises to a $\Ret$ or a $\Vis$ whose continuations are all contained in $\mathcal{R}$. Then, \skey{div-free} is the largest set consisting of all sets $\mathcal{R} = \{P | P \SEarrow \mathcal{R}\}$ and is coinductively defined. If we can find an $\mathcal{R}$ such that for every $P \in \mathcal{R}$, it follows that $P \SEarrow \mathcal{R}$, that is $\mathcal{R}$ is closed under stabilisation, then any $P \in \mathcal{R}$ is divergence-free. Essentially, $\mathcal{R}$ needs to enumerate the symbolic post-stable states of an ITree; for example, $\mathcal{R} = \{\skey{run}~E\}$ satisfies the provisos and so $\skey{run}~E$ is divergence-free. We have proved that $P \in \skey{div-free} \iff (\nexists s @ P \xrightarrow{s} \skey{div})$, which gives the operational meaning.

With our transition relation, we can define Roscoe's step relation, which links the operational and denotational semantics of CSP~\cite[Section~9.5]{Roscoe2010-UCS}. The utility of this definition and the following theorems is to permit symbolic verification of CSP processes by calculating their set-based characterisation.

\begin{definition}[Roscoe's Step Relation] \isalink{https://github.com/isabelle-utp/interaction-trees/blob/4bdea2d0a52341e7a19abc3950a3bcdd4b65e7fd/ITree_FDSem.thy\#L46}
$$(P \xRightarrow{s} P') \defs ((\exists t \in \Sigma\,\skey{list} @ s = t \append [\ret{x}] \land P \xrightarrow{t} \ret{x} \land P' = \skey{stop}) \lor (set(s) \subseteq \Sigma \land P \xrightarrow{s} P'))$$
\end{definition}
Here, $set(s)$ extracts the set of elements from a list. The step relation is similar to $\xrightarrow{s}$, except that the event type is adjoined with a special termination event $\ret{}$. We define the enlarged set $\Sigma^\checkmark \defs \Sigma \cup \{ \ret{x} | x \in \src\}$, which adds a family of events parametrised by return values, as in the semantics of Occam~\cite{Roscoe1984-Occam}, which derives from CSP. A termination is signalled when the transition relation reaches a $\Ret~x$ in the ITree, where the trace is augmented with $\ret{x}$ and the successor state is set to $\skey{stop}$. We often use a condition of the form $set(s) \subseteq \Sigma$ to mean that no $\ret{x}$ event is in $s$. We can now define the sets of traces, failures, and divergences~\cite{Roscoe2010-UCS}:

\begin{definition}[Traces, Failures, and Divergences] \isalink{https://github.com/isabelle-utp/interaction-trees/blob/4bdea2d0a52341e7a19abc3950a3bcdd4b65e7fd/ITree_FDSem.thy\#L77}
  \begin{align*}
    \skey{traces}(P) &\defs \{s | set(s) \subseteq \Sigma^\checkmark \land (\exists P' @ P \xRightarrow{s} P')\} \\
    P \mathop{\skey{ref}} E &\defs ((\exists F @ P = \Vis~F \land E \cap \dom(F) = \emptyset) \lor (\exists x @ P = \Ret~x \land \ret{x} \notin E)) \\
    \skey{failures}(P) &\defs \left\{(s, X) | set(s) \subseteq \Sigma^\checkmark \land (\exists Q @ P \xRightarrow{s} Q \land Q \mathop{\skey{ref}} X)\right\} \\
    \skey{divergences}(P) &\defs \{s \append t | set(s) \subseteq \Sigma \land set(t) \subseteq \Sigma \land (\exists Q @ P \xRightarrow{s} Q \land \divergent{Q})\}
  \end{align*}
\end{definition}
\noindent The set $\skey{traces}(P)$ is the set of all possible event sequences that $P$ can perform. For $\skey{failures}(P)$, we need to determine the set of events that an ITree is refusing, $P \mathop{\skey{ref}} E$. If $P$ is a visible event, $\Vis~F$, then any set of events $E$ outside of $\dom(F)$ is refused. If $P$ is a return event, $\Ret~x$, then every event other than $\ret{x}$ is refused. With this, we can implement Roscoe's form for the failures. Finally, the divergences is simply a trace $s$ leading to a divergent state $\divergent{Q}$, followed by any trace $t$. We exemplify these definitions with two calculations of failures:
\begin{align*}
  \skey{failures}(\skey{inp}~c~A) &=
    \begin{array}{l}
      \{([], E) | \forall x \in A @ c.x \notin E \} \cup \{([c.x], E) | x \in A \land \ret{} \notin E\} \\
      \cup~ \{([c.x, \ret{()}], E) | x \in A\}
    \end{array} \\[1ex]
  \skey{failures}(P \mbind Q) &=
    \begin{array}{l}
      \{(s, X) | set(s) \subseteq \Sigma \land (s, X \cup \{\ret{x} | x \in \src\}) \in \skey{failures}(P)\} \\
      \cup~ \{(s \append t, X) | \exists v @ s \append [\ret{v}] \in \skey{traces}(P) \land (t, X) \in \skey{failures}(Q(v))\}
    \end{array}
\end{align*}
The failures of $\skey{inp}~c~A$ consist of
\begin{inparaenum}
\item the empty trace, where no valid input on $c$ is refused;
\item the trace where an input event $c.x$ occurred, and $\ret{()}$ is not being refused; and
\item the trace where both $c.x$ and $\ret{()}$ occurred, and every event is refused.
\end{inparaenum}
The failures of $P \mbind Q$ consist of
\begin{inparaenum}
\item the failures of $P$ that do not reach a return, and
\item the terminating traces of $P$, ending in $\ret{v}$ appended with a failure of $Q(v)$, the continuation.
\end{inparaenum}
With the help of Isabelle's simplifier, these equations can be used to calculate the failures and divergences automatically, which can be easier to reason with than directly applying coinduction.

We conclude this section with some important properties of our semantic model:

\begin{theorem}[Semantic Model Properties] \isalink{https://github.com/isabelle-utp/interaction-trees/blob/4bdea2d0a52341e7a19abc3950a3bcdd4b65e7fd/ITree_FDSem.thy\#L363}
  \begin{align*}
    & (s, X) \in \skey{failures}(P) \land (Y \cap \{x | s \append [x] \in \skey{traces}(P)\} = \emptyset) \implies (s, X \cup Y) \in \skey{failures}(P) \\
    & s \in \skey{divergences}(P) \land set(t) \subseteq \Sigma \implies s \append t \in \skey{divergences}(P) \\
    & P \approx Q \implies (\skey{failures}(P) = \skey{failures}(Q) \land \skey{divergences}(P) = \skey{divergences}(Q)) \\
    & P \in \skey{div-free} \iff \skey{divergences}(P) = \emptyset \\
    & P \in \skey{div-free} \implies (\forall s~a @ s \append [a] \in \skey{traces}(P) \implies (s, \{a\}) \notin \skey{failures}(P))
  \end{align*}
\end{theorem}
The first two are standard healthiness conditions of the failures-divergences model~\cite{Roscoe2010-UCS}, called \textit{\textbf{F3}} and \textit{\textbf{D1}}, respectively. \textit{\textbf{F3}} states that if $(s, X)$ is a failure of $P$ then any event that cannot subsequently occur after $s$, according to the \skey{traces}, must also be refused. \textit{\textbf{D1}} states that the set of divergences is extension closed. We have also proved that two weakly bisimilar processes have the same divergences and failures.  %
The following result links the coinductive definition of divergence freedom and the set of divergences. The final result demonstrates that ITrees satisfy Roscoe's definition of determinism for CSP~\cite{Roscoe2010-UCS}. If an ITree $P$ is divergence-free, there is no trace after which an event can be accepted and refused.

Finally, we have stronger results relating weak bisimulation with the trace and divergence semantics.

\begin{theorem}
  $P \approx Q \iff (\skey{traces}(P) = \skey{traces}(Q) \land \skey{divergences}(P) = \skey{divergences}(Q))$ \isalink{https://github.com/isabelle-utp/interaction-trees/blob/6be22531983c2020595fba6b67fb1910399d601f/UTP/ITree_FDSem.thy\#L786}
\end{theorem}

\noindent We can prove a weak bisimulation between $P$ and $Q$ by showing that these processes have the same traces and divergences. We do not need to consider the refusals because this level has no nondeterminism. Alternatively, we could consider nondeterminism similarly to that shown in \S\ref{sec:nondet} by introducing a distinguished event that the semantic model abstracts. In this case, the refusal information is vital, and this particular result would no longer hold.

\section{Animation by Code Generation}
\label{sec:animation}
This section shows how ITrees can be animated by code generation and develops a command called \lstinline{animate}. This command can be used to execute and probe the behaviour of an ITree-based model. In contrast to the \lstinline{execute} command of \S\ref{sec:rel}, it is interactive and requires the user to select a visible event to proceed.

The Isabelle code generator~\cite{Haftman2010-CodeGen, Haftmann2013-DataRefinement} can be used to extract code from (co)datatypes, functions, and other constructs to functional languages like SML, Haskell, and Scala. Although ITrees can be infinite, this is not a problem for languages with lazy evaluation so that we can step through the ITree's behaviour. Code generation then allows us to support the generation of verified animators and provides a potential route to correct implementations.

The main challenge is to find a suitable computable representation of partial functions. Whilst $A \pfun B$ is partly computable, we can only apply it to a value and see whether it yields an output. For animations and implementations, however, we typically want to determine a menu of enabled events for the user to select. Moreover, calculating semantics for CSP operators like $\extchoice$ and $\parallel$ requires us to compute with partial functions. For this, we need a way of calculating values for functions $\dom$, $\dres$, and $\oplus$, which is impossible for arbitrary partial functions. Instead, we need a concrete implementation and a data refinement~\cite{Haftmann2013-DataRefinement}. We choose associative lists as an implementation, $A \pfun B \approx (A \times B)~\skey{list}$, which limits us to finite constructions. However, it has the benefit of being easily printed, making the animator easier to implement. ITrees then have the following representation in Haskell:

\begin{figure}
  
  \centering
    \framebox{
      \includegraphics[width=13cm]{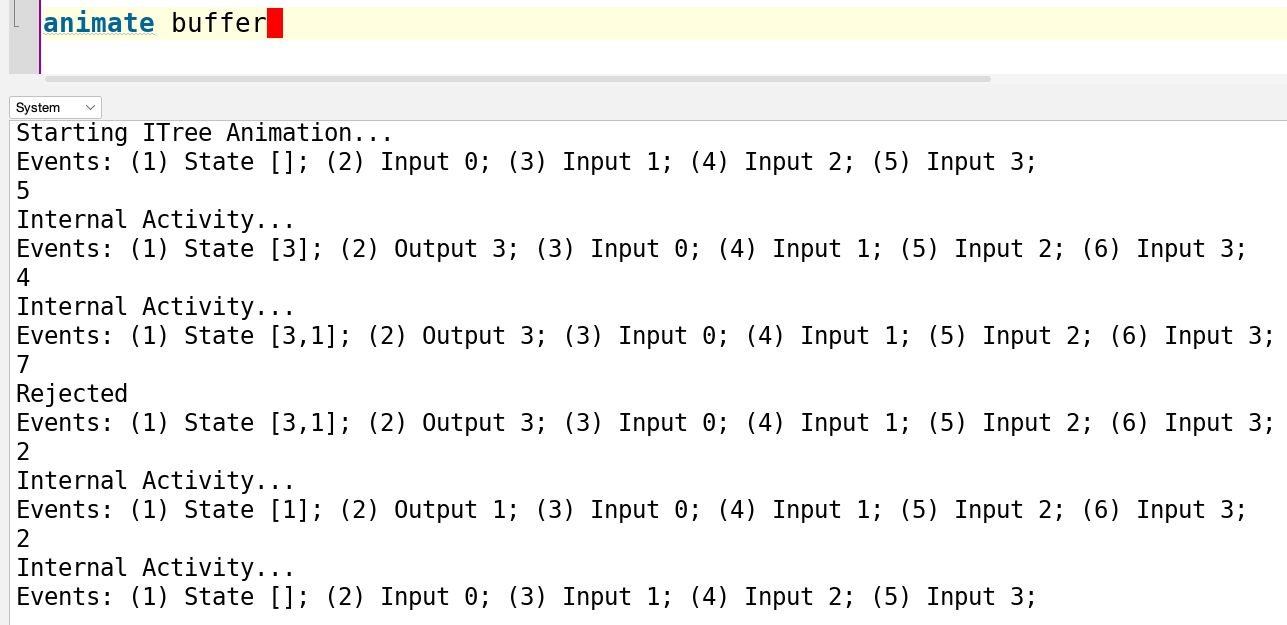}
      }

  \caption{Animating the CSP buffer}
  \label{fig:buffer}

  \vspace{-3ex}
\end{figure}

\lstset{language=Haskell}
\begin{lstlisting}
data Pfun a b = Pfun_alist [(a, b)];
data Itree a b = Ret b | Sil(Itree a b) | Vis(Pfun a (Itree a b))
\end{lstlisting}
Each of the semantic definitions detailed in sections \ref{sec:rel} and \ref{sec:csp-circus}, including corecursive functions, automatically map to Haskell functions operating over this structure. For constructs like $\skey{inp}$ (\cref{def:basic-constructs}), there is more work to support code generation since these can potentially produce an infinite number of events which an associative list cannot capture. Consider, for example, $\skey{inp}~c~\{0..\}$, for $c : \nat \pto E$, which can produce any event $c.i$ for $i \ge 0$. We can code generate this by limiting the value set to be finite, for example, $\{0..3\}$. Then, the code generator maps this to a list $[0,1,2,3]$, which is computable.

The code for the animator steps through $\tau$s until it reaches either a $\ret{x}$, in which case we terminate, or a $\Vis$, in which case the user can choose an option. Since divergence is a possibility, we limit the number of $\tau$s that will be skipped. The user can continue or abort the animation after $n = 20$ $\tau$ steps. If an empty event choice is encountered, the animation terminates due to deadlock. Otherwise, it displays a menu of events, allows the user to choose one, and recurses following the given continuation.

\lstset{language=Isabelle}

We only need to augment the generated code for a particular ITree with the animator code to generate an animator. We develop a command \lstinline{animate} (\href{https://github.com/isabelle-utp/interaction-trees/blob/master/simulation/ITree_Simulation.thy}{\isalogo}), which inputs a defined ITree and performs an animation. The command (1) runs the code generator, (2) adds the animator code, (3) compiles the code using the Glasgow Haskell Compiler (GHC), and (4) finally runs the binary on a console. This required us to modify Isabelle to add functionality in the PIDE editor interface to start the animation. Technically, this is provided by a new ``active area''\footnote{Please see \texttt{src/Pure/PIDE/active.ML} in the Isabelle source code for more information.}, which is a clickable part of the Output tab in the interface. When the user places their cursor over the \lstinline{animate} command in the editor, a ``Start animation'' link is shown, which the user can click to start the animation using the jEdit command-line console.

\cref{fig:buffer} shows an animation of the CSP buffer in \S\ref{sec:csp-circus}, with the possible inputs limited to $\{0.. 3\}$. We provide an empty list as a parameter for the initial state. The animator tells us the events enabled and allows us to pick one by typing a number into the console. Since lenses and expressions can also be code generated, we can also animate the \Circus version of the buffer with very similar output.

As a more sophisticated example, we have implemented a distributed ring buffer adapted from the original \Circus paper~\cite{Woodcock2001-Circus}. The idea is to represent a buffer as a ring of one-place cells and a controller that manages reading and writing to the ring.

It has the following form: \isalink{https://github.com/isabelle-utp/interaction-trees/blob/9bfc25ecdce8ccb5d400c71f89921c145bd00e63/examples/RingBuffer.thy} $$(Controller \parallel[\{rd.c, wrt.c | c \in \nat\}] \left(\Interleave i\in\{0..maxbuf\} @ Cell(i))\right) \hide \{rd.c, wrt.c | c \in \nat\}$$ where $rd.c$ and $wrt.c$ are internal channels for the controller to communicate with the ring, hidden in the overall process network. The individual cells do not communicate with each other, hence the use of interleaving $\interleave$, but the controller communicates with all cells. Channel $rd.c$ is used to read the current value of cell $c$, and $wrt.c$ is used to write a value. Each cell is a single place buffer with a state variable $val$ and has the following form:
\begin{definition}[Ring Buffer Cell]
$$Cell(i) \defs wrt?v \then val := v \relsemi loop~(wrt?v \then val := v \extchoice rd!val \then \skey{Skip})$$
\end{definition}
\noindent Initially, the cell is empty, awaiting a write command over channel $wrt$. Following this, the cell can either overwrite its current value or advertise its current value over channel $rd$. The cells are arranged through indexed interleaving, and the buffer size is $maxbuf + 1$. The channels $input$ and $output$ communicate with the overall buffer.

The controller has four state variables: (1) $sz :: \nat$, the current buffer size; (2) $rtop :: \nat$ a pointer to the next available cell; (3) $rbot :: \nat$ the index of the first value stored; (4) $cache :: \int$ the cached first element of the buffer. The controller is described using the following actions:

\begin{definition}[Ring Buffer Controller]
\begin{align*}
InputCtrl &\defs
\begin{array}{l}
sz < maxbuf \guard input?x \then \\
\left(\begin{array}{l}
sz = 0 \guard sz := 1 \relsemi cache := x \\
\extchoice sz > 0 \guard wrt.rtop!x \then sz := sz + 1 \relsemi rtop := (rtop + 1) \mathop{\textrm{mod}} maxring
\end{array}\right)
\end{array} \\ \\
OutputCtrl &\defs 
\begin{array}{l}
sz > 0 \guard output!cache \then \\
\left(\begin{array}{l}
sz > 1 \guard rd.rbot?x \then \begin{array}{l}sz := sz - 1 \relsemi cache := x \relsemi \\ rbot := (rbot + 1) \mathop{\textrm{mod}} maxring \end{array} \\
\extchoice sz = 1 \guard sz := 0
\end{array} \right)
\end{array} \\ \\
Controller &\defs  sz := 0 \relsemi rtop := 0 \relsemi rbot := 0 \relsemi loop~(InputCtrl \extchoice OutputCtrl)
\end{align*}
\end{definition}
\noindent $InputCtrl$ represents a controller input. An input can be accepted if the size is less than $maxbuf$. If the buffer is empty ($sz = 0$), the element is placed in the $cache$. Otherwise, it is sent to the next available cell at $rtop$. The index $rtop$ is updated using modulo arithmetic to characterise the circular nature of the buffer. $OutputCtrl$ represents a controller output. If the buffer is non-empty, then the buffer can output the cached head. Following this, if there is more than one element, the controller retrieves the element at $rbot$, decreases the buffer size, updates that cache, and finally updates the $rbot$ index. If the buffer only had one element, there is no buffer head to cache. The overall behaviour of the controller is to start empty, with both $rtop$ and $rbot$ pointing to index $0$, and then to iterate a choice between $InputCtrl$ and $OutputCtrl$.

We tested the animator on the ring buffer, using an Apple M3 Pro with 18GB of memory as the platform. We set up the example so that we can vary $maxbuf$ to observe the scalability of the animator. On this platform, we can efficiently animate this example for a relatively small ring of 100 cells, with a similar output to Figure~\ref{fig:buffer}, which is a very satisfying result. 

We were also able to animate with a much larger ring with 850 cells, which requires about 3 seconds to compute the next step. With 5000 cells, the animator takes around 50 seconds to calculate the next transition. The highest number of cells we could reasonably animate is around 2000. However, we have not attempted to optimise the code, and several data types could be replaced with efficient implementations to improve scalability. Thus, this approach to animation and potential implementation is very promising.

\section{System Modelling with Z-Machines}
\label{sec:zmachines}
In this section, we apply our ITree and \Circus library to create a formal modelling language and tool called ``Z-Machines'', which is in the style of the Z specification language~\cite{Spivey89, Woodcock96-UsingZ}, and B method~\cite{Abrial96BBook}. Z-Machines are a form of abstract machine, similar to B machines~\cite{Abrial96BBook}, that use our Z toolkit as the underlying expression language. The implementation of Z-Machines acts as a case study for our ITrees library by demonstrating its applicability in creating accessible verification tools. We implement several Isabelle commands for creating Z-Machine artefacts and a technique for verifying invariants. We illustrate these commands using a variant of the buffer example (\cref{ex:buffer}), which is bounded to a specific size and can be considered as a specification for the ring buffer in \S\ref{sec:animation}.

A Z-Machines consists of a set of operations that act over a state, formally:

\begin{definition}[Z-Machine]
  A Z-Machine consists of (1) a state space type $S$; (2) a set of state invariants $P_i : S \to \bool$; (3) a set of operations $Op_j : T_j \to ((), S)\skey{htree}$, each parametrised by $T_j$; and (4) an initialisation $I : S$. 
\end{definition}

\noindent Operations are used to update the value of state variables deterministically and can optionally take inputs and produce outputs. Each operation in a Z-Machine is given a semantics as a parametric homogeneous Kleisli tree. The parameters are used to encode inputs and outputs for the operation. The operations are composed in an action system~\cite{Back1989ActionSystems} to produce the overall Z-Machine ITree, which can be verified and animated.

We consider each command to create the Z-Machine components and their formal semantics. Each command is interpreted as a set of updates on the Isabelle document model, which creates definitions and other formal artefacts.

We use the \lstinline{zstore} command (\S\ref{sec:model-imper}), to create the bounded buffer state space:

\lstset{language=Isabelle}
\begin{example}[Bounded Buffer Store] $ $%

\begin{lstlisting}
  consts MAX_SIZE :: "nat" and VAL :: "int set"
  
  zstore Buffer = 
    sz  :: "nat"
    buf :: "int list"
  where 
    "sz = length buf"
    "sz ~≤~ MAX_SIZE"
\end{lstlisting}

\vspace{-2ex}
\end{example}

\noindent This creates two variables: $sz$ and $buf$, and links them using an invariant. Variable $sz$ represents the size of the buffer, and $buf$ is its contents. Using the \lstinline{consts} command, we also declare two abstract constants. \textit{MAX\_SIZE} characterises the maximum size of the buffer. \textit{VAL} is a finite set of integers representing the possible values we can insert into the buffer. These abstract constants can be assigned concrete definitions later for animation and verification. The store also has two invariants, requiring (1) that the buffer size is the same as the length and (2) that the size is no greater than the maximum size.

Operations are defined using the \lstinline{zoperation} command, which has the following syntax:

\begin{definition}[Operation Syntax] \isalink{https://github.com/isabelle-utp/Z_Machines/blob/5a9553830cebdc145e2d1267b4a0a6b215b1fa44/Z_Operations.thy}
\begin{align*}
  \textit{param} &::= name ~\bm{\in}~ term \\
  \textit{operation} &::= \textbf{zoperation}~name ~~ \textbf{params}~param^{*} ~~ \textbf{pre}~term ~~ \textbf{update}~assignment
\end{align*}
\end{definition}

\noindent An operation consists of a name, a set of parameters, a precondition term, and an update. A parameter consists of a name and a term, characterising the set of values from which the parameter is drawn. We use this set to bound the possible inputs to an operation to allow animation, similar to \S\ref{sec:animation}. The precondition acts as a guard for the operation, which must be satisfied for the operation to be executed. The update is a sequence of simultaneous assignments to variables in state space. 

As part of generating the Z-Machine semantics, each operation is assigned a unique event channel, $Op_j^c : T_j \pto E$, sharing the same name and as part of a generated channel type $E$. The semantics of an operation is shown below:

\begin{definition}[Operation Semantics] $ $%
\vspace{1ex}
  
$\left\llbracket
  \begin{array}{l}
    \textbf{zoperation}~Op_j \\
    ~~ \textbf{params}~x_1\!\bm{\in}\!A_1 ~\cdots~ x_n\!\bm{\in}\!A_n \\
    ~~ \textbf{pre}~P \\
    ~~ \textbf{update}~\sigma
  \end{array}\right\rrbracket = \left(Op_j \defs Op_j^c?\vec{x}\in \vec{A} \,|\, P(\vec{x}) \then \sigma(\vec{x})\right)
  $

$\text{where } \vec{x} = (x_1, \cdots, x_n) \text{ and } \vec{A} = A_1 \times \cdots \times A_n$
\end{definition}

\noindent Here, $A \times B$ denotes the Cartesian product of the two sets $A$ and $B$. An operation accepts parameters that inhabit the corresponding parameter sets ($\vec{A}$) and satisfies the precondition $P$ in
the context of the current state. The operation update is executed when such parameters are provided, with the parameters as inputs ($\vec{x}$). Parameter sets are specified as expressions, meaning
the acceptable parameters can vary from state to state. When the Z-Machine is in a state that satisfies the precondition $P$ of an operation for a particular valuation of parameters ($\vec{x}$), the event $Op_j^c.\vec{x}$ is enabled. If the current state cannot satisfy the precondition, an operation's behaviour is $\skey{Stop}$.

Below, we define three operations for the bounded buffer:

\begin{example}[Buffer Operations] $ $%

\begin{lstlisting}
  zoperation Input =
    params v~∈~VAL
    pre "sz < MAX_SIZE"
    update "[ sz' = sz + 1, buf' = buf @ [v] ]"

  zoperation Output =
    params v~∈~VAL
    pre "sz > 0" "v = hd buf"
    update "[ sz' = sz - 1, buf' = tl buf ]"

  zoperation Size =
    emit sz
\end{lstlisting}

\vspace{-2ex}
\end{example}

\noindent The first operation, \textit{Input}, adds a value to the buffer. It has one parameter, \textit{v}, drawn from \textit{VAL}, to enumerate the possible events. The precondition requires that the size is strictly less than the maximum size. The update increases the size and adds the new value to the buffer. The second operation, \textit{Output}, likewise has a single parameter, but this time represents an output. The precondition requires that the buffer is non-empty and that the output value is at the buffer's head. The update decreases the size and removes the head from the buffer. The third operation, \textit{Size}, allows us to view the current size of the buffer. It uses the keyword \lstinline{emit}, which is shorthand for an operation of a single output parameter equal to the given expression, in this case \textit{sz}. 

Once semantics have been assigned for each of the operations, we can give the overall semantics for the Z-Machine itself:

\begin{definition}[Z-Machine Syntax and Semantics] \isalink{https://github.com/isabelle-utp/Z_Machines/blob/5a9553830cebdc145e2d1267b4a0a6b215b1fa44/Z_Machine.thy}$ $%
\begin{align*}
  \textit{zmachine} &::= \textbf{zmachine}~name ~~ \textbf{init}~term ~~ \textbf{operations}~name^{*}
\end{align*}

  \vspace{1ex}

$\left\llbracket
  \begin{array}{l}
    \textbf{zmachine}~M \\
    ~~ \textbf{init}~~ \sigma \\
    ~~ \textbf{operations}~ Op_1 \cdots Op_n
  \end{array}\right\rrbracket = (M \defs \assigns{\sigma} \fatsemi loop (Op_1 \extchoice Op_2 \extchoice \cdots \extchoice Op_n))$
\end{definition}

\noindent A Z-Machine consists of an initialisation assignment ($\sigma$) and a set of operations. The Z-Machine initialises the state using $\sigma$ and then enters a loop where the user can choose an enabled operation for execution. Below is the Z-Machine for the bounded buffer:

\begin{example}[Bounded Buffer Z-Machine] $ $%

\begin{lstlisting}
zmachine Bounded_Buffer =
  init "[sz' = 0, buf' = []]"
  operations Input Output Size
\end{lstlisting}
  
\end{example}
\noindent This Z-Machine initialises $sz$ and $buf$, and collects together the operations.

\begin{figure}
  \centering
    \framebox{
      \includegraphics[width=9cm]{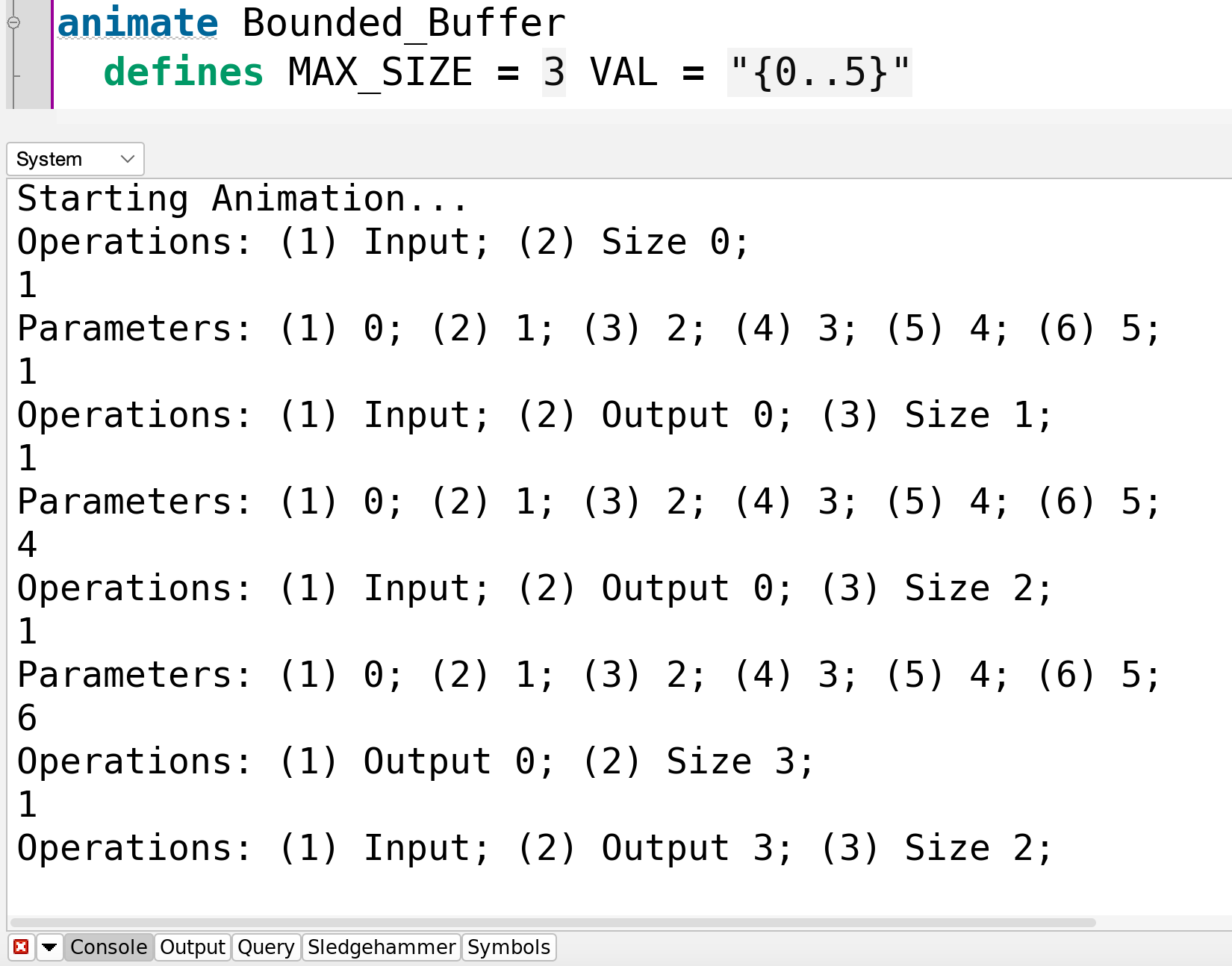}
      }

  \caption{Animating the Bounded Buffer Z-Machine}
  \label{fig:boundedbuffer}

\end{figure}

Z-Machines can be animated using the \lstinline{animate} command developed in \S\ref{sec:animation}, provided each operation draws parameters from finite sets or enumerable types. If this is not the case, the code generator will give an error message, and similarly, if any operation uses an undefined abstract constant. An example animation is shown in Figure~\ref{fig:boundedbuffer}. The animator displays the enabled operations and parameter combinations at each point, and the user can select one. Since the animator needs to enumerate all possibilities, we supply a finite set \textit{VAL} and limit the buffer to size 3.

Verification of Z-Machines involves identifying invariants that characterise specific critical properties. We then need to show that the initialisation establishes the invariants and that each operation preserves them, meaning that the machine satisfies them in any reachable state. Using the weakest preconditions, we can calculate the proof obligations (POs) necessary for an operation to maintain invariants. Specifically, we need to show that $P \implies \skey{wp}~Op_j~P$ for each operation. Typically, this involves showing that if the invariant holds for some state, then once all the updates of an operation are applied, the invariant still holds.

For example, if we consider operation \texttt{Input}, one PO is to show that if \texttt{sz = length buf} then also \texttt{sz + 1 = length (buf @ [v])}, after the updates for \texttt{sz} and \texttt{buf} are applied. This follows, since \texttt{sz + 1 = length buf + 1 = length (buf @ [v])}, a proof that can be automated by the simplifier.

We contribute a proof method called \texttt{zpog\_full} (Z proof obligation generator), which generates the set of POs required for an operation to satisfy the specification (\href{https://github.com/isabelle-utp/Z_Machines/blob/aabf191cd9a73b148c162c5f37265a66515f9dd2/Z_Operations.thy\#L185}{\isalogo}). This method is implemented by application of the weakest precondition laws using the simplifier (see \S\ref{subsec:hoarelogic}), and by Isabelle's automated deduction methods. The POs can typically be discharged with the help of the \lstinline{sledgehammer} tool. Below, we verify that the bounded buffer Z-Machine satisfies the invariants.

\begin{example}[Bounded Buffer Invariant Verification] $ $ %

\begin{lstlisting}
  lemma Init_correct: "Init establishes Buffer_inv"
    by zpog_full

  lemma Size_correct: "Size(n) preserves Buffer_inv"
    by zpog_full

  lemma Input_correct: "Input(v) preserves Buffer_inv"
    by zpog_full

  lemma Output_correct: "Output(v) preserves Buffer_inv"
    apply zpog_full
    apply (metis diff_le_self dual_order.trans)
    done
\end{lstlisting}

\vspace{-2ex}
\end{example}

\noindent Here, $C~\skey{establishes}~P$ is short-hand for the Hoare triple $\lamporttriple{True}{C}{P}$, and $C~\skey{preserves}~P$ is short-hand for $\lamporttriple{P}{C}{P}$. The proofs proceed by application of the weakest precondition laws by \texttt{zpog\_full}. For the first three proofs, the simplifier can solve the POs automatically. For the fourth proof obligation, some arithmetic reasoning is required, which we automate with a call to \lstinline{sledgehammer}. This produces a proof using the resolution prover \lstinline{metis} and several laws relating to the order on natural numbers.

\section{Related Work}
\label{sec:related}
Infinite trees are a ubiquitous model for concurrency~\cite{Glabbeek1997CCS-CSP}. In particular, ITrees can be seen as a restricted encoding of Milner's synchronisation trees~\cite{Milner1980, Winskel1984STrees, Milner1989}. In contrast to ITrees, synchronisation trees allow multiple events from each node, including visible and $\tau$ events. They have seen several generalisations, most recently by Ferlez et al.~\cite{Ferlez2014-GSTrees}, who formalise Generalized Synchronisation Trees based on partial orders, define bisimulation relations~\cite{Ferlez2018-BisimGSTrees}, and apply them to hybrid systems. Our work differs because ITrees use explicit computation and corecursion, but mutual insights will likely be gained.

ITrees~\cite{ITrees2019}, and their mechanisation in Coq, have been applied in various projects as a way of defining abstract yet executable semantics~\cite{KLL+19, ZHHZ20, MHA20, ZZF20, ZHK+21, LPZ21, SZ21}. They have been used to verify C programs~\cite{KLL+19} and an HTTP key-value server~\cite{LPZ21}.
Chappe et al.~\cite{Chappe2023CTrees} introduce Choice Trees as a conservative extension of ITrees. The main innovation is to add nondeterminism support through constructors $\skey{brS}$ and $\skey{brD}$, which replace the $\Sil$ constructor. Whereas $\Sil$ is deterministic, these two constructors allow a finite number of internal choices. Constructor $\skey{brS}$ represents ``stepped'' branching, where a $\tau$ transition accompanies the resolution of an internal choice. In contrast, $\skey{brD}$ is ``delayed'' branching, where deadlocked (``stuck'' in \cite{Chappe2023CTrees}) branches are eliminated from the choice. The latter choice is similar to the external choice since deadlocked branches are likewise pruned but more closely resembles angelic nondeterminism~\cite{Ribeiro2019Angelic}.

The Coq mechanisation of ITrees uses features unavailable in Isabelle, notably type constructor variables (rank-$n$ polymorphism). Though this is an apparent weakness, the quest to implement ITrees in the more restrictive type system of Isabelle/HOL has entailed several unique advantages. In \cite{ITrees2019}, the $\Vis$ constructor has two parameters, rather than one, for the output event $e : \mathcal{E}~\mathcal{A}$ and $k : \mathcal{A} \rightarrow itree~~\mathcal{E}~~\mathcal{R}$, a total function, for the continuation. There, $\mathcal{E}$ is a type constructor representing the output sent to the environment, which is parametric over $\mathcal{A}$, the type of answers received back from the environment. In contrast, our work instead (1) fixes a non-parametric event universe $E$; (2) uses a partial function of type $E \pfun (E, R)~\skey{itree}$; and (3) uses prisms~\cite{Pickering2017-Optics} to characterise channels.

Our one-parameter version of $\skey{Vis}$ is, in some respects, more general than the two-parameter one since it allows a variety of communication paradigms and a natural encoding of external choice~\cite{Brookes1984}. In \cite{ITrees2019}, the focus is on a communication scheme where (1) the process sends an output to the environment in $\mathcal{E}~\mathcal{A}$, and (2) the environment then answers back with a value in $\mathcal{A}$. In CSP, such a scheme can be encoded either with a single event (e.g. $c!y?x \then P(x)$), where the outputs and inputs are present as parameters, or via two separate events for the input and output communication (e.g. $c!y \then d?x \then P(x)$).

Additionally, Coq ITrees~\cite{ITrees2019} only permit choices to be made by the environment when the answer is returned, not in the output event itself. CSP's external choice operator has yet to be encoded in Coq ITrees, and \cite{Chappe2023CTrees} has only nondeterministic (i.e. internal) choice. Such an encoding could be achieved via a special event in $choose : \mathcal{E}~\mathcal{I}$ to represent that the ITree is asking the environment to resolve a choice, with $\mathcal{I}$ being the possible inputs. Similarly, modelling deadlock could use a special event $deadlock : \mathcal{E}~\emptyset$ with an empty return type. The disadvantage of such an encoding is that a custom notion of equality is required to reproduce algebraic laws such as $P \extchoice \skey{stop} = P$. We see our natural encoding of external choice as the central contribution of our work. At the same time, this additional generality comes at a cost since the interpretation combinator of the original works~\cite{ITrees2019}, which harnesses the output-input pattern to interpret events as monadic actions, requires more effort to encode.

The use of partial functions in our work means that external choice operators can be straightforwardly implemented by the composition of the underlying choice functions (e.g. using $P \odot Q$). This, in turn, means that the algebraic properties of the choice combinator lift to ITrees directly and allow flexible algebraic semantics. One technical exception to the generality of our work is the situation where an empty answer is used ($\mathcal{A} = \emptyset$), which in the Coq work allows a $\Vis$ that performs an output but has no continuations. Isabelle/HOL has no empty type since all types must exhibit one element and cannot support empty answers. This behaviour requires a slightly different encoding where the output is sent, and the process immediately deadlocks, though this requires two $\Vis$ operators rather than one. Aside from this situation, we can usually encode the two-parameter version $\Vis~e~k$ as $\vbar\, x\!\in\! \dom(\pmatch_e) \then k(\pmatch_e(x))$ with $e : A \pto E$.

A further benefit of having a fixed $E$ is that ITrees become simpler semantic objects. For example, traces can be represented simply as lists of events rather than the bespoke type used in \cite{ITrees2019}. These are amenable to first-order automated proof~\cite{Blanchette2016Hammers}, which has allowed us to develop our library quickly and with minimal effort.

Previously, we have demonstrated an Isabelle-based theory library and verification tool for reactive systems~\cite{Foster17c, Foster2021-JLAMP}. This supports verification and step-wise development of nondeterministic and infinite-state systems based on the CSP~\cite{Brookes1984, Hoare85} and \Circus~\cite{Woodcock2001-Circus} process languages. This includes a specification mechanism called reactive contracts and a calculational proof strategy. Extensions of our theory support reasoning about hybrid dynamical systems, which makes it ideal for verifying autonomous robots.

The Z notation has been implemented in a HOL-based theorem prover several times, notably in ProofPower-Z~\cite{Arthan2004ProofPower} and HOL-Z~\cite{Brucker2003}. HOL-Z is also implemented in Isabelle/HOL and includes a parser for Z schemas, formal semantics, and proof support. Our implementation is less advanced, does not have the Z schema calculus, and uses types rather than sets to characterise the hierarchy of data structures in Z. This aids proof automation through the type system but at the expense of fidelity to the Z standard. Nevertheless, our implementation of Z-Machines provides proof and animation support, provided that a Z model can be encoded in our restricted subset.

Several tools are available to analyse CSP processes, including the FDR refinement checker~\cite{fdr}, and the PAT model checker~\cite{SunLDP09}. These tools offer a more automated approach to analysis than formal proof, though they require the generation of explicitly labelled transition systems, which is hampered by the state explosion problem. FDR has a simple tool, ProBE, for exploring process behaviour by stepping through their definitions. PAT has a more feature-rich CSP simulator that allows users to perform various simulation tasks:
\begin{inparaenum}[(1)]
\item complete finite-state generation based on the execution graph;
\item automatic random simulation;
\item user interactive simulation with step-by-step execution and trace display and replay.
\end{inparaenum}
These tools are valuable complements that co-exist in the verification ecosystem. Our focus is on symbolic analysis of processes using proof, though our ITree animator can be applied to search-based analysis, similar to model checking. Moreover, tools like \lstinline{sledgehammer} increase automation, making verification via proof a realistic possibility.

Recently, the set-based theory of CSP has also been mechanised in the Isabelle-based HOL-CSP tool~\cite{Taha2020CSP-Isabelle,Crisafulli2023HOL-CSP}, which is also based on the failures-divergences model~\cite{Roscoe2005-TPC,Roscoe2010-UCS}. They have used their library to verify the deadlock-freedom of the famous ``dining philosophers'' example for an arbitrary number of philosophers $N \ge 2$. A further application is modelling and verifying autonomous vehicles, including the continuous dynamics~\cite{Crisafulli2023HOL-CSP}. This work is complementary to our library since HOL-CSP is not limited to deterministic constructions, but on the other hand, HOL-CSP process specifications are not executable. We hope to combine these libraries to realise these mutual benefits.

\section{Conclusions}
\label{sec:concl}
In this paper, we have demonstrated how Interaction
Trees~\cite{ITrees2019} can be used to unify animation and deductive
verification of software models, from high-level system models to
lower-level program models, in Isabelle/HOL. Our approach harnesses
the codatatype package~\cite{Blanchette2014BNF} to encode infinite
transition systems, and the code generator~\cite{Haftman2010-CodeGen,
Haftmann2013-DataRefinement} to provide animation and execution. Our
results indicate that the technique provides both tractable
verification, with the help of Isabelle's proof
automation~\cite{Blanchette2016Hammers}, and efficient
execution. Though ITrees are intrinsically deterministic, we have
shown how to model nondeterministic behaviour using special
events. We applied our technique to simple imperative programs, the
CSP and \Circus process languages~\cite{Oliveira&09}, and to an abstract machine notation
based on Z~\cite{Spivey89}. We note, however, that our library applies to various
other process algebraic and modelling languages.

Our work has many practical applications in producing verified
simulations, and we have several associated lines of ongoing work. In
a parallel paper~\cite{Ye2024-ITreeRoboChart}, we have used our ITree
library to mechanise semantics for the RoboChart
language~\cite{Miyazawa2019-RoboChart}, a formal UML-like language for
modelling robots with denotational semantics based on CSP. However,
this semantics does not yet consider the real-time operators, which
will require us to consider discrete time, which we believe can be
supported using a dedicated time event in ITrees, similar to
tock-CSP~\cite{Roscoe2005-TPC}, a timed version of CSP. This will build on our colleagues'
work with $\checkmark$-$\textit{tock}$~\cite{Baxter2021-TickTock}, a
new semantics for tock-CSP.

Separately, we have also used our Z-Machine formalism to give a
simplified semantics to RoboChart state
machines~\cite{Yan2023-ZMachineRoboChart}, for the purpose of
compositional invariant-based verification, including deadlock
checking. We plan to link Isabelle/HOL with the Eclipse-based RoboTool
modelling environment to allow seamless verification and feedback
for software engineers. This link will open up a pathway from
graphical models to verified implementations of autonomous robotic
controllers. In concert with this, we will also explore links to our
other theories for hybrid
systems~\cite{Foster2020-dL,Foster19b-HybridRelations}, to allow
verification of controllers in the presence of a continuously evolving
environment.

The work described in this paper has also been used pedagogically to support
two courses on assured software engineering, one for third-year undergraduates
and one for external industrial participants. Our courses use our implementation
of imperative programs and Hoare logic to teach program verification and Z-Machines
to teach Z-based formal specification. The benefit of this approach is that students
need only learn a single tool (Isabelle) to support the different pedagogical goals. Moreover,
Isabelle's document model has allowed us to create DSLs that support appropriate abstraction
levels to minimise the technical detail we expose to students. Our students' feedback
has been universally positive, and we plan to report further on this when we have more data.

In the future, we plan to link ITrees to our formalisation of
reactive contracts~\cite{Foster17c,Foster2021-JLAMP}, which provides
both denotational semantics for \Circus and a refinement calculus for
reactive systems, building on our link with failures-divergences. We will also continue to expand our imperative program verification
tool by considering more advanced concepts, like memory management and associated separation logics.
We will also consider generating imperative code using a suitable mechanised semantics
for a language target. Finally, we will provide a more
user-friendly interface for our animator, as found in tools like
FDR4's probe tool~\cite{fdr} and ProB~\cite{ProB} for Event-B.

\bibliography{references}

\begin{thebibliography}{10}

\bibitem{Abrial96BBook}
Jean-Raymond Abrial.
\newblock {\em The {B-Book}: assigning programs to meanings}.
\newblock Cambridge University Press, 1996.

\bibitem{Haftmann2012NBE}
K.~Aehlig, F.~Haftmann, and T.~Nipkow.
\newblock A compiled implementation of normalisation by evaluation.
\newblock {\em Journal of Functional Programming}, 22(1):9--30, 2012.

\bibitem{Armstrong2015}
A.~Armstrong, V.~Gomes, and G.~Struth.
\newblock Building program construction and verification tools from algebraic
  principles.
\newblock {\em Formal Aspects of Computing}, 28(2), 2015.

\bibitem{Arthan2004ProofPower}
R.~Arthan.
\newblock On formal specification of a proof tool.
\newblock In {\em Formal Software Development Methods}, volume 551 of {\em
  LNCS}. Springer, 1991.

\bibitem{Back1989ActionSystems}
R.-J. Back and R.~Kurki-Suonio.
\newblock Decentralization of process nets with centralized control.
\newblock {\em Distributed Computing}, 3:73--87, June 1989.

\bibitem{Baxter2021-TickTock}
J.~Baxter, P.~Ribeiro, and A.~Cavalcanti.
\newblock Sound reasoning in {tock-CSP}.
\newblock {\em Acta Informatica}, April 2021.
\newblock \href {https://doi.org/10.1007/s00236-020-00394-3}
  {\path{doi:10.1007/s00236-020-00394-3}}.

\bibitem{Blanchette2017Corec}
J.~C. Blanchette, A.~Bouzy, A.~Lochbihler, A.~Popescu, and D.~Traytel.
\newblock {Friends with Benefits: Implementing Corecursion in Foundational
  Proof Assistants}.
\newblock In {\em {Programming Languages and Systems, 26th European Symposium
  on Programming (ESOP)}}, April 2017.

\bibitem{Blanchette2014BNF}
J.~C. Blanchette, J.~H{\"o}lzl, A.~Lochbihler, L.~Panny, A.~Popescu, and
  D.~Traytel.
\newblock Truly modular (co)datatypes for {Isabelle/HOL}.
\newblock In Gerwin Klein and Ruben Gamboa, editors, {\em 5th Intl. Conf. on
  Interactive Theorem Proving (ITP)}, volume 8558 of {\em LNCS}, pages 93--110.
  Springer, 2014.

\bibitem{Blanchette2016Hammers}
J.~C. Blanchette, C.~Kaliszyk, L.~C. Paulson, and J.~Urban.
\newblock Hammering towards {QED}.
\newblock {\em Journal of Formalized Reasoning}, 9(1), 2016.
\newblock \href {https://doi.org/10.6092/issn.1972-5787/4593}
  {\path{doi:10.6092/issn.1972-5787/4593}}.

\bibitem{Blanchette2015ExtCorec}
J.~C. Blanchette, A.~Popescu, and D.~Traytel.
\newblock Foundational extensible corecursion: a proof assistant perspective.
\newblock In {\em 20th Intl. Conf. on Functional Programming (ICFP)}, pages
  192--204. ACM, August 2015.
\newblock \href {https://doi.org/10.1145/2858949.2784732}
  {\path{doi:10.1145/2858949.2784732}}.

\bibitem{Blanchette2017Coinductive}
J.~C. Blanchette, A.~Popescu, and D.~Traytel.
\newblock Soundness and completeness proofs by coinductive methods.
\newblock {\em Journal of Automated Reasoning}, 58:149--179, 2017.
\newblock \href {https://doi.org/10.1007/s10817-016-9391-3}
  {\path{doi:10.1007/s10817-016-9391-3}}.

\bibitem{Brookes1984}
S.~D. Brookes, C.~A.~R. Hoare, and A.~W. Roscoe.
\newblock A theory of communicating sequential processes.
\newblock {\em Journal of the ACM}, 31(3):560--599, 1984.
\newblock \href {https://doi.org/10.1145/828.833} {\path{doi:10.1145/828.833}}.

\bibitem{Brucker2003}
A.~D. Brucker, F.~Rittinger, and B.~Wolff.
\newblock Hol-z 2.0: A proof environment for z-specifications.
\newblock {\em Journal of Universal Computer Science}, 9(2), February 2003.

\bibitem{Cavalcanti04}
A.~Cavalcanti and J.~Woodcock.
\newblock A tutorial introduction to designs in unifying theories of
  programming.
\newblock In {\em Proc. 4th Intl. Conf. on Integrated Formal Methods (IFM)},
  volume 2999 of {\em LNCS}, pages 40--66. Springer, 2004.

\bibitem{Chappe2023CTrees}
N.~Chappe, P.~He, L.~Henrio, Y.~Zakowski, and S.~Zdancewic.
\newblock Choice trees: Representing nondeterministic, recursive, and impure
  programs in {Coq}.
\newblock In {\em Proc. ACM Programming Lang. (POPL)}, volume~61. ACM, January
  2023.

\bibitem{Crisafulli2023HOL-CSP}
P.~Crisafulli, S.~Taha, and B.~Wolff.
\newblock Modeling and analysing cyber-physical systems in hol-csp.
\newblock {\em Robotics and Autonomous Systems}, 170, 2023.

\bibitem{Dijkstra75}
E.~W. Dijkstra.
\newblock Guarded commands, nondeterminacy and formal derivation of programs.
\newblock {\em Communications of the ACM}, 18(8):453--457, 1975.

\bibitem{Feiler2012MBE}
P.~H. Feiler and D.~P. Gluch.
\newblock {\em Model-Based Engineering with {AADL}: An Introduction to the
  {SAE} {Architecture Analysis \& Design Language}}.
\newblock SEI Series in Software Engineering. Addison-Wesley Professional,
  2012.

\bibitem{Ferlez2014-GSTrees}
J.~Ferlez, R.~Cleaveland, and S.~Marcus.
\newblock Generalized synchronization trees.
\newblock In {\em Proc. 17th Intl. Conf. on Foundations of Software Science and
  Computation Structures (FOSSACS)}, volume 8412 of {\em LNCS}, pages 304--319.
  Springer, 2014.
\newblock \href {https://doi.org/10.1007/978-3-642-54830-7_20}
  {\path{doi:10.1007/978-3-642-54830-7_20}}.

\bibitem{Ferlez2018-BisimGSTrees}
J.~Ferlez, R.~Cleaveland, and S.~I. Marcus.
\newblock Bisimulation in behavioral dynamical systems and generalized
  synchronization trees.
\newblock In {\em Proc. 2018 IEEE Conf. on Decision and Control (CDC)}, pages
  751--758. IEEE, 2018.
\newblock \href {https://doi.org/10.1109/CDC.2018.8619607}
  {\path{doi:10.1109/CDC.2018.8619607}}.

\bibitem{Foster09}
J.~Foster.
\newblock {\em Bidirectional programming languages}.
\newblock PhD thesis, University of Pennsylvania, 2009.

\bibitem{Foster19b-HybridRelations}
S.~Foster.
\newblock Hybrid relations in {Isabelle/UTP}.
\newblock In {\em 7th Intl. Symp. on Unifying Theories of Programming (UTP)},
  volume 11885 of {\em LNCS}, pages 130--153. Springer, 2019.

\bibitem{Foster2020-IsabelleUTP}
S.~Foster, J.~Baxter, A.~Cavalcanti, J.~Woodcock, and F.~Zeyda.
\newblock Unifying semantic foundations for automated verification tools in
  {Isabelle/UTP}.
\newblock {\em Science of Computer Programming}, 197, October 2020.
\newblock \href {https://doi.org/10.1016/j.scico.2020.102510}
  {\path{doi:10.1016/j.scico.2020.102510}}.

\bibitem{Foster17c}
S.~Foster, A.~Cavalcanti, S.~Canham, J.~Woodcock, and F.~Zeyda.
\newblock Unifying theories of reactive design contracts.
\newblock {\em Theoretical Computer Science}, 802:105--140, January 2020.
\newblock \href {https://doi.org/10.1016/j.tcs.2019.09.017}
  {\path{doi:10.1016/j.tcs.2019.09.017}}.

\bibitem{Foster2021-ITrees}
S.~Foster, C.-K. Hur, and J.~Woodcock.
\newblock Formally verified simulations of state-rich processes using
  interaction trees in {Isabelle/HOL}.
\newblock In {\em 32nd Intl. Conf. on Concurrency Theory (CONCUR)}, volume 203
  of {\em LIPIcs}. Schloss Dagstuhl -- Leibniz-Zentrum f{\"u}r Informatik,
  2021.

\bibitem{Foster2021-JLAMP}
S.~Foster, K.~Ye, A.~Cavalcanti, and J.~Woodcock.
\newblock Automated verification of reactive and concurrent programs by
  calculation.
\newblock {\em Journal of Logical and Algebraic Methods in Programming}, 121,
  June 2021.
\newblock \href {https://doi.org/10.1016/j.jlamp.2021.100681}
  {\path{doi:10.1016/j.jlamp.2021.100681}}.

\bibitem{fdr}
T.~Gibson-Robinson, P.~Armstrong, A.~Boulgakov, and A.~W. Roscoe.
\newblock {FDR3 --- A Modern Refinement Checker for CSP}.
\newblock In Erika Ábrahám and Klaus Havelund, editors, {\em Tools and
  Algorithms for the Construction and Analysis of Systems}, volume 8413 of {\em
  LNCS}, pages 187--201, 2014.

\bibitem{Gleirscher2018-NewOpportunitiesIntegrated}
M.~Gleirscher, S.~Foster, and J.~Woodcock.
\newblock New opportunities for integrated formal methods.
\newblock {\em ACM Comput. Surv.}, 52(6), 2019.

\bibitem{Gomes2016}
V.~B.~F Gomes and G.~Struth.
\newblock Modal {Kleene} algebra applied to program correctness.
\newblock In {\em Formal Methods}, volume 9995 of {\em LNCS}, pages 310--325.
  Springer, 2016.

\bibitem{Haftmann2013-DataRefinement}
F.~Haftmann, A.~Krauss, O.~Kuncar, and T.~Nipkow.
\newblock Data refinement in {Isabelle/HOL}.
\newblock In {\em Proc. 4th Intl. Conf. on Interactive Theorem Proving (ITP)},
  volume 7998 of {\em LNCS}, pages 100--115. Springer, 2013.

\bibitem{Haftman2010-CodeGen}
F.~Haftmann and T.~Nipkow.
\newblock Code generation via higher-order rewrite systems.
\newblock In {\em 10th Intl. Symp. on Functional and Logic Programming
  (FLOPS)}, volume 6009 of {\em LNCS}, pages 103--117. Springer, 2010.

\bibitem{Hennessy1995TPL}
Matthew Hennessy and Tim Regan.
\newblock A process algebra for timed systems.
\newblock {\em Information and Computation}, 117(2):221--239, 1995.

\bibitem{Hoare85}
C.~A.~R. Hoare.
\newblock {\em {Communicating Sequential Processes}}.
\newblock Prentice-Hall, 1985.

\bibitem{Hoare87}
C.~A.~R. Hoare, I.~Hayes, J.~He, C.~Morgan, A.~Roscoe, J.~Sanders,
  I.~S{\o}rensen, J.~Spivey, and B.~Sufrin.
\newblock The laws of programming.
\newblock {\em Communications of the ACM}, 30(8):672--687, August 1987.

\bibitem{Hoare&98}
C.~A.~R. Hoare and J.~He.
\newblock {\em Unifying {Theories} of {Programming}}.
\newblock Prentice-Hall, 1998.

\bibitem{KLL+19}
Nicolas Koh, Yao Li, Yishuai Li, Li~yao Xia, Lennart Beringer, Wolf Honor\'{e},
  William Mansky, Benjamin~C. Pierce, and Steve Zdancewic.
\newblock {From C to Interaction Trees: Specifying, Verifying, and Testing a
  Networked Server}.
\newblock In {\em Proc. 8th ACM SIGPLAN International Conference on Certified
  Programs and Proofs (CPP)}, 2019.
\newblock \href {https://doi.org/10.1145/3293880.3294106}
  {\path{doi:10.1145/3293880.3294106}}.

\bibitem{ProB}
M.~Leuschel and M.~Butler.
\newblock {ProB}: an automated analysis toolset for the {B} method.
\newblock {\em Int J Softw Tools Technol Transf}, 10:185--203, 2008.
\newblock \href {https://doi.org/10.1007/s10009-007-0063-9}
  {\path{doi:10.1007/s10009-007-0063-9}}.

\bibitem{LPZ21}
Yishuai Li, Benjamin~C. Pierce, and Steve Zdancewic.
\newblock Model-based testing of networked applications.
\newblock In {\em Proc. 30th ACM SIGSOFT International Symposium on Software
  Testing and Analysis (ISSTA)}, 2021.

\bibitem{MHA20}
William Mansky, Wolf Honor\'{e}, and Andrew~W. Appel.
\newblock Connecting higher-order separation logic to a first-order outside
  world.
\newblock In {\em Proc. 29th European Symposium on Programming (ESOP)}, 2020.

\bibitem{Milner1980}
Robin Milner.
\newblock {\em A Calculus of Communicating Systems}, volume~92 of {\em Lecture
  Notes in Computer Science}.
\newblock Springer, 1980.

\bibitem{Milner1989}
Robin Milner.
\newblock {\em Communication and Concurrency}.
\newblock Prentice Hall, 1989.

\bibitem{Miyazawa2019-RoboChart}
A.~Miyazawa, P.~Ribeiro, W.~Li, A.~Cavalcanti, J.~Timmis, and J.~Woodcock.
\newblock {RoboChart}: modelling and verification of the functional behaviour
  of robotic applications.
\newblock {\em Software and Systems Modelling}, January 2019.
\newblock \href {https://doi.org/10.1007/s10270-018-00710-z}
  {\path{doi:10.1007/s10270-018-00710-z}}.

\bibitem{Foster2020-dL}
J.~H.~Y. Munive, G.~Struth, and S.~Foster.
\newblock Differential {Hoare} logics and refinement calculi for hybrid systems
  with {Isabelle/HOL}.
\newblock In {\em RAMiCS}, volume 12062 of {\em LNCS}. Springer, April 2020.
\newblock \href {https://doi.org/10.1007/978-3-030-43520-2_11}
  {\path{doi:10.1007/978-3-030-43520-2_11}}.

\bibitem{Oliveira&09}
M.~Oliveira, A.~Cavalcanti, and J.~Woodcock.
\newblock {A UTP semantics for {C}ircus}.
\newblock {\em Formal Aspects of Computing}, 21:3--32, 2009.
\newblock \href {https://doi.org/10.1007/s00165-007-0052-5}
  {\path{doi:10.1007/s00165-007-0052-5}}.

\bibitem{Paige1997FM-IntegratedFormalMethods}
R.~F. Paige.
\newblock A meta-method for formal method integration.
\newblock In {\em Proc. 4th. Intl. Symp. on Formal Methods Europe (FME)},
  volume 1313 of {\em LNCS}, pages 473--494. Springer, 1997.

\bibitem{Pickering2017-Optics}
M.~Pickering, J.~Gibbons, and N.~Wu.
\newblock Profunctor optics: Modular data accessors.
\newblock {\em The Art, Science, and Engineering of Programming}, 1(2), 2017.
\newblock \href {https://doi.org/10.22152/programming-journal.org/2017/1/7}
  {\path{doi:10.22152/programming-journal.org/2017/1/7}}.

\bibitem{Ribeiro2019Angelic}
P.~Ribeiro and A.~Cavalcanti.
\newblock Angelic processes for {CSP} via the {UTP}.
\newblock {\em Theoretical Computer Science}, 2019.

\bibitem{Roscoe1984-Occam}
A.~W. Roscoe.
\newblock Denotational semantics for {Occam}.
\newblock In {\em Intl. Seminar on Concurrency}, volume 197 of {\em LNCS},
  pages 306--329. Springer, 1984.

\bibitem{Roscoe2005-TPC}
A.~W. Roscoe.
\newblock {\em The Theory and Practice of Concurrency}.
\newblock Prentice-Hall, 2005.

\bibitem{Roscoe2010-UCS}
A.~W. Roscoe.
\newblock {\em Understanding Concurrent Systems}.
\newblock Springer, 2010.

\bibitem{SZ21}
Lucas Silver and Steve Zdancewic.
\newblock {Dijkstra} monads forever: {Termination}-sensitive specifications for
  {Interaction} {Trees}.
\newblock {\em Proceedings of the ACM on Programming Languages}, 5(POPL),
  January 2021.
\newblock \href {https://doi.org/10.1145/3434307} {\path{doi:10.1145/3434307}}.

\bibitem{Spivey89}
M.~Spivey.
\newblock {\em The Z-Notation - A Reference Manual}.
\newblock Prentice Hall, Englewood Cliffs, N. J., 1989.

\bibitem{Taha2020CSP-Isabelle}
S.~Taha, B.~Wolff, and L.~Ye.
\newblock Philosophers may dine -- definitively!
\newblock In {\em Proc. 16th Intl. Conf. on Integrated Formal Methods}, LNCS.
  Springer, 2020.
\newblock \href {https://doi.org/10.1007/978-3-030-63461-2_23}
  {\path{doi:10.1007/978-3-030-63461-2_23}}.

\bibitem{Glabbeek1997CCS-CSP}
R.~J. van Glabbeek.
\newblock Notes on the methodology of {CCS} and {CSP}.
\newblock {\em Theoretical Computer Science}, 1997.

\bibitem{Winskel1984STrees}
G.~Winsel.
\newblock Synchronisation trees.
\newblock {\em Theoretical Computer Science}, 34(1-2):33--82, 1984.

\bibitem{Woodcock2001-Circus}
J.~Woodcock and A.~Cavalcanti.
\newblock A concurrent language for refinement.
\newblock In A.~Butterfield, G.~Strong, and C.~Pahl, editors, {\em Proc. 5th
  Irish Workshop on Formal Methods (IWFM)}, Workshops in Computing. BCS, July
  2001.

\bibitem{Woodcock96-UsingZ}
J.~Woodcock and J.~Davies.
\newblock {\em Using Z: Specification, Refinement, and Proof}.
\newblock Prentice-Hall, 1996.

\bibitem{ITrees2019}
L.-Y. Xia, Y.~Zakowski, P.~He, C.-K. Hur, G.~Malecha, B.~C. Pierce, and
  S.~Zdancewic.
\newblock Interaction trees: Representing recursive and impure programs in
  {Coq}.
\newblock In {\em Proc. 47th ACM SIGPLAN Symposium on Principles of Programming
  Languages (POPL)}. ACM, 2020.
\newblock \href {https://doi.org/10.1145/3371119} {\path{doi:10.1145/3371119}}.

\bibitem{Yan2023-ZMachineRoboChart}
F.~Yan, S.~Foster, and I.~Habli.
\newblock Automated compositional verification for robotic state machines using
  {Isabelle/HOL}.
\newblock In {\em Proc. 27th Intl. Conf. on Engineering of Complex Computer
  Systems (ICECCS)}. IEEE, June 2023.
\newblock \href {https://doi.org/10.1109/ICECCS59891.2023.00029}
  {\path{doi:10.1109/ICECCS59891.2023.00029}}.

\bibitem{Ye2024-ITreeRoboChart}
K.~Ye, S.~Foster, and J.~Woodcock.
\newblock Formally verified animation for {RoboChart} using interaction trees.
\newblock {\em Journal of Logical and Algebraic Methods in Programming}, 137,
  February 2024.

\bibitem{ZHHZ20}
Yannick Zakowski, Paul He, Chung-Kil Hur, and Steve Zdancewic.
\newblock An equational theory for weak bisimulation via generalized
  parameterized coinduction.
\newblock In {\em Proc. 9th ACM SIGPLAN International Conference on Certified
  Programs and Proofs (CPP)}, 2020.
\newblock \href {https://doi.org/10.1145/3372885.3373813}
  {\path{doi:10.1145/3372885.3373813}}.

\bibitem{ZZF20}
Vadim Zaliva, Ilia Zaichuk, and Franz Franchetti.
\newblock Verified translation between purely functional and imperative domain
  specific languages in {HELIX}.
\newblock In {\em Proc. 12th International Conference on Verified Software:
  Theories, Tools, Experiments (VSTTE)}, 2020.

\bibitem{ZHK+21}
Hengchu Zhang, Wolf Honor\'{e}, Nicolas Koh, Yao Li, Yishuai Li, Li-Yao Xia,
  Lennart Beringer, William Mansky, Benjamin~C. Pierce, and Steve Zdancewic.
\newblock Verifying an {HTTP} key-value server with {Interaction} {Trees} and
  {VST}.
\newblock In {\em Proc. 12th International Conference on Interactive Theorem
  Proving (ITP)}, 2021.
\newblock \href {https://doi.org/10.4230/LIPIcs.ITP.2021.32}
  {\path{doi:10.4230/LIPIcs.ITP.2021.32}}.

\end{thebibliography}


\begin{thebibliography}{10}

\bibitem{Abrial96BBook}
Jean-Raymond Abrial.
\newblock {\em The {B-Book}: {A}ssigning programs to meanings}.
\newblock Cambridge University Press, 1996.

\bibitem{Haftmann2012NBE}
K.~Aehlig, F.~Haftmann, and T.~Nipkow.
\newblock A compiled implementation of normalisation by evaluation.
\newblock {\em Journal of Functional Programming}, 22(1):9--30, 2012.

\bibitem{Armstrong2015}
A.~Armstrong, V.~Gomes, and G.~Struth.
\newblock Building program construction and verification tools from algebraic
  principles.
\newblock {\em Formal Aspects of Computing}, 28(2), 2015.

\bibitem{Arthan2004ProofPower}
R.~Arthan.
\newblock On formal specification of a proof tool.
\newblock In {\em Formal Software Development Methods}, volume 551 of {\em
  LNCS}. Springer, 1991.

\bibitem{Back1989ActionSystems}
R.-J. Back and R.~Kurki-Suonio.
\newblock Decentralization of process nets with centralized control.
\newblock {\em Distributed Computing}, 3:73--87, June 1989.

\bibitem{Baxter2021-TickTock}
J.~Baxter, P.~Ribeiro, and A.~Cavalcanti.
\newblock Sound reasoning in {tock-CSP}.
\newblock {\em Acta Informatica}, April 2021.
\newblock \href {https://doi.org/10.1007/s00236-020-00394-3}
  {\path{doi:10.1007/s00236-020-00394-3}}.

\bibitem{Bertot2004Coq}
Y.~Bertot and P.~Cast{\'e}ran.
\newblock {\em Coq'Art: the calculus of inductive constructions}.
\newblock Springer, 2004.

\bibitem{Blanchette2017Corec}
J.~C. Blanchette, A.~Bouzy, A.~Lochbihler, A.~Popescu, and D.~Traytel.
\newblock {Friends with Benefits: Implementing Corecursion in Foundational
  Proof Assistants}.
\newblock In {\em {Programming Languages and Systems, 26th European Symposium
  on Programming (ESOP)}}, April 2017.

\bibitem{Blanchette2014BNF}
J.~C. Blanchette, J.~H{\"o}lzl, A.~Lochbihler, L.~Panny, A.~Popescu, and
  D.~Traytel.
\newblock Truly modular (co)datatypes for {Isabelle/HOL}.
\newblock In Gerwin Klein and Ruben Gamboa, editors, {\em 5th Intl. Conf. on
  Interactive Theorem Proving (ITP)}, volume 8558 of {\em LNCS}, pages 93--110.
  Springer, 2014.

\bibitem{Blanchette2016Hammers}
J.~C. Blanchette, C.~Kaliszyk, L.~C. Paulson, and J.~Urban.
\newblock Hammering towards {QED}.
\newblock {\em Journal of Formalized Reasoning}, 9(1), 2016.
\newblock \href {https://doi.org/10.6092/issn.1972-5787/4593}
  {\path{doi:10.6092/issn.1972-5787/4593}}.

\bibitem{Blanchette2015ExtCorec}
J.~C. Blanchette, A.~Popescu, and D.~Traytel.
\newblock Foundational extensible corecursion: a proof assistant perspective.
\newblock In {\em 20th Intl. Conf. on Functional Programming (ICFP)}, pages
  192--204. ACM, August 2015.
\newblock \href {https://doi.org/10.1145/2858949.2784732}
  {\path{doi:10.1145/2858949.2784732}}.

\bibitem{Blanchette2017Coinductive}
J.~C. Blanchette, A.~Popescu, and D.~Traytel.
\newblock Soundness and completeness proofs by coinductive methods.
\newblock {\em Journal of Automated Reasoning}, 58:149--179, 2017.
\newblock \href {https://doi.org/10.1007/s10817-016-9391-3}
  {\path{doi:10.1007/s10817-016-9391-3}}.

\bibitem{Boulton1993}
R.~Boulton, A.~Gordon, M.~Gordon, J.~Harrison, J.~Herbert, and J.~van Tassel.
\newblock Experience with embedding hardware description languages in hol.
\newblock In {\em Proc. IFIP Intl. Conf. on Theorem Provers in Circuit Design},
  pages 129--156, 1993.

\bibitem{Bousse2016Execution}
E.~Bousse, T.~Degueule, D.~Vojtisek, T.~Mayerhofer, J.~Deantoni, and
  B.~Combemale.
\newblock Execution framework of the {GEMOC} studio (tool demo).
\newblock In {\em Proceedings of the 2016 {ACM} {SIGPLAN} {International}
  {Conference} on {Software} {Language} {Engineering}}, {SLE} 2016, pages
  84--89. Association for Computing Machinery, October 2016.
\newblock \href {https://doi.org/10.1145/2997364.2997384}
  {\path{doi:10.1145/2997364.2997384}}.

\bibitem{Brookes1984}
S.~D. Brookes, C.~A.~R. Hoare, and A.~W. Roscoe.
\newblock A theory of communicating sequential processes.
\newblock {\em Journal of the ACM}, 31(3):560--599, 1984.
\newblock \href {https://doi.org/10.1145/828.833} {\path{doi:10.1145/828.833}}.

\bibitem{Brucker2003}
A.~D. Brucker, F.~Rittinger, and B.~Wolff.
\newblock {HOL-Z 2.0}: {A} proof environment for {Z}-specifications.
\newblock {\em Journal of Universal Computer Science}, 9(2), February 2003.

\bibitem{Cavalcanti04}
A.~Cavalcanti and J.~Woodcock.
\newblock A tutorial introduction to designs in unifying theories of
  programming.
\newblock In {\em Proc. 4th Intl. Conf. on Integrated Formal Methods (IFM)},
  volume 2999 of {\em LNCS}, pages 40--66. Springer, 2004.

\bibitem{Chappe2023CTrees}
N.~Chappe, P.~He, L.~Henrio, Y.~Zakowski, and S.~Zdancewic.
\newblock Choice trees: Representing nondeterministic, recursive, and impure
  programs in {Coq}.
\newblock In {\em Proc. ACM Programming Lang. (POPL)}, volume~61. ACM, January
  2023.

\bibitem{Ciccozzi2019Execution}
F.~Ciccozzi, I.~Malavolta, and B.~Selic.
\newblock Execution of {UML} models: {A} systematic review of research and
  practice.
\newblock {\em Software \& Systems Modeling}, 18(3):2313--2360, June 2019.
\newblock \href {https://doi.org/10.1007/s10270-018-0675-4}
  {\path{doi:10.1007/s10270-018-0675-4}}.

\bibitem{Crisafulli2023HOL-CSP}
P.~Crisafulli, S.~Taha, and B.~Wolff.
\newblock Modeling and analysing cyber-physical systems in {HOL-CSP}.
\newblock {\em Robotics and Autonomous Systems}, 170, 2023.

\bibitem{Demoura2015Lean}
L.~de~Moura, S.~Kong, J.~Avigad, F.~van Doorn, and J.~von Raumer.
\newblock The {Lean} theorem prover (system description).
\newblock In {\em Proc. 25th Intl. Conf. on Automated Deduction (CADE)}, 2015.

\bibitem{Dijkstra75}
E.~W. Dijkstra.
\newblock Guarded commands, nondeterminacy and formal derivation of programs.
\newblock {\em Communications of the ACM}, 18(8):453--457, 1975.

\bibitem{Feiler2012MBE}
P.~H. Feiler and D.~P. Gluch.
\newblock {\em Model-Based Engineering with {AADL}: An Introduction to the
  {SAE} {Architecture Analysis \& Design Language}}.
\newblock SEI Series in Software Engineering. Addison-Wesley Professional,
  2012.

\bibitem{Ferlez2014-GSTrees}
J.~Ferlez, R.~Cleaveland, and S.~Marcus.
\newblock Generalized synchronization trees.
\newblock In {\em Proc. 17th Intl. Conf. on Foundations of Software Science and
  Computation Structures (FOSSACS)}, volume 8412 of {\em LNCS}, pages 304--319.
  Springer, 2014.
\newblock \href {https://doi.org/10.1007/978-3-642-54830-7_20}
  {\path{doi:10.1007/978-3-642-54830-7_20}}.

\bibitem{Ferlez2018-BisimGSTrees}
J.~Ferlez, R.~Cleaveland, and S.~I. Marcus.
\newblock Bisimulation in behavioral dynamical systems and generalized
  synchronization trees.
\newblock In {\em Proc. 2018 IEEE Conf. on Decision and Control (CDC)}, pages
  751--758. IEEE, 2018.
\newblock \href {https://doi.org/10.1109/CDC.2018.8619607}
  {\path{doi:10.1109/CDC.2018.8619607}}.

\bibitem{Foster09}
J.~Foster.
\newblock {\em Bidirectional programming languages}.
\newblock PhD thesis, University of Pennsylvania, 2009.

\bibitem{Foster19b-HybridRelations}
S.~Foster.
\newblock Hybrid relations in {Isabelle/UTP}.
\newblock In {\em 7th Intl. Symp. on Unifying Theories of Programming (UTP)},
  volume 11885 of {\em LNCS}, pages 130--153. Springer, 2019.

\bibitem{Foster2020-IsabelleUTP}
S.~Foster, J.~Baxter, A.~Cavalcanti, J.~Woodcock, and F.~Zeyda.
\newblock Unifying semantic foundations for automated verification tools in
  {Isabelle/UTP}.
\newblock {\em Science of Computer Programming}, 197, October 2020.
\newblock \href {https://doi.org/10.1016/j.scico.2020.102510}
  {\path{doi:10.1016/j.scico.2020.102510}}.

\bibitem{Foster17c}
S.~Foster, A.~Cavalcanti, S.~Canham, J.~Woodcock, and F.~Zeyda.
\newblock Unifying theories of reactive design contracts.
\newblock {\em Theoretical Computer Science}, 802:105--140, January 2020.
\newblock \href {https://doi.org/10.1016/j.tcs.2019.09.017}
  {\path{doi:10.1016/j.tcs.2019.09.017}}.

\bibitem{Foster2021-ITrees}
S.~Foster, C.-K. Hur, and J.~Woodcock.
\newblock Formally verified simulations of state-rich processes using
  interaction trees in {Isabelle/HOL}.
\newblock In {\em 32nd Intl. Conf. on Concurrency Theory (CONCUR)}, volume 203
  of {\em LIPIcs}. Schloss Dagstuhl -- Leibniz-Zentrum f{\"u}r Informatik,
  2021.

\bibitem{Foster2021-IsaSACM}
S.~Foster, Y.~Nemouchi, M.~Gleirscher, R.~Wei, and T.~Kelly.
\newblock Integration of formal proof into unified assurance cases with
  {Isabelle/SACM}.
\newblock {\em Formal Aspects of Computing}, 2021.

\bibitem{Foster2021-JLAMP}
S.~Foster, K.~Ye, A.~Cavalcanti, and J.~Woodcock.
\newblock Automated verification of reactive and concurrent programs by
  calculation.
\newblock {\em Journal of Logical and Algebraic Methods in Programming}, 121,
  June 2021.
\newblock \href {https://doi.org/10.1016/j.jlamp.2021.100681}
  {\path{doi:10.1016/j.jlamp.2021.100681}}.

\bibitem{foster2020formal}
Simon Foster, Yakoub Nemouchi, Colin O'Halloran, Karen Stephenson, and Nick
  Tudor.
\newblock Formal model-based assurance cases in {Isabelle/SACM}: An autonomous
  underwater vehicle case study.
\newblock In {\em Proceedings of the 8th International Conference on Formal
  Methods in Software Engineering}, pages 11--21, 2020.

\bibitem{FosterMGS21}
Simon Foster, Jonathan Juli{\'{a}}n~Huerta y~Munive, Mario Gleirscher, and
  Georg Struth.
\newblock Hybrid systems verification with {I}sabelle/{HOL}: Simpler syntax,
  better models, faster proofs.
\newblock In {\em {FM} 2021}, volume 13047 of {\em LNCS}, pages 367--386,
  Heidelberg, 2021. Springer.
\newblock \href {https://doi.org/10.1007/978-3-030-90870-6\_20}
  {\path{doi:10.1007/978-3-030-90870-6\_20}}.

\bibitem{Gibbons2014}
J.~Gibbons and N.~Wu.
\newblock Folding domain-specific languages: deep and shallow embeddings.
\newblock In {\em Proc. 19th Intl. Conf. on Functional Programming (ICFP)},
  pages 339--347. ACM, 2014.

\bibitem{fdr}
T.~Gibson-Robinson, P.~Armstrong, A.~Boulgakov, and A.~W. Roscoe.
\newblock {FDR3 --- A Modern Refinement Checker for CSP}.
\newblock In Erika Ábrahám and Klaus Havelund, editors, {\em Tools and
  Algorithms for the Construction and Analysis of Systems}, volume 8413 of {\em
  LNCS}, pages 187--201, 2014.

\bibitem{Gleirscher2018-NewOpportunitiesIntegrated}
M.~Gleirscher, S.~Foster, and J.~Woodcock.
\newblock New opportunities for integrated formal methods.
\newblock {\em ACM Comput. Surv.}, 52(6), 2019.

\bibitem{Gomes2016}
V.~B.~F Gomes and G.~Struth.
\newblock Modal {Kleene} algebra applied to program correctness.
\newblock In {\em Formal Methods}, volume 9995 of {\em LNCS}, pages 310--325.
  Springer, 2016.

\bibitem{Haftmann2013-DataRefinement}
F.~Haftmann, A.~Krauss, O.~Kuncar, and T.~Nipkow.
\newblock Data refinement in {Isabelle/HOL}.
\newblock In {\em Proc. 4th Intl. Conf. on Interactive Theorem Proving (ITP)},
  volume 7998 of {\em LNCS}, pages 100--115. Springer, 2013.

\bibitem{Haftman2010-CodeGen}
F.~Haftmann and T.~Nipkow.
\newblock Code generation via higher-order rewrite systems.
\newblock In {\em 10th Intl. Symp. on Functional and Logic Programming
  (FLOPS)}, volume 6009 of {\em LNCS}, pages 103--117. Springer, 2010.

\bibitem{Hennessy1995TPL}
Matthew Hennessy and Tim Regan.
\newblock A process algebra for timed systems.
\newblock {\em Information and Computation}, 117(2):221--239, 1995.

\bibitem{Hoare85}
C.~A.~R. Hoare.
\newblock {\em {Communicating Sequential Processes}}.
\newblock Prentice-Hall, 1985.

\bibitem{Hoare87}
C.~A.~R. Hoare, I.~Hayes, J.~He, C.~Morgan, A.~Roscoe, J.~Sanders,
  I.~S{\o}rensen, J.~Spivey, and B.~Sufrin.
\newblock The laws of programming.
\newblock {\em Communications of the ACM}, 30(8):672--687, August 1987.

\bibitem{Hoare&98}
C.~A.~R. Hoare and J.~He.
\newblock {\em Unifying {Theories} of {Programming}}.
\newblock Prentice-Hall, 1998.

\bibitem{Klein2009}
G.~Klein et~al.
\newblock {seL4}: Formal verification of an {OS} kernel.
\newblock In {\em Proc. 22nd Symp. on Operating Systems Principles (SOSP)},
  pages 207--220. ACM, 2009.

\bibitem{KLL+19}
Nicolas Koh, Yao Li, Yishuai Li, Li~yao Xia, Lennart Beringer, Wolf Honor\'{e},
  William Mansky, Benjamin~C. Pierce, and Steve Zdancewic.
\newblock {From C to Interaction Trees: Specifying, Verifying, and Testing a
  Networked Server}.
\newblock In {\em Proc. 8th ACM SIGPLAN International Conference on Certified
  Programs and Proofs (CPP)}, 2019.
\newblock \href {https://doi.org/10.1145/3293880.3294106}
  {\path{doi:10.1145/3293880.3294106}}.

\bibitem{Kolovos2008Epsilon}
Dimitrios~S. Kolovos, Richard~F. Paige, and Fiona A.~C. Polack.
\newblock The {Epsilon} {Transformation} {Language}.
\newblock In Antonio Vallecillo, Jeff Gray, and Alfonso Pierantonio, editors,
  {\em Theory and {Practice} of {Model} {Transformations}}, pages 46--60,
  Berlin, Heidelberg, 2008. Springer.
\newblock \href {https://doi.org/10.1007/978-3-540-69927-9_4}
  {\path{doi:10.1007/978-3-540-69927-9_4}}.

\bibitem{ProB}
M.~Leuschel and M.~Butler.
\newblock {ProB}: {An} automated analysis toolset for the {B} method.
\newblock {\em Int J Softw Tools Technol Transf}, 10:185--203, 2008.
\newblock \href {https://doi.org/10.1007/s10009-007-0063-9}
  {\path{doi:10.1007/s10009-007-0063-9}}.

\bibitem{LPZ21}
Yishuai Li, Benjamin~C. Pierce, and Steve Zdancewic.
\newblock Model-based testing of networked applications.
\newblock In {\em Proc. 30th ACM SIGSOFT International Symposium on Software
  Testing and Analysis (ISSTA)}, 2021.

\bibitem{MHA20}
William Mansky, Wolf Honor\'{e}, and Andrew~W. Appel.
\newblock Connecting higher-order separation logic to a first-order outside
  world.
\newblock In {\em Proc. 29th European Symposium on Programming (ESOP)}, 2020.

\bibitem{Milner1980}
Robin Milner.
\newblock {\em A Calculus of Communicating Systems}, volume~92 of {\em Lecture
  Notes in Computer Science}.
\newblock Springer, 1980.

\bibitem{Milner1989}
Robin Milner.
\newblock {\em Communication and Concurrency}.
\newblock Prentice Hall, 1989.

\bibitem{Miyazawa2019-RoboChart}
A.~Miyazawa, P.~Ribeiro, W.~Li, A.~Cavalcanti, J.~Timmis, and J.~Woodcock.
\newblock {RoboChart}: {Modelling} and verification of the functional behaviour
  of robotic applications.
\newblock {\em Software and Systems Modelling}, January 2019.
\newblock \href {https://doi.org/10.1007/s10270-018-00710-z}
  {\path{doi:10.1007/s10270-018-00710-z}}.

\bibitem{Foster2020-dL}
J.~H.~Y. Munive, G.~Struth, and S.~Foster.
\newblock Differential {Hoare} logics and refinement calculi for hybrid systems
  with {Isabelle/HOL}.
\newblock In {\em RAMiCS}, volume 12062 of {\em LNCS}. Springer, April 2020.
\newblock \href {https://doi.org/10.1007/978-3-030-43520-2_11}
  {\path{doi:10.1007/978-3-030-43520-2_11}}.

\bibitem{Nipkow2002HoareLogic}
T.~Nipkow.
\newblock Hoare logics in isabelle/hol.
\newblock In {\em Proof and System-Reliability}. Springer, 2002.

\bibitem{Nipkow2002Isabelle}
T.~Nipkow, M.~Wenzel, and L.~C. Paulson.
\newblock {\em Isabelle/HOL: a proof assistant for higher-order logic}.
\newblock Springer, 2002.

\bibitem{Oliveira&09}
M.~Oliveira, A.~Cavalcanti, and J.~Woodcock.
\newblock {A UTP semantics for {C}ircus}.
\newblock {\em Formal Aspects of Computing}, 21:3--32, 2009.
\newblock \href {https://doi.org/10.1007/s00165-007-0052-5}
  {\path{doi:10.1007/s00165-007-0052-5}}.

\bibitem{Paige1997FM-IntegratedFormalMethods}
R.~F. Paige.
\newblock A meta-method for formal method integration.
\newblock In {\em Proc. 4th. Intl. Symp. on Formal Methods Europe (FME)},
  volume 1313 of {\em LNCS}, pages 473--494. Springer, 1997.

\bibitem{Pickering2017-Optics}
M.~Pickering, J.~Gibbons, and N.~Wu.
\newblock Profunctor optics: Modular data accessors.
\newblock {\em The Art, Science, and Engineering of Programming}, 1(2), 2017.
\newblock \href {https://doi.org/10.22152/programming-journal.org/2017/1/7}
  {\path{doi:10.22152/programming-journal.org/2017/1/7}}.

\bibitem{Ribeiro2019Angelic}
P.~Ribeiro and A.~Cavalcanti.
\newblock Angelic processes for {CSP} via the {UTP}.
\newblock {\em Theoretical Computer Science}, 2019.

\bibitem{Roscoe1984-Occam}
A.~W. Roscoe.
\newblock Denotational semantics for textsf{occam}.
\newblock In {\em Intl. Seminar on Concurrency}, volume 197 of {\em LNCS},
  pages 306--329. Springer, 1984.

\bibitem{Roscoe2005-TPC}
A.~W. Roscoe.
\newblock {\em The Theory and Practice of Concurrency}.
\newblock Prentice-Hall, 2005.

\bibitem{Roscoe2010-UCS}
A.~W. Roscoe.
\newblock {\em Understanding Concurrent Systems}.
\newblock Springer, 2010.

\bibitem{SZ21}
Lucas Silver and Steve Zdancewic.
\newblock {Dijkstra} monads forever: {Termination}-sensitive specifications for
  {Interaction} {Trees}.
\newblock {\em Proceedings of the ACM on Programming Languages}, 5(POPL),
  January 2021.
\newblock \href {https://doi.org/10.1145/3434307} {\path{doi:10.1145/3434307}}.

\bibitem{Spivey89}
M.~Spivey.
\newblock {\em The Z Notation --- A Reference Manual}.
\newblock Prentice Hall, Englewood Cliffs, N. J., 1989.

\bibitem{SunLDP09}
Jun Sun, Yang Liu, Jin~Song Dong, and Jun Pang.
\newblock Pat: Towards flexible verification under fairness.
\newblock {\em Proceedings of the 21th International Conference on Computer
  Aided Verification (CAV'09)}, 5643:709--714, 2009.

\bibitem{Taha2020CSP-Isabelle}
S.~Taha, B.~Wolff, and L.~Ye.
\newblock Philosophers may dine --- definitively!
\newblock In {\em Proc. 16th Intl. Conf. on Integrated Formal Methods}, LNCS.
  Springer, 2020.
\newblock \href {https://doi.org/10.1007/978-3-030-63461-2_23}
  {\path{doi:10.1007/978-3-030-63461-2_23}}.

\bibitem{Glabbeek1997CCS-CSP}
R.~J. van Glabbeek.
\newblock Notes on the methodology of {CCS} and {CSP}.
\newblock {\em Theoretical Computer Science}, 1997.

\bibitem{Wei2024ACCESS}
Ran Wei, Simon Foster, Haitao Mei, Fang Yan, Ruizhe Yang, Ibrahim Habli, Colin
  O’Halloran, Nick Tudor, Tim Kelly, and Yakoub Nemouchi.
\newblock Access: Assurance case centric engineering of safety–critical
  systems.
\newblock {\em Journal of Systems and Software}, 213, 2024.
\newblock URL:
  \url{https://www.sciencedirect.com/science/article/pii/S0164121224000773},
  \href {https://doi.org/10.1016/j.jss.2024.112034}
  {\path{doi:10.1016/j.jss.2024.112034}}.

\bibitem{Wenzel2019-Isar}
M.~Wenzel.
\newblock Interaction with formal mathematical documents in {Isabelle/PIDE}.
\newblock In {\em CICM}, LNCS 11617, pages 1--15. Springer, 2019.

\bibitem{Wenzel2007}
M.~Wenzel and B.~Wolff.
\newblock Building formal method tools in the {Isabelle/Isar} framework.
\newblock In {\em TPHOLs}, volume 4732 of {\em LNCS}. Springer, 2007.

\bibitem{Winskel1984STrees}
G.~Winsel.
\newblock Synchronisation trees.
\newblock {\em Theoretical Computer Science}, 34(1-2):33--82, 1984.

\bibitem{Woodcock2001-Circus}
J.~Woodcock and A.~Cavalcanti.
\newblock A concurrent language for refinement.
\newblock In A.~Butterfield, G.~Strong, and C.~Pahl, editors, {\em Proc. 5th
  Irish Workshop on Formal Methods (IWFM)}, Workshops in Computing. BCS, July
  2001.

\bibitem{Woodcock96-UsingZ}
J.~Woodcock and J.~Davies.
\newblock {\em Using Z: Specification, Refinement, and Proof}.
\newblock Prentice-Hall, 1996.

\bibitem{ITrees2019}
L.-Y. Xia, Y.~Zakowski, P.~He, C.-K. Hur, G.~Malecha, B.~C. Pierce, and
  S.~Zdancewic.
\newblock Interaction trees: Representing recursive and impure programs in
  {Coq}.
\newblock In {\em Proc. 47th ACM SIGPLAN Symposium on Principles of Programming
  Languages (POPL)}. ACM, 2020.
\newblock \href {https://doi.org/10.1145/3371119} {\path{doi:10.1145/3371119}}.

\bibitem{Xia2022ITrees}
Li-yao Xia.
\newblock {\em Executable {Denotational} {Semantics} with {Interaction}
  {Trees}}.
\newblock Ph.{D}., University of Pennsylvania, United States -- Pennsylvania,
  2022.
\newblock ISBN: 9798351434490.

\bibitem{Yan2023-ZMachineRoboChart}
F.~Yan, S.~Foster, and I.~Habli.
\newblock Automated compositional verification for robotic state machines using
  {Isabelle/HOL}.
\newblock In {\em Proc. 27th Intl. Conf. on Engineering of Complex Computer
  Systems (ICECCS)}. IEEE, June 2023.
\newblock \href {https://doi.org/10.1109/ICECCS59891.2023.00029}
  {\path{doi:10.1109/ICECCS59891.2023.00029}}.

\bibitem{Ye2024-ITreeRoboChart}
K.~Ye, S.~Foster, and J.~Woodcock.
\newblock Formally verified animation for {RoboChart} using interaction trees.
\newblock {\em Journal of Logical and Algebraic Methods in Programming}, 137,
  February 2024.

\bibitem{ZHHZ20}
Yannick Zakowski, Paul He, Chung-Kil Hur, and Steve Zdancewic.
\newblock An equational theory for weak bisimulation via generalized
  parameterized coinduction.
\newblock In {\em Proc. 9th ACM SIGPLAN International Conference on Certified
  Programs and Proofs (CPP)}, 2020.
\newblock \href {https://doi.org/10.1145/3372885.3373813}
  {\path{doi:10.1145/3372885.3373813}}.

\bibitem{ZZF20}
Vadim Zaliva, Ilia Zaichuk, and Franz Franchetti.
\newblock Verified translation between purely functional and imperative domain
  specific languages in {HELIX}.
\newblock In {\em Proc. 12th International Conference on Verified Software:
  Theories, Tools, Experiments (VSTTE)}, 2020.

\bibitem{ZHK+21}
Hengchu Zhang, Wolf Honor\'{e}, Nicolas Koh, Yao Li, Yishuai Li, Li-Yao Xia,
  Lennart Beringer, William Mansky, Benjamin~C. Pierce, and Steve Zdancewic.
\newblock Verifying an {HTTP} key-value server with {Interaction} {Trees} and
  {VST}.
\newblock In {\em Proc. 12th International Conference on Interactive Theorem
  Proving (ITP)}, 2021.
\newblock \href {https://doi.org/10.4230/LIPIcs.ITP.2021.32}
  {\path{doi:10.4230/LIPIcs.ITP.2021.32}}.

\end{thebibliography}

\end{document}